\catcode`\@=11
\newif\if@fewtab\@fewtabtrue
{\count255=\time\divide\count255 by 60
\xdef\hourmin{\number\count255}
\multiply\count255 by-60\advance\count255 by\time
\xdef\hourmin{\hourmin:\ifnum\count255<10 0\fi\the\count255}}
\def\ps@draft{\let\@mkboth\@gobbletwo
    \def\@oddfoot{\hbox to 7 cm{\tiny \versionno
       \hfil}\hskip -7cm\hfil\rm\thepage \hfil {\tiny\draftdate}}
    \def\@oddhead{}
    \def\@evenhead{}\let\@evenfoot\@oddfoot}
\def\draftdate{\number\month/\number\day/\number\year\ \ \ \hourmin }

\def\citen#1{\if@filesw \immediate\write \@auxout {\string\citation{#1}}\fi%
\@tempcntb\m@ne \let\@h@ld\relax \def\@citea{}%
\@for \@citeb:=#1\do {\@ifundefined {b@\@citeb}%
    {\@h@ld\@citea\@tempcntb\m@ne{\bf ?}%
    \@warning {Citation `\@citeb ' on page \thepage \space undefined}}%
    {\@tempcnta\@tempcntb \advance\@tempcnta\@ne
    \setbox\z@\hbox\bgroup\ifcat0\csname b@\@citeb \endcsname \relax
    \egroup \@tempcntb\number\csname b@\@citeb \endcsname \relax
    \else \egroup \@tempcntb\m@ne \fi \ifnum\@tempcnta=\@tempcntb
    \ifx\@h@ld\relax \edef \@h@ld{\@citea\csname b@\@citeb\endcsname}%
    \else \edef\@h@ld{\hbox{--}\penalty\@highpenalty
    \csname b@\@citeb\endcsname}\fi
    \else \@h@ld\@citea\csname b@\@citeb \endcsname \let\@h@ld\relax \fi}%
\def\@citea{,\penalty\@highpenalty\hskip.13em plus.13em minus.13em}}\@h@ld}
\def\@citex[#1]#2{\@cite{\citen{#2}}{#1}}%
\def\@cite#1#2{\leavevmode\unskip\ifnum\lastpenalty=\z@\penalty\@highpenalty\fi%
  \ [{\multiply\@highpenalty 3 #1%
  \if@tempswa,\penalty\@highpenalty\ #2\fi}]}   %
\makeatother 
\catcode`\@=12

\newcommand\almtps[6]{K^{#1#2#3}_{#4#5#6}}
\newcommand\Almtps[6]{\mathcal K^{#1#2#3}_{#4#5#6}}
\newcommand\AlmtpsO[6]{\widehat{\mathcal K}^{#1#2#3}_{#4#5#6}}
\def\arc           {\ell}
\def\arcS          {\ensuremath{\arc_S}}
\def\arcT          {\ensuremath{\arc_\T}}
\def\bb            {{\mathscr B}}
\def\be            {\begin{equation}}
\def\bearl         {\begin{array}{l}}
\def\bearll        {\begin{array}{ll}}
\def\blue          {\color{mydarkblue}}

\def\Bnii          {\ensuremath{B(U_{\ia_1},U_{\ib_1},...\,,U_{\ia_n},U_{\ib_n})}}
\def\Bnij          {\ensuremath{B(U_{\ia_1},U_{\ja_1},...\,,U_{\ia_n},U_{\ja_n})}}

\def\BOnii         {\ensuremath{B_\circ(U_{\ia_1},U_{\ib_1},...\,,U_{\ia_n},U_{\ib_n})}}
\def\BOnij         {\ensuremath{B_\circ(U_{\ia_1},U_{\ja_1},...\,,U_{\ia_n},U_{\ja_n})}}

\def\BOnn          {\ensuremath{B_\circ(U_{\ia_1},...\,,U_{\ia_{2n}})}}
\def\C             {\ensuremath{\mathcal C}}
\def\CA            {\ensuremath{\mathscr A}}
\newcommand\CAC[6] {C_{#1#2,#3#4}^{~~~#5#6}}
\newcommand\CACl[6]{C_{#1#2,#3#4,#5#6}}
\def\cbulk         {c^{\text{bulk}}}

\def\cir           {\,{\circ}\,}

\def\Cob           {\ensuremath{\mathcal C\hspace{-1.1pt}\mbox{\sl ob}}}
\def\complex       {{\ensuremath{\mathbbm C}}}

\newcommand\coronebd[2]{c(#1;#2)}
\newcommand\coronebD[2]{b(#1;#2)} 
\newcommand\ctpd[3]{\ensuremath{c(#1,#2;#3)}}
\newcommand\ctpD[3]{\ensuremath{b(#1,#2;#3)}}
\newcommand\ctpdc[6]{\ctpd{#1}{#2}{#3}_{#4,#5#6}}
\newcommand\ctpdo[6]{\ctpd{#1}{#2}{#3}_{#4}}
\newcommand\ctpdO[6]{\ctpD{#1}{#2}{#3}_{#4}}
\def\CX            {\ensuremath{c_\X}}
\def\dim           {\mathrm{dim}}
\def\dimc          {\dim_\complex}
\def\dsty          {\displaystyle }
\def\ee            {\end{equation}}

\def\eear          {\end{array}}
\def\End           {\mathrm{End}}
\def\eps           {\varepsilon}
\def\eq            {\,{=}\,}
\newcommand\erf[1] {(\ref{#1})}
\newcommand\F[9]   {{\sf F}_{\!{\sss#6}#4{\sss#7},{\sss#8}#5{\sss#9}}
                   ^{\,({#1}\,{#2})\,{#3}}}
\newcommand\FA[9]  {\mbox{\sf F}[A]^{(#1#2#3#4)}_{#5#6#7,#8#9}}
\def\FF            {{\sf F}}
\def\FFA           {{\sf F}$[A]$}
\def\findim        {fi\-ni\-te-di\-men\-si\-o\-nal}
\newcommand\fourpb[7]{\bb(#1#2#3#4)_{#5,#6#7}}

\def\GammaM        {\ensuremath{\Gamma_{\!\!\M}}}

\def\gree          {\color{ForestGreen}}
\def\hom           {\mathrm{Hom}}
\newcommand\Hom[2] {\ensuremath{\hom(#1,#2)}}

\def\HomA          {\hom_{\!A}}
\newcommand\Homaa[2]{\ensuremath{\HomAA(#1,#2)}}
\def\HomAA         {\hom_{\!A|A}}

\def\I             {\ensuremath{\mathcal I}}
\def\ia            {{\ensuremath{\imath}}}
\def\ib            {{\ensuremath{{\bar\imath}}}}
\def\id            {\mbox{\sl id}}

\newcommand\IFAAnumb[5]{\varpi^{#1#2#3}_{#4#5}}

\def\imbed         {\mbox{\sc i}}  
\def\iN            {\,{\in}\,}

\def\Inv           {^{-1}}
\def\itx           {\item[\raisebox{.08em}{\rule{.44em}{.44em}}]}
\def\ja            {{\ensuremath{\jmath}}}
\def\jb            {{\ensuremath{{\bar\jmath}}}}
\def\kb            {\ensuremath{{\bar k}}}
\newcommand\labl[1]{\label{#1}\ee}

\def\longrsqarrow  {\scalebox{.38}{\includegraphics{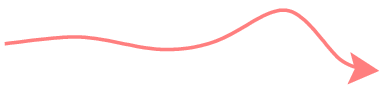}}}
\def\lsqarrow      {\scalebox{.38}{\includegraphics{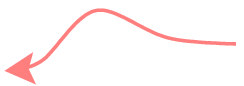}}}

\def\M             {{\mathscr M}}
\def\Mb            {\overline{\mathscr M}}
\def\MX            {\ensuremath{{\mathscr M}_{\rm X}}}
\newcommand\N[3]   {{N_{#1#2}}^{\!\!#3}}
\def\nl            {{\circ}}
\newcommand\nxl[1] {\\[#1mm]}
\newcommand\Nxl[1] {\\[-1.3em]\\[#1mm]}
\def\nxtp          {\nxl{3}\raisebox{.08em}{\rule{.41em}{.41em}}~\,}

\def\one           {{\ensuremath{\mathbf 1}}}
\def\onedim        {one-dimen\-sio\-nal}
\newcommand\onefsc[3]{b^{#1,#2}_{#3}}
\def\orr           {\mbox{\sl or\/}}
\newcommand\ot[1]  {\,{\otimes^{#1}}\,}
\newcommand\oT[1]  {\,{\otimes^{#1}}}

\def\oti           {\,{\otimes}\,}
\def\otic          {\,{\otimes_\complex}\,}

\def\otim          {\,{\otimes^-}\,}
\def\otip          {\,{\otimes^+}}
\newcommand\pb[1]  {\sse$\blue #1$}
\newcommand\pg[1]  {\sse$\gree #1$}
\def\Phi           {\varPhi}
\newcommand\PHi[1] {\Phi_{\!#1}}
\def\proj          {\pi_\circ}
\def\Psi           {\varPsi}
\newcommand\pl[1]  {\sse$#1$}
\def\qb            {{\bar q}}

\def\qquand        {\qquad{\rm and}\qquad}
\newcommand\R[5]   {{\sf R}^{(#1\,#2)#3}_{#4\,#5}}
\def\rep           {representation}

\def\rsqarrow      {\scalebox{.38}{\includegraphics{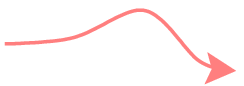}}}
\def\sse           {\scriptsize}
\def\sss           {\scriptscriptstyle}

\def\T             {{\mathscr T}}
\def\tftC          {\ensuremath{\mbox{\tt tft}_\C}}
\newcommand\tildeI[4]{K^{#1#2}_{#3#4}}
\def\Times         {\,{\times}\,}
\def\To            {\,{\to}\,} 
\def\vio           {\color{DarkViolet}}
\def\X             {{\rm X}}
\def\Xh            {\ensuremath{{\widehat{\X}}}}
\def\YS            {\ensuremath{{\mathrm Y}_{\!S}}}
\def\YT            {\ensuremath{{\mathrm Y}_{\!\T}}}
\def\Z             {{\mathrm Z}}
\def\zet           {\ensuremath{\mathbb Z}}

\newcommand\includepicclal[1] {{\begin{picture}(0,0)(0,0)
                   \scalebox{.31}{\includegraphics{imgs/pic_clal_#1.eps}}\end{picture}}}
\newcommand\Includepicclal[1] {{\begin{picture}(0,0)(0,0)
                   \scalebox{.38}{\includegraphics{imgs/pic_clal_#1.eps}}\end{picture}}}
\newcommand\InCludepicclal[1] {{\begin{picture}(0,0)(0,0)
                   \scalebox{.435}{\includegraphics{imgs/pic_clal_#1.eps}}\end{picture}}}
\newcommand\INcludepicclal[1] {{\begin{picture}(0,0)(0,0)
                   \scalebox{.456}{\includegraphics{imgs/pic_clal_#1.eps}}\end{picture}}}
\newcommand\eqpic[4]{\begin{eqnarray}
                   \begin{picture}(#2,#3){}\end{picture}\nonumber\\
                   \raisebox{-#3pt}{ \begin{picture}(#2,#3) #4 \end{picture} }
                   \label{#1} \\~\nonumber \end{eqnarray} }

\documentclass[12pt]{article}
\usepackage{amsthm,latexsym,amsmath,amssymb,amsfonts,color,colordvi,bbm}
\usepackage[mathscr]{eucal}
\usepackage{graphics}
\usepackage{rotating}
\usepackage[pagewise,modulo]{lineno}   
\numberwithin{equation}{section}

\definecolor{DarkViolet} {rgb}  {0.580392,0.000000,0.827450}
\definecolor{ForestGreen}{rgb}  {0.100000,0.408823,0.100000}
\definecolor{green3}     {rgb}  {0.000000,0.803921,0.000000}
\definecolor{mydarkblue} {rgb}  {0.282352,0.239215,0.803921}
\definecolor{OrangeRed}  {rgb}  {1.000000,0.270588,0.000000}
\definecolor{red3}       {rgb}  {0.803921,0.000000,0.000000}
\def\hom           {\mathrm{Hom}} 
\begin{document}
\begin{flushright}
   {\sf Hamburger$\;$Beitr\"age$\;$zur$\;$Mathematik$\;$Nr.$\;$336}\\[2mm]
   July 2009
\end{flushright}
\vskip 2.5em
\begin{center}\Large
THE THREE-DIMENSIONAL ORIGIN \\ OF THE CLASSIFYING ALGEBRA
\end{center}\vskip 2.1em
\begin{center}
  ~J\"urgen Fuchs\,$^{\,a}$,~
  ~Christoph Schweigert\,$^{\,b}$,~
  ~Carl Stigner\,$^{\,a}$
\end{center}

\vskip 9mm

\begin{center}\it$^a$
  Teoretisk fysik, \ Karlstads Universitet\\
  Universitetsgatan 21, \ S\,--\,651\,88\, Karlstad
\end{center}
\begin{center}\it$^b$
  Organisationseinheit Mathematik, \ Universit\"at Hamburg\\
  Bereich Algebra und Zahlentheorie\\
  Bundesstra\ss e 55, \ D\,--\,20\,146\, Hamburg
\end{center}
\vskip 3.5em
\noindent{\sc Abstract}
\\[3pt]
It is known that reflection coefficients for bulk fields of a rational
conformal field theory in the presence of an elementary boundary condition
can be obtained as representation matrices of irreducible representations of
the classifying algebra, a semisimple commutative associative complex algebra.
\\
We show how this algebra arises naturally from the three-dimensional geometry
of factorization of correlators of bulk fields on the disk. This allows us 
to derive explicit expressions for the structure constants of the classifying
algebra as invariants of ribbon graphs in the three-manifold $S^2\Times S^1$.
Our result unravels a precise relation between intertwiners of
the action of the mapping class group on spaces of conformal blocks
and boundary conditions in rational conformal field theories.

\newpage

\section{Introduction}

Structure constants of operator product expansions (OPEs) have played an important 
role in shaping our understanding of correlation functions in two-dimensional
conformal field theory. In fact, one approach to conformal field theory has been 
to identify a subset of fundamental correlators -- which are ultimately encoded 
in appropriate OPEs -- from which all other correlators can be obtained by 
sewing. Since a given correlation function can typically be constructed 
from the fundamental correlators by sewing in several distinct ways,
the uniqueness of the correlators imposes various
necessary conditions on the sewing procedure. These conditions are known as
sewing constraints, or factorization constraints \cite{sono2,lewe3,prss3}.
There are actually two different types of factorizations: 
\\
--- \emph{Boundary factorization}, involving a cutting of the world sheet along
    an interval that connects two points on its boundary,
    yields a correlator with two additional insertions of boundary fields. 
\\
--- \emph{Bulk factorization}, for which the cutting is along a circle in the
    interior of the world sheet, yields a correlator with two additional
    bulk field insertions. 

In the guise of associativity of the OPE, constraints from bulk factorization 
have been central in the understanding of the OPE of bulk fields \cite{bepz}. 
Much later it was realized \cite{bppz2,fffs} that the constraints on structure 
constants for boundary fields preserving a given boundary condition have in 
fact a simpler structure and give rise 
to an associative algebra in the tensor category of chiral data.

The OPE coefficients of bulk fields in the presence of a boundary are amenable as 
well. The one-point correlators of bulk fields on a disk, which may be collected 
in so-called boundary states, contain significant information of much interest
for applications, like ground state degeneracies \cite{aflu} or Ramond-Ramond 
charges of string compactifications \cite{bDlr}. Moreover, they provide essential 
information about annulus partition functions and thus encode the spectrum of
boundary fields. Based on an analysis of specific classes of models, it was found 
that the reflection coefficients for bulk fields in the presence of an elementary 
boundary condition can be formulated in terms of representation matrices of 
irreducible representations of a \emph{classifying algebra} \cite{prss3,fuSc5}.\,%
 \footnote{~The same algebra had arisen in the study of integrable
 lattice models \cite{pasq3}.}
In particular, the elementary boundary conditions are in bijection with the
irreducible representations of the classifying algebra. Further, there is evidence 
that the structure constants of the classifying algebra can be expressed in 
terms of traces of intertwiners for the mapping class group action on spaces of 
conformal blocks on the sphere \cite{scfu}, whereby they are related to the 
subbundle structure of the bundles of conformal blocks \cite{fuSc8}. These ideas 
have lead to concrete formulas for operator product coefficients 
\cite{fuSc10,fuSc11,bifs,fhssw,gaRw} with a wide range of uses; 
see e.g.\ \cite{dihs2,hosom,ibsU,kisT} for some applications in string theory.

\medskip

More recently, the {\em TFT approach\/} to the correlation functions of rational
conformal field theories has provided a much more satisfactory understanding of 
RCFT correlators. The main idea of this approach can be summarized as follows. The 
chiral data of a conformal field theory are described by the structure of a modular 
tensor category \cite{mose3,baKir}, and a full local conformal field theory 
based on these chiral data corresponds to (a Morita class of) a symmetric special 
Frobenius algebra in that category \cite{fuRs,fuRs4,fjfrs2,kong6,koRu}. 
A modular tensor category also gives rise to a three-dimensional
topological field theory \cite{retu2}. The TFT approach uses this 
topological field theory to construct the correlators of the local conformal
field theory as invariants of ribbon graphs in three-manifolds with 
boundary. It has been shown \cite{fjfrs} that the so obtained correlators are 
invariant under the mapping class group and obey all factorization constraints. 

The purpose of this paper is to establish the existence of a classifying algebra 
for any RCFT -- an associative commutative algebra \CA\ over the complex numbers
with the property that the homomorphisms given by its irreducible representations 
give the bulk reflection coefficients in the presence of elementary boundary 
conditions. We also show that this algebra is related to traces over intertwiners 
for mapping class group \rep s. These results are obtained with the help of 
manipulations of suitable three-manifolds, making use of the TFT approach and in 
particular of the results of \cite{fjfrs} about factorization. An additional 
ingredient is the identification of a distinguished subspace \BOnii\ of the space 
\Bnii\ of $2n$-point conformal blocks on the sphere that comes with a natural 
projection $\proj{:}~ \Bnii\To \BOnii$, see the picture \erf{basis_B_0} below. 
Such a subspace exists because the modular tensor category has a duality.

It is worth pointing out that the TFT approach allows one to compute arbitrary 
correlation functions of an RCFT. Indeed, together with the chiral symmetry, the 
structural data of the modular tensor category \C\ and of the Frobenius algebra $A$ 
in \C\ determine the RCFT completely. In particular, the boundary conditions of the 
CFT are given by the modules over the algebra $A$. The virtue of the classifying 
algebra is to transfer the problem of doing \rep\ theory for an algebra $A$ in an 
abstract tensor category to the problem of doing \rep\ theory for an algebra \CA\ 
over \complex, keeping at the same time the information about bulk reflection 
coefficients. Once the structure of \CA\ is known, the boundary conditions of the 
CFT can be studied without further reference to the category \C. In particular, 
coefficients of boundary states and thus annulus coefficients are encoded in the 
classifying algebra.

\medskip

Our general strategy is as follows. We compare boundary and bulk factorization 
for correlators of bulk fields on a disk. Such correlators are elements of a space 
of conformal blocks on the sphere. In the TFT approach, they are described by
ribbon graphs in a full three-ball $B^3$.
   
We think about this ball $B^3$ as a quotient of an interval 
bundle over the disk, which is obtained in the following manner. The disk is 
embedded in the equatorial plane of $B^3$. Intervals piercing the interior of 
the disk connect points on the boundary sphere $\partial B^3$ of equal longitude 
and opposite latitude, while the boundary of the disk is connected to the equator 
of the boundary sphere by little intervals in the equatorial plane. This is 
illustrated in the pictures \erf{Sphere_CI}; the first shows the ball $B^3$ 
together with the embedded world sheet (the shaded region) and some of the 
connecting intervals (intervals intersecting the interior $\X{\setminus}\partial\X$ 
of the world sheet shown on the left, and intervals containing a boundary point of
$\X$ shown on the right), while the second picture shows the intersection of
the ball with a vertical plane containing the north and south poles. 
  \eqpic{Sphere_CI} {355} {78} {
  \put(0,0)     {\includepicclal{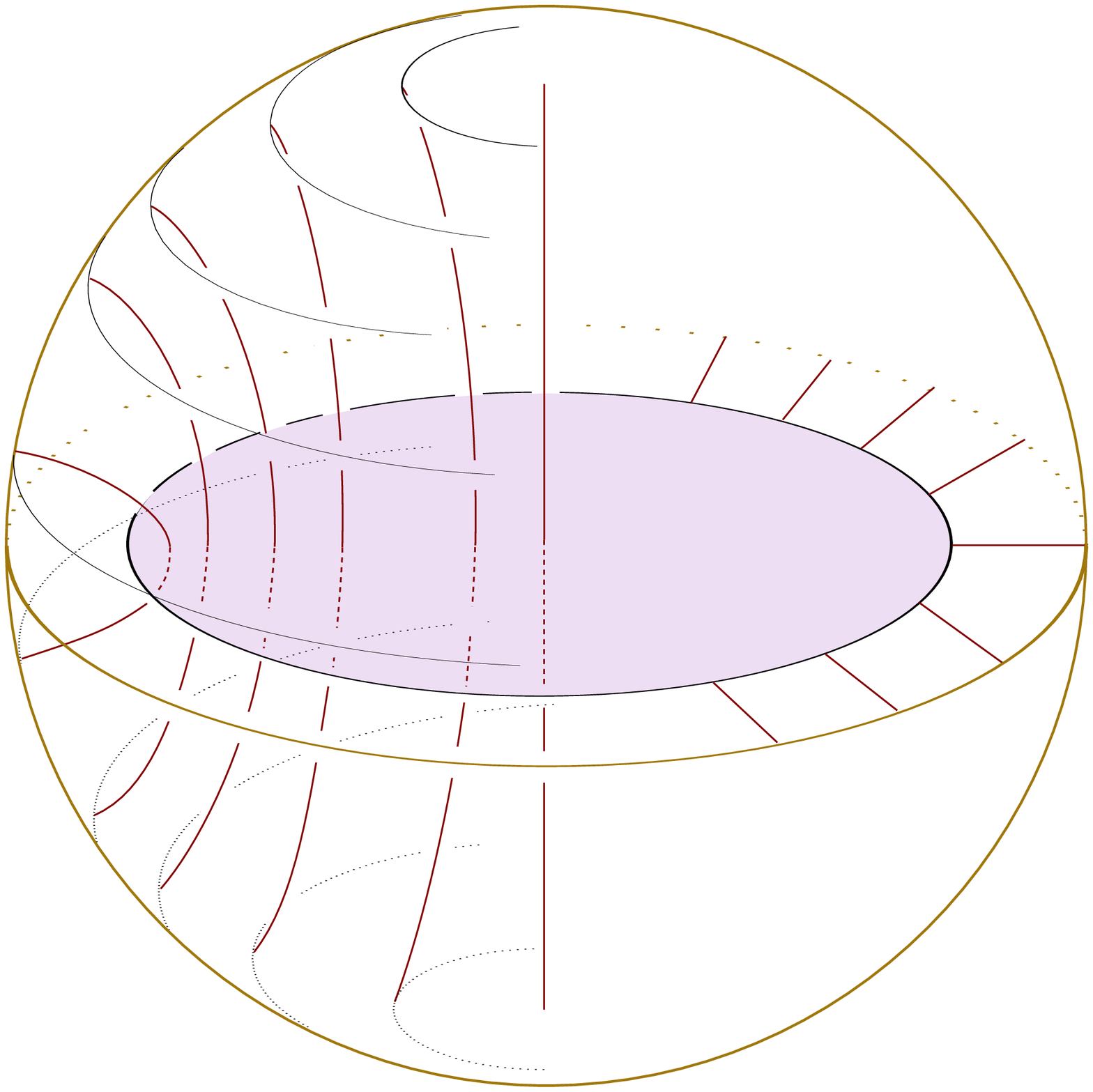}}
  \put(210,0)  {\Includepicclal{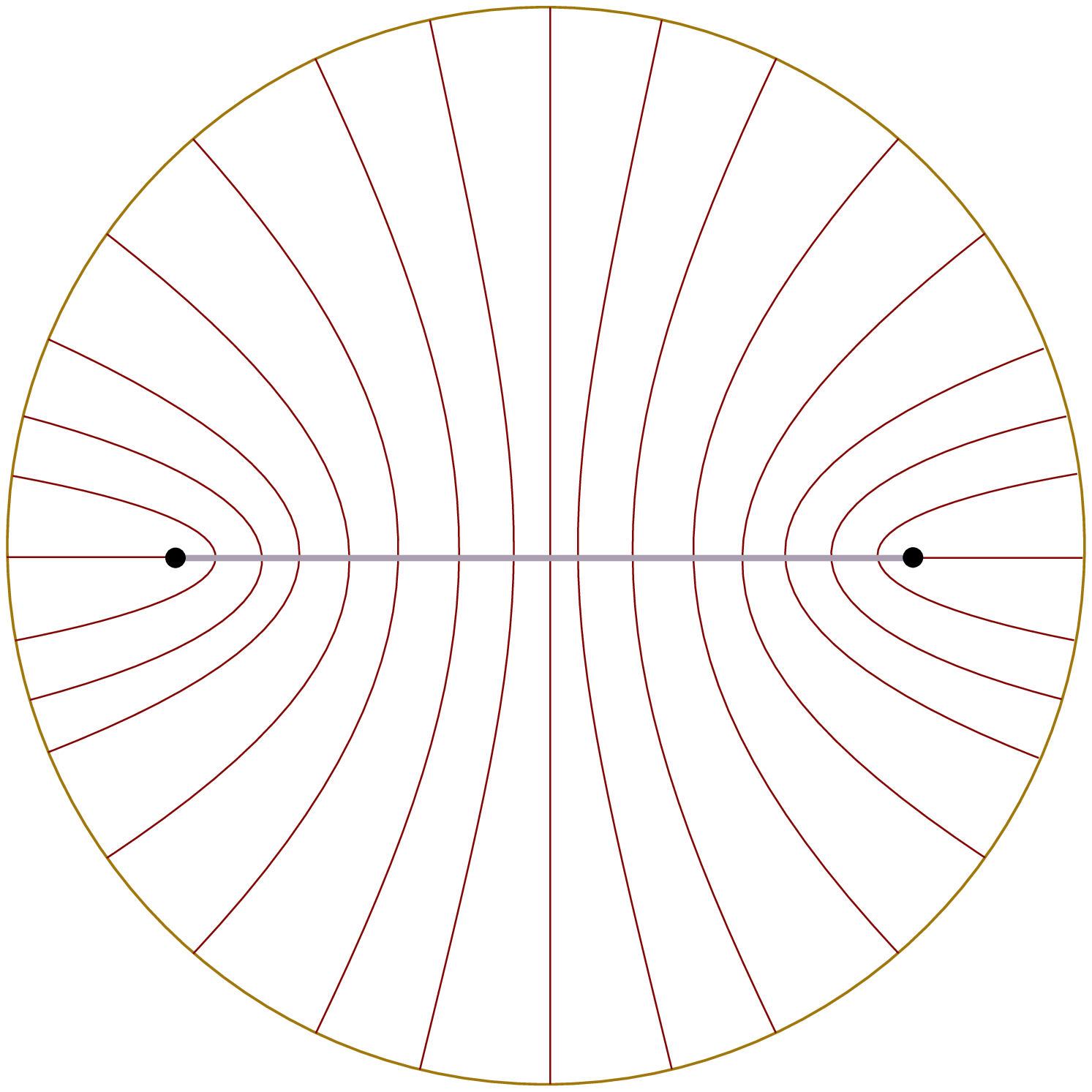}}
 } 

After including, in accordance with the rules (see appendix \ref{app_tft}, or
appendix B of \cite{fjfrs}) of the TFT approach, the appropriate 
ribbon graph in this three-ball, one takes the invariant of the resulting
three-manifold. Thereby the correlator is expressed as an element of the relevant 
space $B$ of conformal blocks on the two-sphere $S^2$. From the correlator one can
obtain the corresponding structure constants by gluing to this ball another full 
ball that contains a standard 
ribbon graph describing an element of the dual basis in the dual space $B^*$.
They are thus given as invariants of ribbon graphs in the three-sphere $S^3$. 

More specifically, we consider a disk with the insertion of an arbitrary number 
$n$ of bulk fields. Performing $n\,{-}\,1$ boundary factorizations and projecting 
on the subspace \BOnn, we find a contribution that is proportional to the product 
$b_{\ia_1}b_{\ia_2}{\cdots}\,b_{\ia_n}$ of $n$ reflection coefficients.
Bulk factorization is more subtle. Indeed, the geometry of three-manifolds arising 
in the TFT approach reveals that such a factorization involves the interchange of 
a contractible and a non-contractible cycle in a full torus, and thus a modular 
$S$-transformation. This is sometimes expressed by saying that the $S$-operation 
interchanges the open and closed channels of the correlator.
Cutting out a solid torus from $S^3$ and gluing it back with an $S$-transformation
yields the closed manifold $S^2\Times S^1$. 

The invariant of a ribbon graph in $S^2\Times S^1$ is the trace of an
endomorphism of the relevant space of conformal blocks on $S^2$.
This is the geometric origin of the fact that traces on spaces of conformal blocks 
on the sphere enter in OPE coefficients. Moreover, after the bulk factorization
a single reflection coefficient $b_k$ remains as a factor in the expression for the 
correlator. By extracting its coefficient in the distinguished subspace \BOnii, one 
arrives at a ribbon graph whose invariant constitutes an intertwiner for the action of 
the mapping class group on the space $B(U_{\ia_1},U_{\ia_2},...\,,U_{\ia_n},U_{k})$
of conformal $n{+}1$-point blocks on the sphere. For $n\eq2$ this intertwiner is 
the invariant of the
following ribbon graph embedded in the three-manifold $S^2\Times [-1,1]$:
  \eqpic{I_qdelta} {240} {116} {
  \put(-4,118)  {$ \AlmtpsO \ia\ja k\alpha\beta\gamma ~= $}
  \put(65,0)    {\Includepicclal{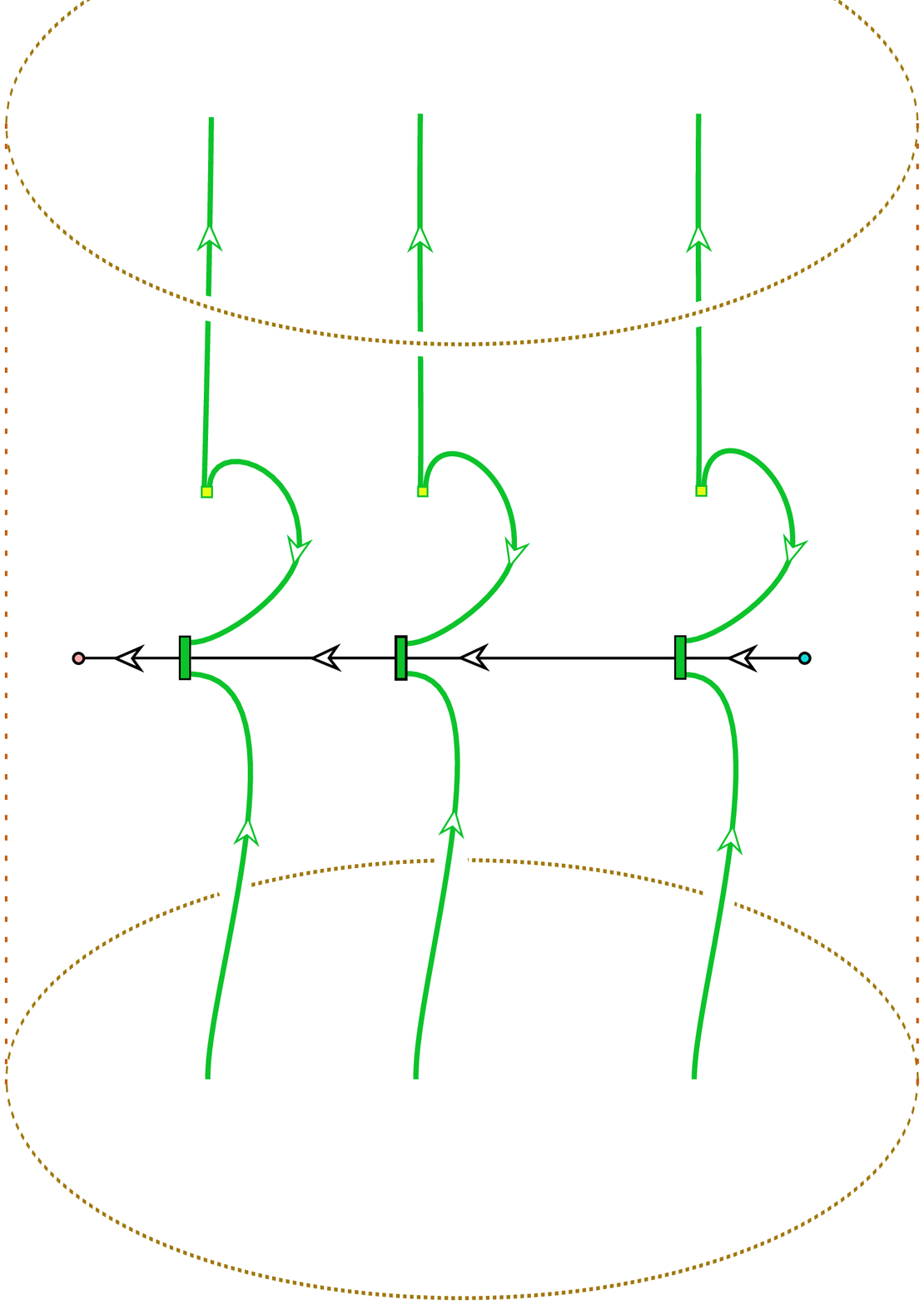}}
   \put(127,-45){
  \put(65,90)    {\pg k}
  \put(78,193)   {\pg \kb}
  \put(56,149)   {\pl {\phi_{\!\gamma}^{}}}
  \put(-11,193)  {\pg \ib}
  \put(29,193)   {\pg \jb}
  \put(-22.5,90) {\pg \ia}
  \put(15,90)    {\pg \ja}
  \put(5,149)    {\pl {\phi_{\!\beta}^{}}}
  \put(-33,149)  {\pl {\phi_{\!\alpha}^{}}}
  \put(38,163)   {\pl A}
 } } 
Here the interval $[-1,1]$ is displayed vertically, while the factor $S^2$ of the 
three-manifold corresponds to horizontal planes (for more details see section 
\ref{s2}). Also, the morphism $\phi_\alpha$ is an element of the space of bulk fields
with chiral labels $\ia$ and $\ib$, and analogously for $\phi_\beta$ and $\phi_\gamma$; 
for any choice of these bulk fields one deals with an intertwiner of the mapping class 
group action on $B(U_i,U_j,U_k)$. Note that nontriviality of this intertwiner 
implies reducibility of the mapping class group \rep\ (see \cite{anfj,anfj2}).

Comparing the outcome of boundary and bulk factorization, one arrives in the
case $n\eq 2$ at a formula for the structure constants of the classifying algebra.
The result is then used to show that the classifying algebra is a semisimple
commutative associative algebra over \complex. This provides precise statements 
and proofs of the existing conjectures about the classifying algebra and of the 
relation between boundary states and traces on spaces of conformal blocks.
In the Cardy case, in which the torus partition function is given by charge
conjugation, the classifying algebra reduces to the Verlinde algebra.

\medskip

This paper is organized as follows.
In sections \ref{s1} and \ref{s2} we describe the boundary and bulk factorization,
respectively, for $n$-point correlators on the disk. Via boundary factorization, 
these are expressed as a product of $n$ one-point correlators on the disk, see 
formula \erf{n-bndfact}, while via bulk factorization one arrives at an expression 
involving only a single one-point correlator as a factor, see \erf{ctpdc_2}. By 
equating these results we arrive, in section \ref{s3}, at the classifying algebra 
\CA. As a vector space \CA\ consists of those bulk fields which have non-vanishing 
one-point correlator on the disk. In a basis $\{\phi^{\ia,\alpha}\}$ of this space
the structure constants are given, up to a simple prefactor (see formula \erf{CLA}),
by the invariant of the ribbon graph $\Almtps \ia\ja k\alpha\beta\gamma$ that is 
obtained by taking the trace over the intertwiner in the ribbon graph \erf{I_qdelta}. 
Taking the trace means that top and bottom of the picture \erf{I_qdelta} are 
identified so that each of the $i$-, $j$- and $k$-ribbons forms a loop and one 
deals with a ribbon graph embedded in the closed three-manifold $S^2\Times S^1$, 
with $S^1$ running vertically. Based on this result for the structure constants we 
show that the algebra \CA\ is commutative and associative, has a unit, and is 
semisimple. In particular, the \onedim\ \CA-\rep s furnished by the elementary
boundary conditions exhaust the (isomorphism classes of) irreducible \CA-\rep s.

Throughout this paper methods and results from the TFT approach are used freely.
Some pertinent information about this approach is collected in the appendix.


\section{Boundary factorization}\label{s1}

The world sheet of our interest is an oriented disk with boundary condition 
$M$ and with, for now, an arbitrary number $n$ of bulk field insertions. The 
boundary condition $M$ is also arbitrary, but kept fixed. After equation 
\erf{res_boundfact_anyM} we will, however, restrict to the case that $M$ is an 
{\em elementary\/} boundary condition (or in algebraic terms, that $M$ is a 
simple $A$-module). The goal is to 
identify quantities -- the structure constants of the classifying algebra -- 
in which the dependence on the choice of $M$ has dropped out. 

We denote the world sheet by $\X\,{\equiv}\,\X(\PHi{\alpha_1},\PHi{\alpha_2},
...\,,\PHi{\alpha_n};M)$, where each of the $n$ field insertions
$\PHi{\alpha_1},\,\PHi{\alpha_2},...\,,\PHi{\alpha_n}$ is a bulk field
  \be
  \PHi\alpha \equiv (U_\ia,U_\ja,\phi_\alpha)
  \labl{Phi_PhiUU}
with $U_\ia$ and $U_\ja$ simple objects of \C\ and $\phi_\alpha$ an element
of the space of $\Homaa{U_\ia\otip A\otim U_\ja}A$ of bimodule morphisms, as 
described in appendix \ref{app_fields}. The 
correlator \ctpd{\PHi{\alpha_1}}{\PHi{\alpha_2},...\,,\PHi{\alpha_n}}{M}
for this world sheet can be expressed, according to
  \be
  \ctpd{\PHi{\alpha_1}}{\PHi{\alpha_2},...\,,\PHi{\alpha_n}}{M} = Z(\MX)\,1 \,,
  \labl{ctpd1n}
as the invariant of a cobordism \MX. The construction of this cobordism, the 
\emph{connecting manifold} of \X, is summarized in appendix \ref{app_tft}.
The connecting manifold is to be regarded as a cobordism from $\emptyset$ to 
$\partial\MX$, and $\partial\MX\eq\Xh$ is the \emph{double} of the world sheet \X.
Accordingly $Z(\MX)$ is a linear map from $\complex\eq Z(\emptyset)$ to the space
$Z(\Xh)$ of conformal blocks; in \erf{ctpd1n} this map is applied to
the number $1\iN\complex$.

In the case at hand, i.e.\ for \X\ a disk, the prescription of appendix \ref{app_tft} 
yields for the double \Xh\ a two-sphere, and for the connecting manifold \MX\ a full 
three-ball. This ball contains a ribbon graph, as shown in the following picture:
\eqpic{n_point_cor} {250} {72} {
\setlength\unitlength{1.2pt}
  \put(-43,59)    {$ \MX ~= $}
  \put(0,0)   { {\INcludepicclal{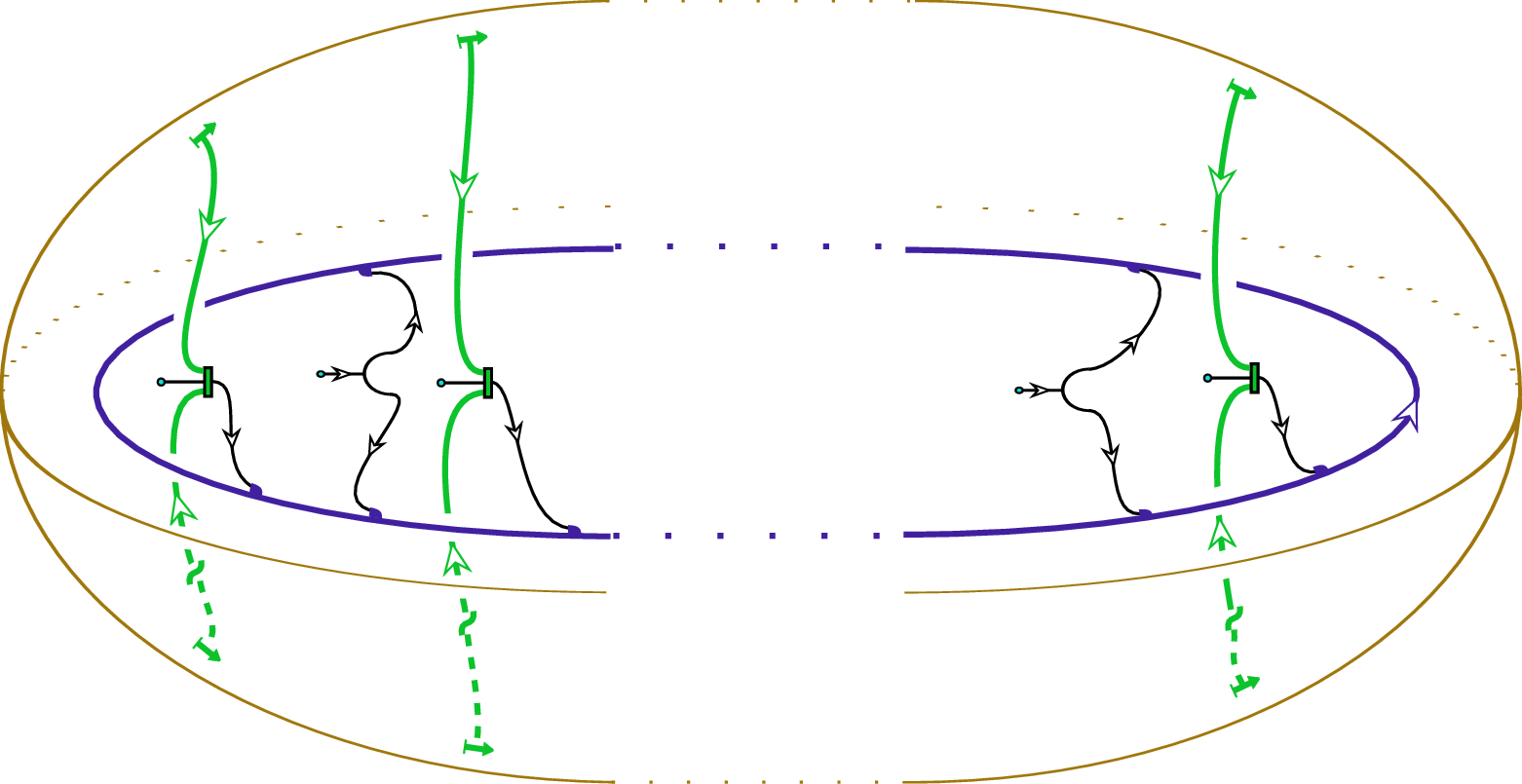}}
  \put(31,68)     {\pl{\phi_{\alpha_1}}}
  \put(75,68)     {\pl{\phi_{\alpha_2}}}
  \put(196,69)    {\pl{\phi_{\alpha_n}}}
  \put(29,10)     {\pg{\ja_1}}
  \put(28,112)    {\pg{\ia_1}}
  \put(72,-4)     {\pg{\ja_2}}
  \put(71,126)    {\pg{\ia_2}}
  \put(193,3)     {\pg{\ja_n}}
  \put(195,118)   {\pg{\ia_n}}
  \put(148,42.5)  {\pb M}
  \put(37.6,53)   {\pl A}
  \put(59.2,50)   {\pl A}
  \put(83.5,53)   {\pl A}
  \put(168.4,50)  {\pl A}
  \put(204.6,54.7){\pl A}
  } }
Here the vertical ribbons are labeled by the objects $U_{\ia_\ell}$ and
$U_{\ja_\ell}$ corresponding to the bulk fields and the annular horizontal ribbon 
is labeled by the $A$-module $M$, while the other horizontal ribbons, which come
from a triangulation of the world sheet, are labeled by the Frobenius algebra $A$
(for more details, compare section 6.3 of \cite{fuRs10}).

By examining a cutting of \X\ along an interval that connects two points on the 
boundary of the disk, in such a way that the resulting two disks contain $\ell$ 
and $n{-}\ell$ bulk insertions, respectively, one obtains the expression 
  \be
  \ctpd{\PHi{\alpha_1}}{\PHi{\alpha_2},...,\PHi{\alpha_n}}{M}
  = \sum_{q\in\I}\sum_{\gamma,\delta}\dim(U_q)\,({c^{\text{bnd}}_{M,M,q}})
  \Inv_{\;\delta\gamma}\, Z(\M_{q\gamma\delta})\,1 \,,
  \ee
for the correlator \erf{ctpd1n} (see formulas (2.37), (2.38) and (4.22) of 
\cite{fjfrs}). Here $c^{\text{bnd}}$ is the boundary two-point function 
\erf{bnd_2pfu}, while for $q$ running over the label set $\I$ for the isomorphism 
classes of simple objects of \C, and for $\gamma$ and $\delta$ running over 
bases of the morphism spaces $\HomA(M\oti U_q,M)$ and $\HomA(M\oti U_\qb,M)$, 
respectively, the cobordism $\M_{q\gamma\delta}$ is given by
  \eqpic{M_qgammadelta} {210} {60} {
  \put(-113,68)   {$ \M_{q\gamma\delta} ~= $}
  \put(-60,10)   { {\Includepicclal{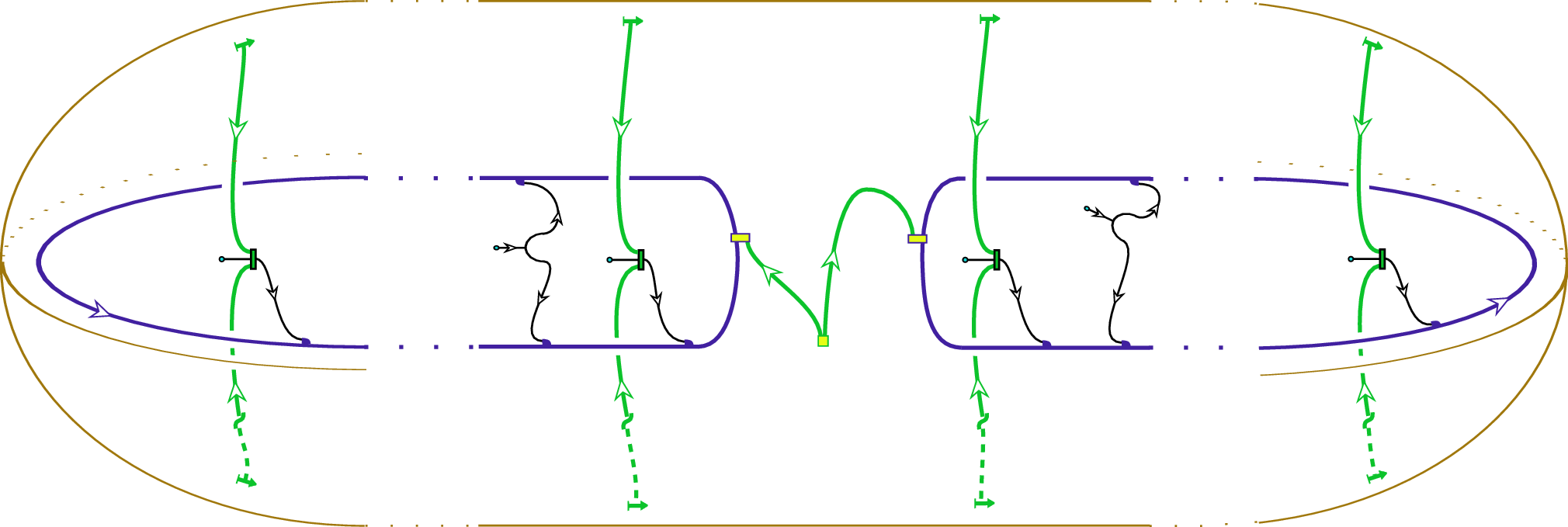}}
  \put(60.3,67)	{\pl{\phi_{\alpha_1}}}
  \put(151,67)	{\pl{\phi_{\alpha_\ell}}}
  \put(235,67)	{\pl{\phi_{\alpha_{\ell+1}}}}
  \put(326,67)	{\pl{\phi_{\alpha_{n}}}} 
  \put(54,124)	{\pg{\ia_1}}
  \put(54,-3)	{\pg{\ja_1}}
  \put(146,128)	{\pg{\ia_\ell}}
  \put(147,-6)	{\pg{\ja_\ell}}
  \put(226,128)	{\pg{\ia_{\ell+1}}}
  \put(225,-6)	{\pg{\ja_{\ell+1}}}
  \put(320,121)	{\pg{\ia_n}}
  \put(321,1)	{\pg{\ja_n}}
  \put(173,73)	{\pl{\psi_\gamma}}
  \put(206,63)	{\pl{\psi_\delta}}
  \put(181,50)	{\pg q}
  \put(197,50)	{\pg\qb}
  \put(347,76)	{\pb M}
  \put(14,76)	{\pb M}
 } }

By construction, both $Z(\MX)\,1$ and $Z(\M_{q\gamma\delta})\,1$ are elements of 
the space $B(U_{\ia_1},U_{\ja_1},...\,,U_{\ia_n},U_{\ja_n})$ of $2n$-point chiral 
blocks on the sphere. A standard basis of this space is given by (compare appendix 
\ref{app_chiral}) the vectors $\bb(\ia_1,\ja_1,...,\ia_n,\ja_n)_{p_1p_2...p_{2n-3},
\alpha_1\alpha_2...\alpha_{2n-2}}$ which are the invariants of the ribbon graphs
  \eqpic{npbasis}{455}{79}{
  \put(-10,85) {$ \widehat\bb(\ia_1,\ja_1,...,\ia_n,\ja_n)_{p_1p_2...p_{2n-3},
                  \alpha_1\alpha_2...\alpha_{2n-2}}~:= $}
  \put(205,0){ \put(0,0){\Includepicclal{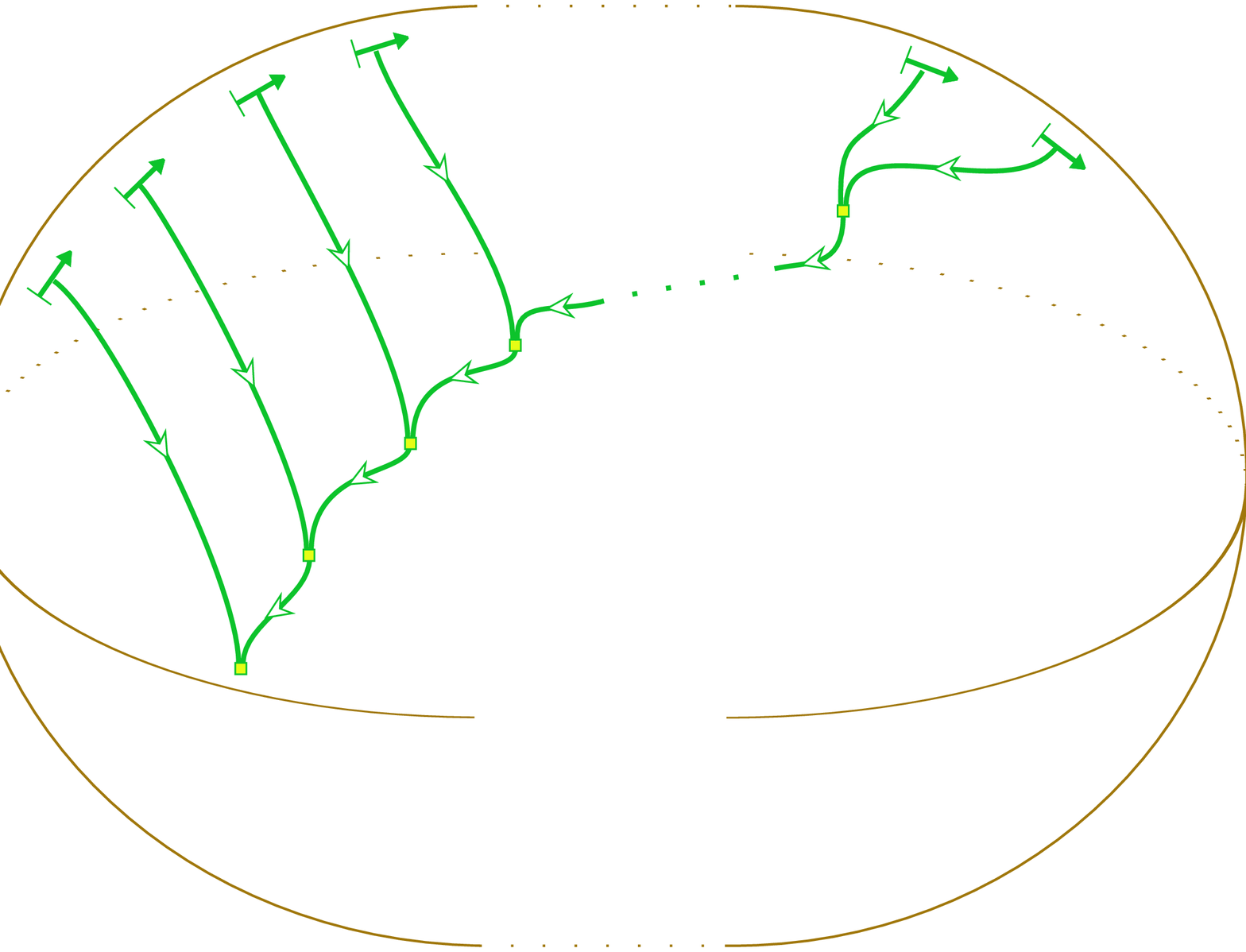}}
  \put(4,133)       {\pg {\ia_1}}
  \put(22,154.8)    {\pg {\ja_1^{}}}
  \put(48,171)      {\pg {\ia_2}}
  \put(73,179.6)    {\pg {\ja_2^{}}}
  \put(181,173)     {\pg {\ia_n}}
  \put(211,156.2)   {\pg {\ja_n^{}}}
  \put(62,59)       {\pg {\ib_1}}
  \put(76,82)       {\pg {p_1^{}}}
  \put(95,101)      {\pg {p_2^{}}}
  \put(164,126)     {\pg {p_{2n-3}^{}}}
  \put(67,72)       {\pl {\alpha_1^{}}}
  \put(86.5,92.5)   {\pl {\alpha_2^{}}}
  \put(106,111.4)   {\pl {\alpha_3^{}}}
  \put(167,137)     {\pl {\alpha_{2n-2}^{}}}
  } }
Here $\alpha_\ell$ labels a basis of the appropriate morphism space, namely 
$\Hom{U_{p_{2s}^{}}{\otimes}\, U_{\ia_{s+1}}}{U_{p_{2s-1}^{}}}$ for $\alpha_{2s}^{}$, and
$\Hom{U_{p_{2s-1}^{}}{\otimes}\, U_{\ja_s}}{U_{p_{2s-2}^{}}}$ for $\alpha_{2s-1}^{}$,
where $s\eq1,2,...\,,n{-}1$, and where we identify $p_0\,{\equiv}\,\ib_1$ and 
$p_{2n-2}\,{\equiv}\,\ja_n$. A vector in $B(U_{\ia_1},U_{\ja_1},...\,,U_{\ia_n},U_{\ja_n})$ can
be expanded with respect to the standard basis \erf{npbasis} by evaluating on it
an element of the dual basis of the
dual space. This amounts to gluing with another three-ball with appropriate ribbon graph
and results in a ribbon graph in the closed three-manifold $S^3$, the invariant of which 
gives the coefficient with respect to the chosen basis.

\medskip

In the following we concentrate on the case $n\eq2$, the case of arbitrary $n$ being
analogous. For $n\eq2$ we denote the two bulk fields by
$\PHi\alpha\,{\equiv}\,(\ia,\ja,\phi_\alpha)$ and 
$\PHi\beta\,{\equiv}\,(k,l,\phi_\beta)$, and expand the correlator as
  \be
  \ctpd{\PHi\alpha}{\PHi\beta}{M} = \sum_{p\in\I}\sum_{\eps_1,\eps_2}
  \ctpdc{\PHi\alpha}{\PHi\beta}{M}p{\eps_1}{\eps_2}\,
  Z(\fourpb \ja\ia lkp{\eps_1}{\eps_2}) \,,
  \labl{C_exp}
where $\fourpb \ja\ia lkp{\eps_1}{\eps_2}$ are basis elements of the form \erf{npbasis} 
for the case of four-point blocks. By evaluating an element of the dual basis on both 
sides of this equality, one obtains the expression
  \be
  \ctpdc{\PHi\alpha}{\PHi\beta}{M}p{\eps_1}{\eps_2}
  = \sum_{q\in\I}\sum_{\gamma,\delta} \dim(U_q)\,
  ({c^{\text{bnd}}_{M,M,q}})\Inv_{\;\delta\gamma}\,
  Z(\M^{q,\gamma\delta}_{p,\eps_1\eps_2})
  \labl{ctpdc_p}
for the coefficients in \erf{C_exp}, with $\M^{q,\gamma\delta}_{p,\eps_1\eps_2}$
the following ribbon graph in $S^3$:
  \eqpic{C_disc_boundfc}{265}{78}{
  \put(0,80)    {$ \M^{q,\gamma\delta}_{p,\eps_1\eps_2} ~= $}
  \put(70,0)   {  {\Includepicclal{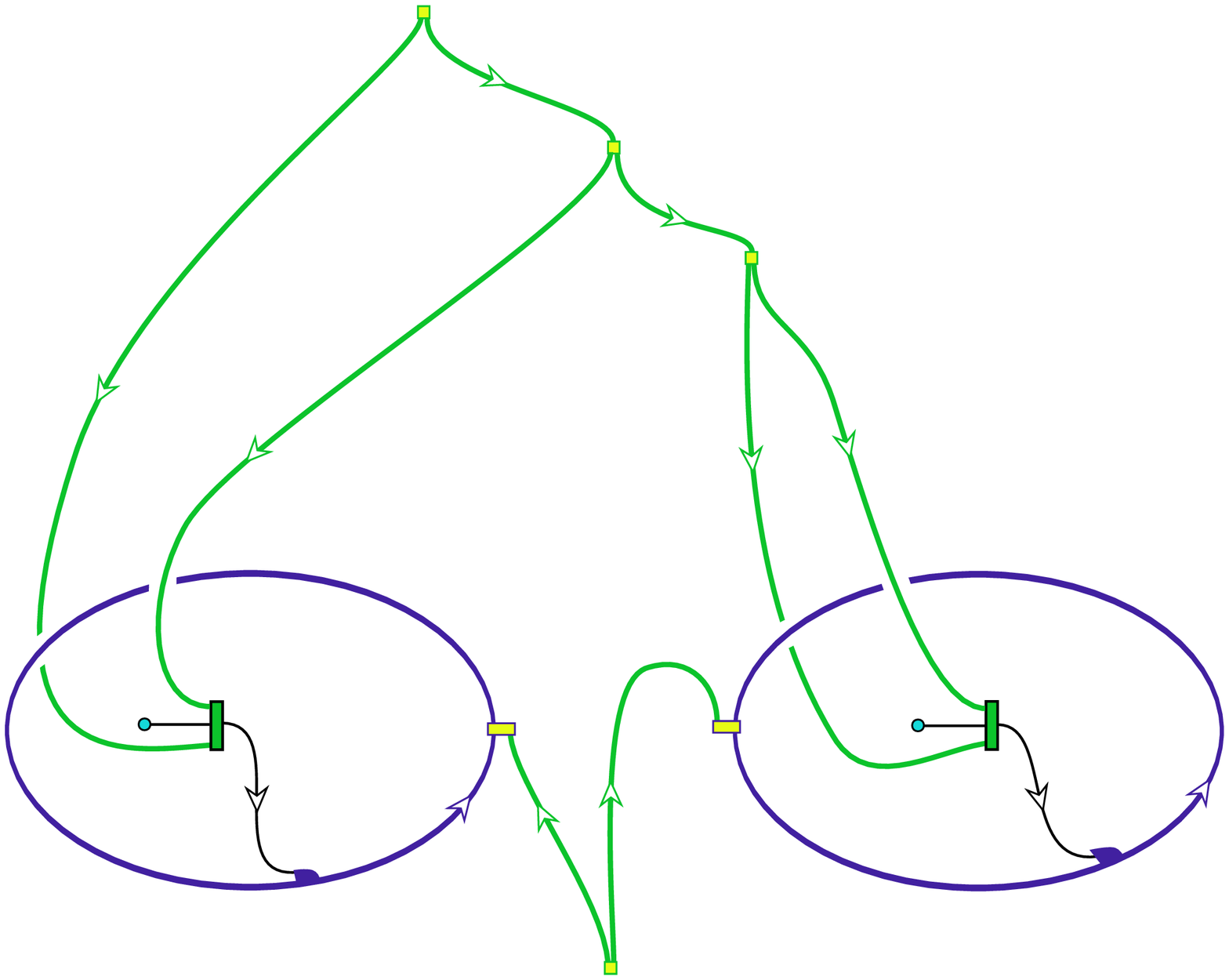}}
  \put(39.7,81)   {\pg \ia}
  \put(155.7,83)  {\pg k}
  \put(12.3,82)   {\pg \ja}
  \put(135.9,82)  {\pg l}
  \put(136,128)   {\pl{\overline{\eps_2}}}
  \put(124,137.8) {\pg p}
  \put(112,147)   {\pl{\overline{\eps_1}}}
  \put(95,160)    {\pg \jb}
  \put(69,69)     {\pb M}
  \put(199,69)    {\pb M}
  \put(93.3,19)   {\pg q}
  \put(109.9,19)  {\pg {\bar q}}
  \put(93,43)     {\pl{\psi_\gamma}}
  \put(133,43)    {\pl{\psi_\delta}}
  \put(40,49)     {\pl{\phi_\alpha}}
  \put(178,49)    {\pl{\phi_\beta}}
  } }
Here the morphisms $\psi_\gamma$ and $\psi_\delta$ correspond to boundary fields as 
described in appendix \ref{app_fields}. The invariant of the graph \erf{C_disc_boundfc}
can be computed by projecting it in a non-singular manner to a plane and interpreting 
the resulting planar picture as a morphism in the category \C. With a suitable choice of
projection the resulting morphism in \Hom\one\one\ looks as follows:
  \eqpic{C_disc_boundmorph} {200} {97} {
  \put(0,96)    {$ Z(\M^{q,\gamma\delta}_{p,\eps_1\eps_2})\,1 ~= $}
  \put(93,0)   {  {\Includepicclal{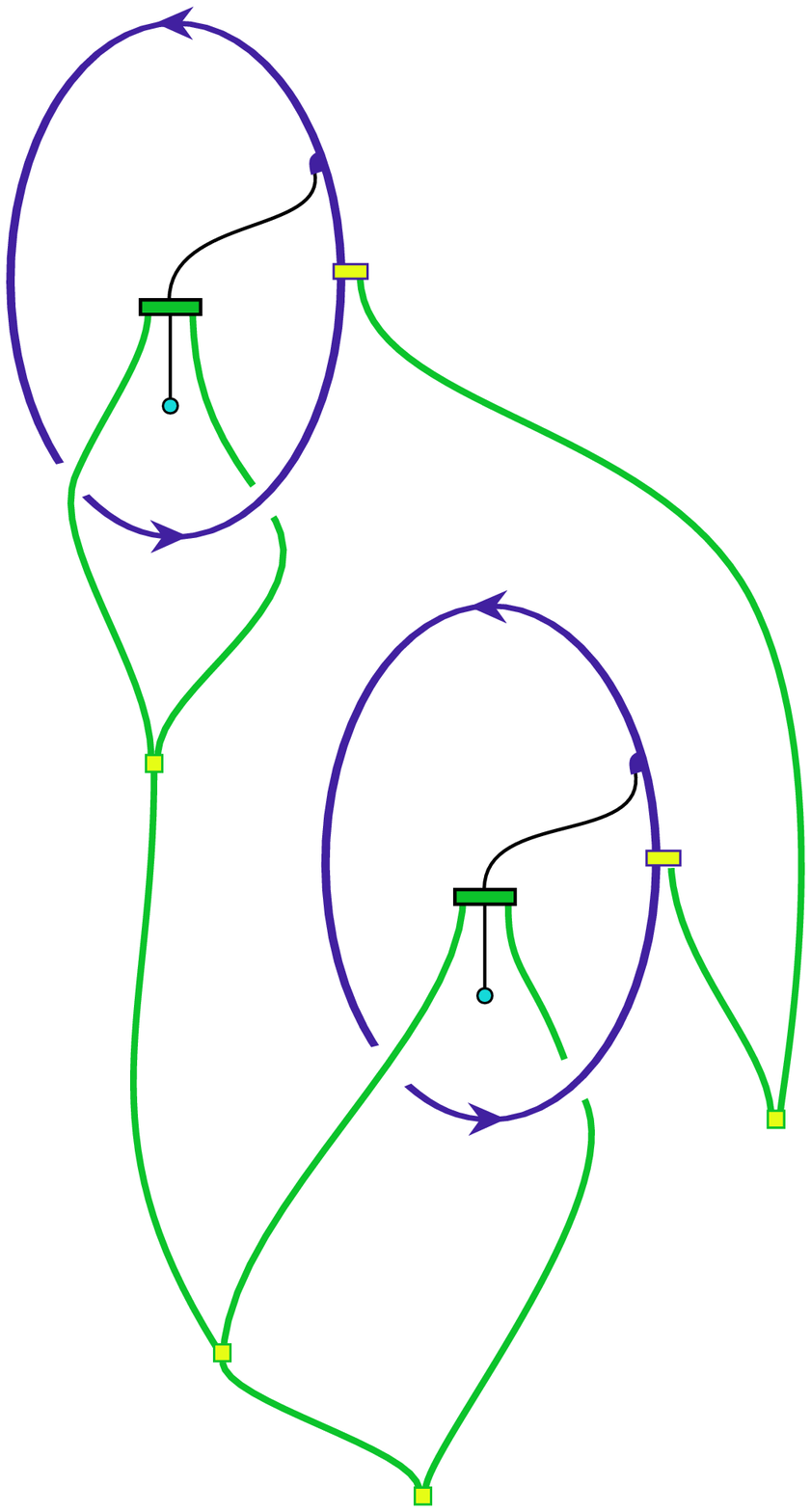}}
\put(4,130)     {\pg k}
\put(38,50)     {\pg \ia}
\put(78,31)     {\pg \ja}
\put(40.4,130)  {\pg l}
\put(24,103)    {\pl{\overline{\eps_2}}}
\put(13,75)     {\pg p}
\put(34,21)     {\pl{\overline{\eps_1}}}
\put(37,5)      {\pg \jb}
\put(87,117)    {\pb M}
\put(40,204)    {\pb M}
\put(96.4,67)   {\pg q}
\put(113,67)    {\pg {\bar q}}
\put(97,91)     {\pl{\psi_\gamma}}
\put(53,174)    {\pl{\psi_\delta}}
\put(73,85.5)   {\pl{\phi_\alpha}}
\put(29,169)    {\pl{\phi_\beta}}
}
}
Using dominance in the space \Hom{U_p\oti U_{\bar q}}{U_p\oti U_{\bar q}} one sees
that only the term with $q\eq p$ contributes in the $q$-summation in 
\erf{C_exp}. Further, for this term one obtains
  \eqpic{C_disc_boundmorph1} {280} {81} {
  \put(0,85)    {$ Z(\M^{p,\gamma\delta}_{p,\eps_1\eps_2})\,1 ~= $}
  \put(175,0)   {  {\Includepicclal{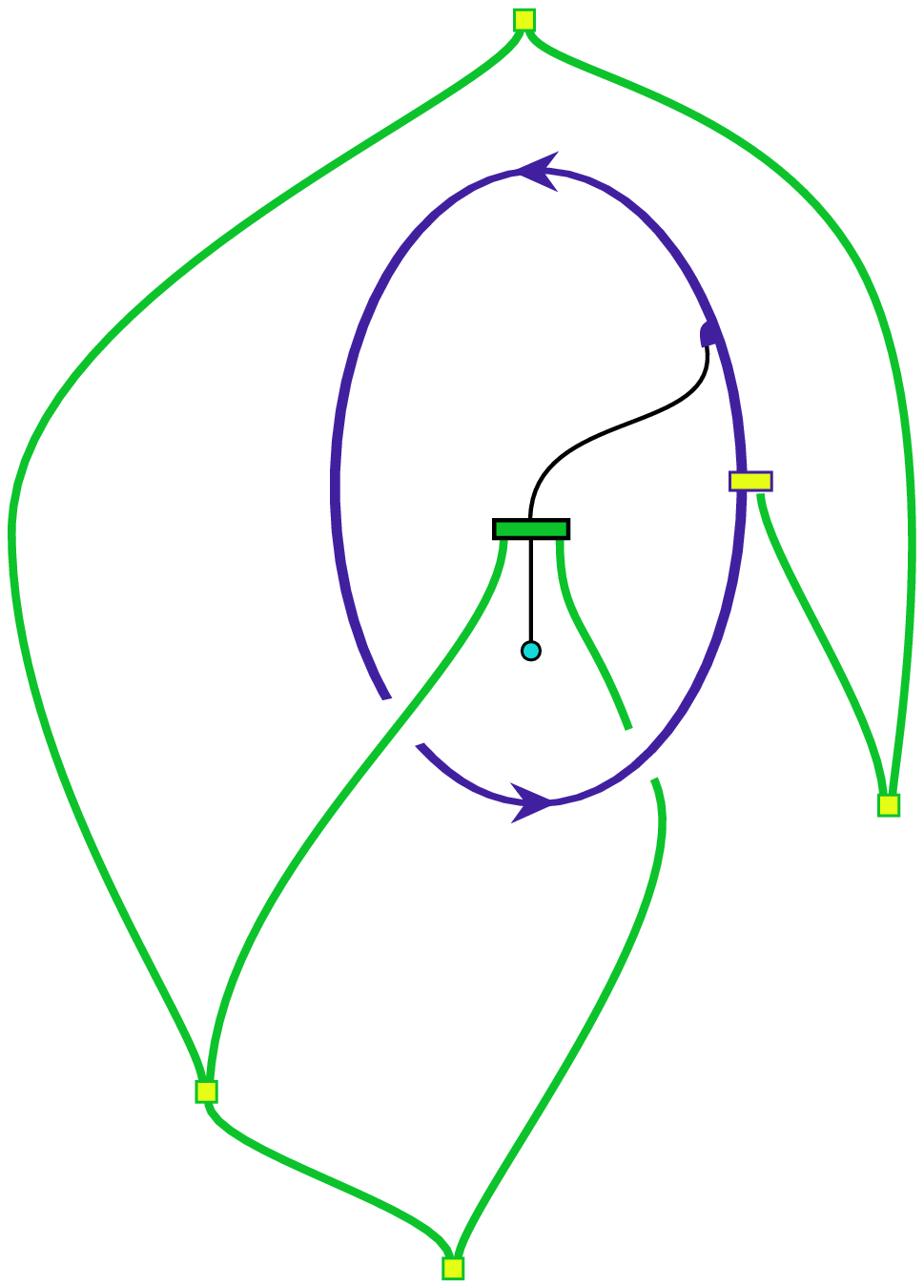}}
  \put(25,45)     {\pg \ia}
  \put(75,40)     {\pg \ja}
  \put(3,88)      {\pg p}
  \put(26.7,21)   {\pl {\overline{\eps_1}}}
  \put(32,4)      {\pg \jb}
  \put(84,107)    {\pb M}
  \put(89.6,66)   {\pg p}
  \put(106,67)    {\pg {\bar p}}
  \put(90,90.5)   {\pl {\psi_\gamma}}
  \put(66,85)     {\pl {\phi_\alpha}}
  }
  \put(96,60)   { {\Includepicclal{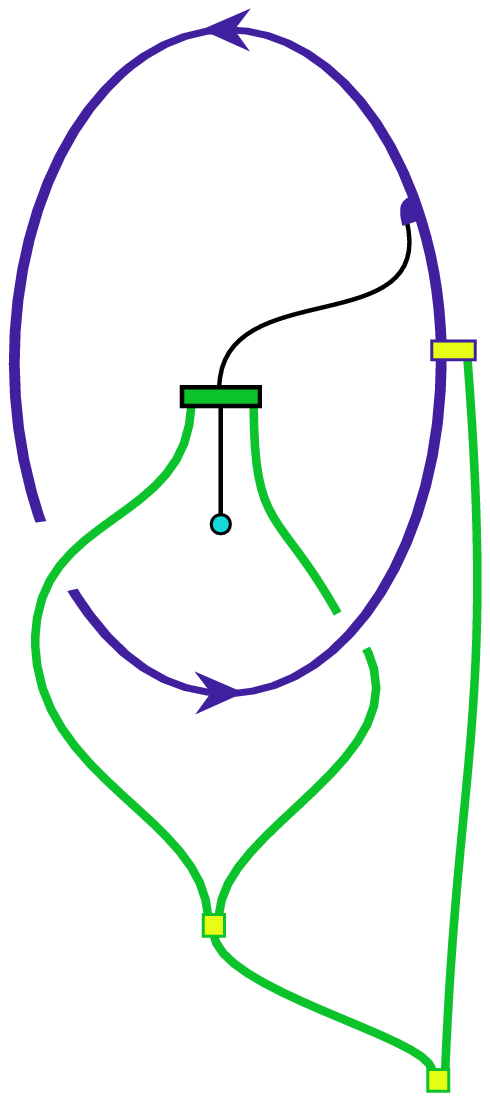}}
   \put(-46,-147){ 	  
  \put(45.5,182)  {\pg k}
  \put(86,182)    {\pg l}
  \put(72,164)    {\pl {\overline{\eps_2}}}
  \put(90,251)    {\pb M}
  \put(97,170)    {\pg {\bar p}}
  \put(75,151)    {\pg p}
  \put(98.5,227.4){\pl {\psi_\delta}}
  \put(75.2,222.3){\pl {\phi_\beta}}
  } } }

Next we note that the summation over the label $p$ in the expression \erf{C_exp} 
for the correlator corresponds to the decomposition
  \be
  \hom(\one,U_\ia\oti U_\ja\oti U_k\oti U_l) \,\cong\, \bigoplus_{p\in\I}
  \hom(U_p^\vee,U_\ia\oti U_\ja) \otic \hom(U_p,U_k\oti U_l)
  \labl{hom1,ijkl}
(which is equivalent to \erf{Hom1,12m} for $m\eq2$). The information we are mainly 
interested in is supplied by the contribution of the \emph{vacuum channel} $p\eq0$ 
to the correlator \erf{C_exp}, corresponding to the summand $p\eq0$ in the 
decomposition \erf{hom1,ijkl}. For any even $m\eq2n$ this contribution can be 
extracted unambiguously because the space \Bnij\ of conformal blocks has a 
distinguished subspace \BOnij\ that corresponds to the vacuum channel; it is obtained 
by choosing in \erf{npbasis} all $p_s^{}$ with odd label $s$ to correspond to the 
identity field, i.e.\ by taking $p_{2s-1}^{}\eq0$ for all $s\eq1,2,...\,,n{-}1$.
The subspace \BOnij\ is zero unless $\ja_{k}\eq\ib_{k}$ for all $k\eq1,2,...\,,n$,
and in the latter case $\BOnij\eq\BOnii$ is one-dimensional, the single contribution 
coming from $p_{2k}\eq\ib_{k}$ for all $k\eq1,2,...\,,n{-}2$ (as well as with all 
multiplicity labels having only a single possible value, which we indicate by the 
symbol `$\circ$'). We denote by
  \be
  \bb(\ia_1\ib_1\ia_2\ib_2...\ia_n\ib_n)_{0\ib_20\ib_3...0\ib_{n-1}0,\nl\nl\dots\nl\nl}
  =: \bb(\ia_1\ib_1\ia_2\ib_2...\ia_n\ib_n)_0
  \labl{bb_0}
the corresponding basis element of the space $B(U_{\ia_1},...,U_{\ib_n})$. In terms 
of the pictorial description \erf{npbasis} of the basis of $B(U_{\ia_1},U_{\ja_1},
...\,,U_{\ia_n},U_{\ja_n})$, the basis elements \erf{bb_0} are given by the 
invariants of the ribbon graphs
  \eqpic{basis_B_0}{440}{80}{
  \put(24,85) {$ \widehat\bb(\ia_1\ib_1\ia_2\ib_2...\ia_n\ib_n)_0~= $}
  \put(160,0){ \put(0,0){\Includepicclal{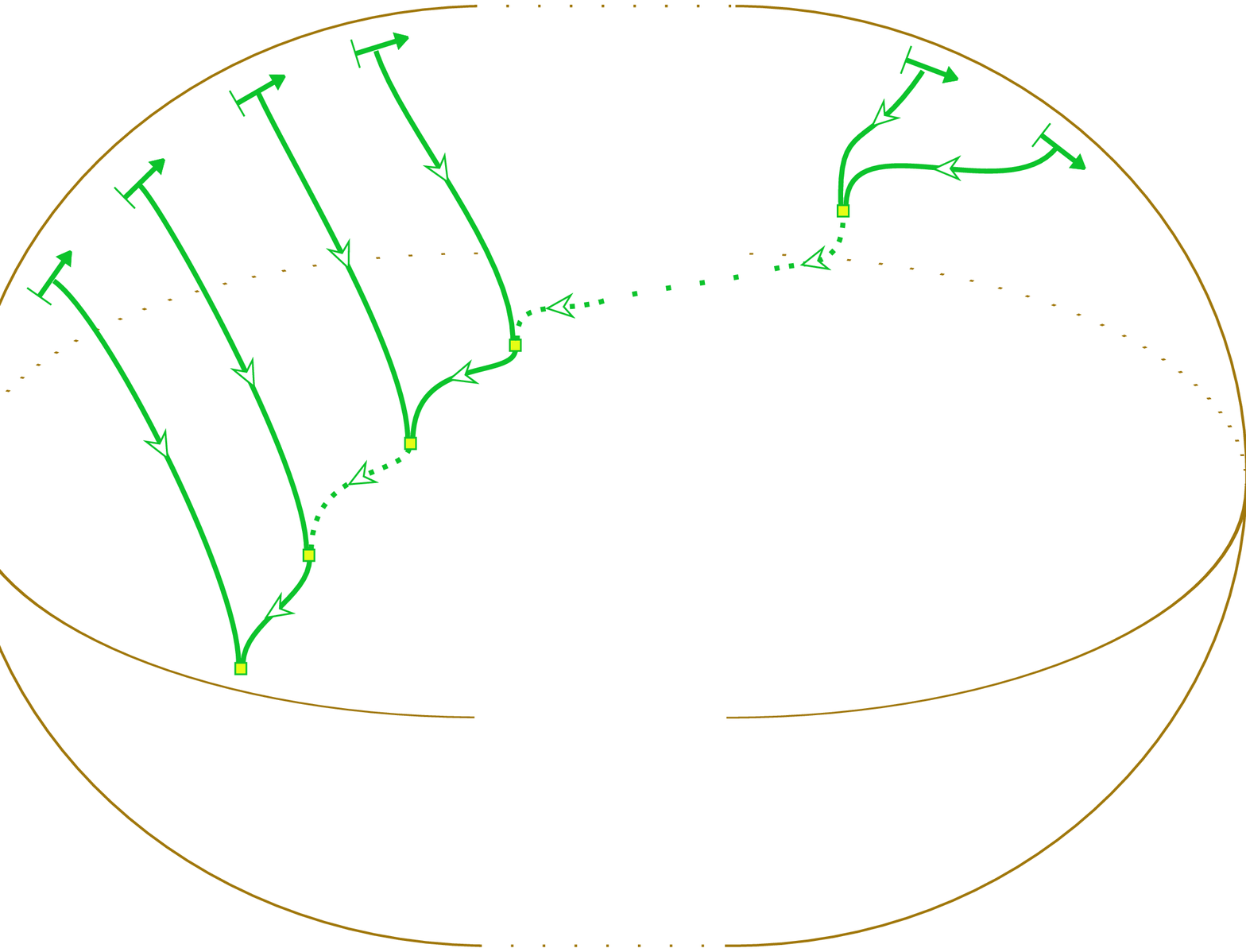}}
  \put(4,133)       {\pg {\ia_1}}
  \put(22,153)      {\pg {\ib_1}}
  \put(49,172.5)    {\pg {\ia_2^{}}}
  \put(73,179.5)    {\pg {\ib_2^{}}}
  \put(181,173)     {\pg {\ia_{n}}}
  \put(211,155)     {\pg {\ib_{n}}}
  \put(63,61)       {\pg {\ib_1^{}}}
  \put(70,79)       {\pg 0}
  \put(89,99)       {\pg {\ib_2^{}}}
  \put(107,113)     {\pg 0}
  \put(166,131)     {\pg 0}
  } }
The subspace \BOnij\ exists as a consequence of the fact that the category \C\ has 
a duality. It comes with a canonical projection $\proj$ from \Bnij\ such that 
$\proj\cir e\eq\id_{\BOnij}$ for $e\colon \BOnij\,{\hookrightarrow}\,\Bnij$. 
Indeed, every basis of \Bnij\ that consists of tensor products of basis elements of
three-point blocks contains the element \erf{bb_0}, and in the basis transformation 
between any two such bases this element is mapped to itself and does not appear at 
all in the transformation of any other basis element.
   
Next we observe, by comparison with formula (4.23) of \cite{fuRs10}, that the two 
factors on the right hand side of \erf{C_disc_boundmorph1} are (apart from some
additional fusing moves of the type displayed in \erf{Fmat} which are implemented
easily) just the structure constants for correlators of a disk with one bulk field 
and one boundary field insertion. For the first factor, the fields are $\Phi_\beta$ 
and $\Psi_\delta$, while for the second they are $\Phi_\alpha$ and $\Psi_\gamma$.
Putting the various ingredients of the coefficients \erf{ctpdc_p} together,
we then arrive at the formula
  \be
  \ctpdo{\PHi\alpha}{\PHi\beta}M0\nl\nl
  \equiv \proj(\ctpd{\PHi\alpha}{\PHi\beta}M)
  = \sum_{\gamma,\delta}(c^{\text{bnd}}_{M,M,0})^{-1}_{\;\delta\gamma}\,
  \coronebd{\PHi\alpha,\Psi_\gamma}M\,\coronebd{\PHi\beta,\Psi_\delta}M
  \labl{res_boundfact_anyM}
for the coefficients with $p\eq0$.

         \medskip

Let us now restrict our attention to the case of a {\em simple\/} $A$-module $M$.
In this case the space $\HomA(M\oti U_0,M)$ in which the morphism $\psi_\gamma$ 
takes values is \onedim, and it has the unit constraint of the category \C\ as a 
canonical basis. As a consequence, for simple $M$ the two disk correlators 
on the right hand side of \erf{res_boundfact_anyM} can be canonically described 
by complex numbers $\coronebd{\PHi\alpha}M$ and $\coronebd{\PHi\beta}M$, and the 
two-point function of boundary fields canonically reduces to a complex number as 
well, see \erf{A13}. Thus for simple $M$  we arrive at the expression 
  \be
  \ctpdo{\PHi\alpha}{\PHi\beta}M0\nl\nl
  = (c^{\text{bnd}}_{M,0})^{-1}_{}\, \coronebd{\PHi\alpha }M\,\coronebd{\PHi\beta }M \,.
  \labl{res_boundfact}

\medskip

The generalization of this result to an arbitrary number $n$ of bulk fields
on the disk is direct. The coefficient $\ctpdo{\PHi{\alpha_1},\PHi{\alpha_2},...\,}
{\PHi{\alpha_n}}M0{\nl\nl\dots}{\nl\nl}$ 
corresponding to the one-dimensional vacuum-channel subspace with basis element 
$\bb(\ib_1\ia_1\ib_2\ia_2...\ib_n\ia_n)_{0\ia_20...\ia_{n-1}0}$ is
  \be
  \ctpdo{\Phi_{\alpha_1},\Phi_{\alpha_2},...\,} {\Phi_{\alpha_n}}M0 
  {\nl\nl\dots}{\nl\nl}
  = (({c^{\text{bnd}}_{M,0}})^{-1})^{n-1}_{} \prod_{i=1}^{n}\coronebd{\Phi_{\alpha_i}}M \,.
  \labl{n-bndfact0}
In short, the vacuum-channel coefficient of an $n$-point correlator
of bulk fields on the disk is the product over the boundary state coefficients
$\coronebd{\Phi_{\alpha_i}}M$ for each of the field insertions, up to a normalization 
factor given by a power of the inverse of the boundary vacuum two-point function.

Now observe that the latter prefactor can actually be absorbed into a simple
change of normalization, namely by dividing the correlators of bulk fields on 
a disk by the boundary vacuum two-point function $c^{\text{bnd}}_{M,0}$:
  \be
  \ctpdO{\PHi{\alpha_1},\PHi{\alpha_2},...\,} {\PHi{\alpha_n}}M0 {\nl\nl\dots}{\nl\nl}
  = \prod_{i=1}^{n}\coronebD{\PHi{\alpha_i}}M 
  \labl{n-bndfact}
with $\ctpdO{\Phi_{\alpha_1},...}{\Phi_{\alpha_n}}M0 {\nl\nl\dots}{\nl\nl} \eq
\ctpdo{\Phi_{\alpha_1},...\,}{\Phi_{\alpha_n}}M0 {\nl\nl\dots}{\nl\nl}/c^{\text{bnd}}_{M,0}$.
Moreover, the re-normalized boundary state coefficients 
  \be
  \onefsc\ia\alpha M \equiv
  \coronebD{\PHi\alpha}M = (c^{\text{bnd}}_{M,0})^{-1}_{}\, \coronebd{\PHi\alpha^{\ia\ib}}M
  \labl{refl_coeff} 
are nothing but the \emph{reflection coefficients} which appear (see \cite{cale}
or section 2.7 of \cite{fuSc6}) in the expansion into boundary fields 
of the bulk field $\Phi_\alpha$ when it comes close to the boundary.

  
\section{Bulk factorization}\label{s2}

Our next goal is to derive another equality for the same disk correlator
\erf{n-bndfact0} for $n$ bulk fields, this time making use of bulk factorization. 
One expects that the operator products of bulk fields enter in the resulting
formulas. In fact, one of the main results of this paper are explicit 
expressions for such a factorization. The result has a
remarkably simple structure, involving traces of intertwiners for an action
of the mapping class group of a sphere with marked points.

Given any world sheet \X, the examination of a cutting of \X\ along a circle $S$
in the interior of \X\ leads to the expression
  \be
  c(\X) = \sum_{q_1,q_2\in\I}\sum_{\gamma,\delta}\dim(U_{q_1})\,\dim(U_{q_2})\,
  (\cbulk_{q_1,q_2})\Inv_{\;\delta\gamma}\, Z(\M_{\X;q_1q_2\gamma\delta})\,1
  \ee
for the correlator of \X. Here $\cbulk$ is the bulk two-point function described in
\erf{bulk_2pfu}, while the labels $\gamma$ and $\delta$ 
run over bases of the bimodule morphism spaces $\Homaa{U_{q_1}\otip A\otim U_{q_2}}A$ 
and $\Homaa{U_{\bar q_1}\otip A\otim U_{\bar q_2}}A$, respectively. For any 
$q_1,q_2\iN\I$ and any value of $\gamma$ and $\delta$ the manifold 
$\M_{\X;q_1q_2\gamma\delta}$ is a cobordism from $\emptyset$ to 
$\partial\M_{\X;q_1q_2\gamma\delta} \eq \Xh$, which is just what is needed for 
$Z(\M_{\X;q_1q_2\gamma\delta})1$ to be an element of the same space of 
conformal blocks as the correlator $c(\X)$.

  \pagebreak 

Implementing the construction summarized in appendix \ref{app_tft}, one obtains the
following prescription for obtaining the cobordism $\M_{\X;q_1q_2\gamma\delta}$:
\nxtp
First, remove from the connecting manifold \MX\ for \X\ the union \YS\ of all 
connecting intervals that intersect the image of the cutting circle $S\,{\subset}\,\X$
under the embedding $\imbed\colon \X\To\MX$. More explicitly, \YS\ is the preimage
  \be
  \YS := \pi_\X^{-1}(\imbed(S)) \, \subset\MX
  \labl{def_YS}
of $\imbed(S)$ under the canonical projection $\pi_\X$ from \MX\ to $\imbed(\X)$; 
by construction, \YS\ is an annulus, and $\partial\YS\,{\subset}\,\partial\MX$.
\nxl2
By a judicious choice of the triangulation $\Gamma$ of \X, without loss of generality 
we can assume that the circle $\imbed(S)$ intersects precisely one of the $A$-ribbons 
that belong to the triangulation, i.e.\ $\imbed(S)\,{\cap}\,\Gamma$ consists of 
a single arc, which we denote by \arcS.
\nxl2
As an illustration, when \X\ is a disk, the annulus \YS\ and arc \arcS\ for one 
possible choice of $S$ and $\Gamma$ look as follows:
  \eqpic{YS_arcS_for_disk} {180} {89} {
  \put(0,0)    {\Includepicclal{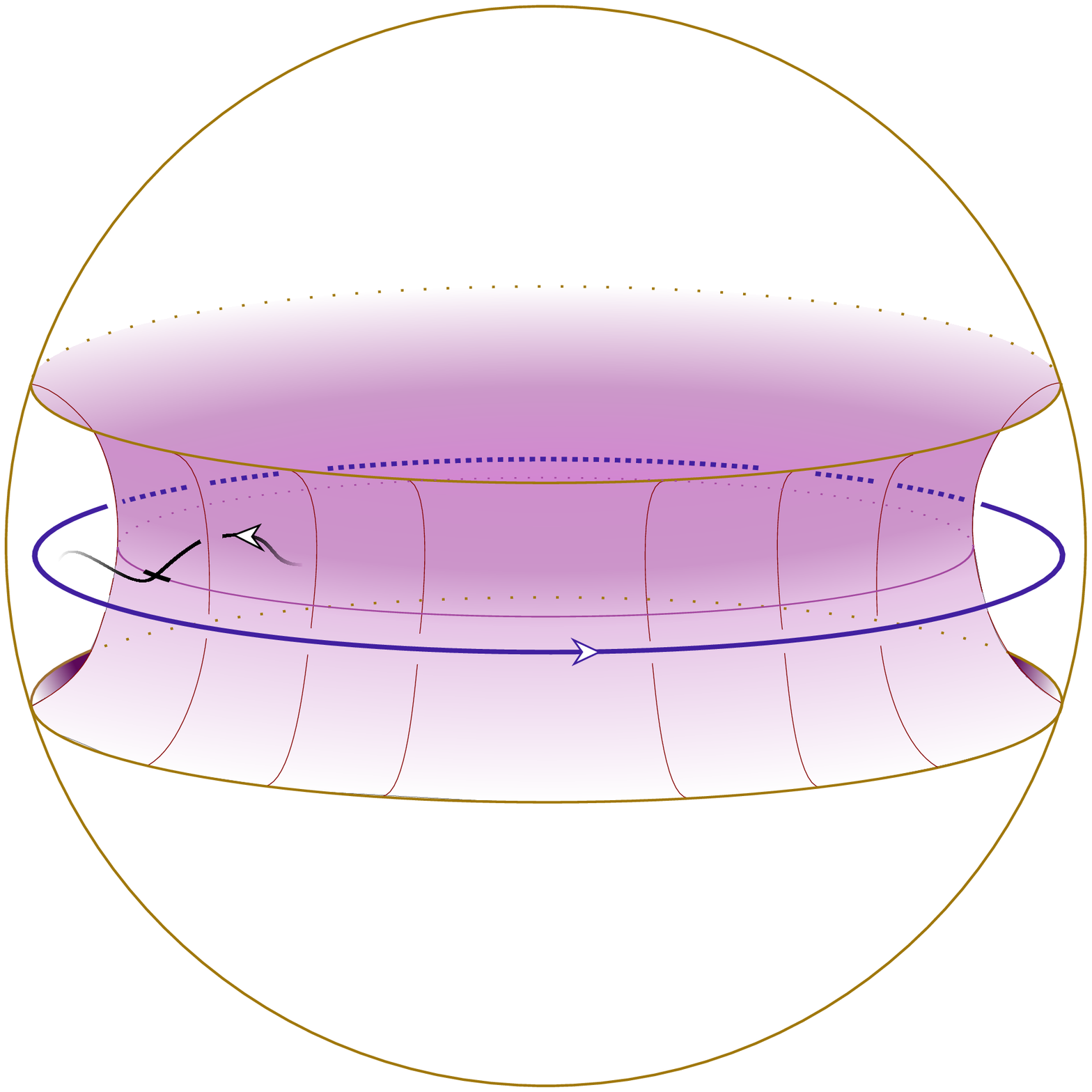}}
  \put(-9,67.5)   {\begin{turn}{35}\longrsqarrow\end{turn}}
  \put(-15,65)    {$ \arcS $}
  \put(81,74)     {\pb M }
  \put(100,92)    {\pl {\imbed(S)} }
  \put(183.6,113) {\lsqarrow}
  \put(210.6,115) {$\vio \YS $}
 }
~\\[-2.9em]
\nxtp
Take the closure of the so obtained manifold. This results in a three-manifold 
(with corners) $\MX^\circ$ whose boundary contains two disjoint copies $\YS^{1,2}$ 
of \YS. (If $\MX^\circ$ is disconnected, then  $\YS^1$ and $\YS^2$ lie in two 
different connected components.) Here the labeling is chosen in such a way that of 
the two copies $\arcS^{1,2}\,{\subset}\,\YS^{1,2}$ of \arcS, the one denoted by 
$\arcS^1$ belongs to an $A$-ribbon whose core is oriented outwards, while the 
one denoted by $\arcS^2$ belongs to an $A$-ribbon with inwards-oriented core.
\nxtp
For $\kappa\iN\{1,2$\}, identify the surfaces $\YS^\kappa\,{\subset}\,\partial
\MX^\circ$ with the distinguished annular part $\YT^\kappa$ of the boundary of 
another three-manifold (with corners) $\T_{\!q_1q_2\gamma\delta}$. This manifold 
$\T_{\!q_1q_2\gamma\delta}$ is a full torus (with corners) containing a specific 
ribbon graph that includes in particular the morphisms $\phi_\gamma$ and $\phi_\delta$ 
for two additional bulk field insertions. Describing the full torus as a cylinder 
with top and bottom identified, $\T_{\!q_1q_2\gamma\delta}$ looks as follows:
\eqpic{stickyft}{270}{142}{
   \put(0,-5){
  \put(-10,146)   {$ \T_{\!q_1q_2\gamma\delta} ~= $}
  \put(75,0)  {  \Includepicclal{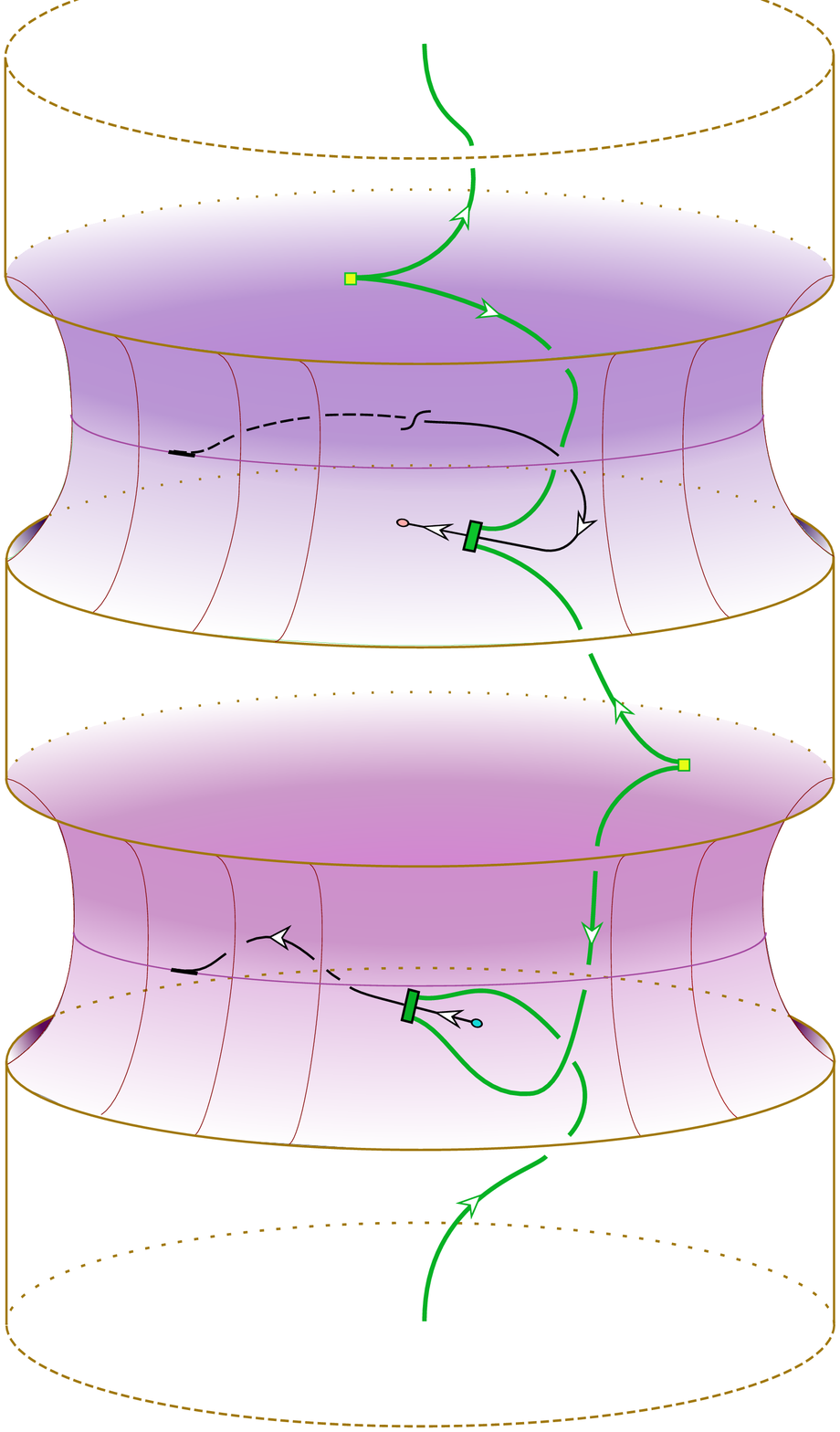} }
  \put(175,0) {
  \put(-5,26)     {\pg {q_2}}
  \put(14,215)    {\pg {\bar q_2}}
  \put(13,171)    {\pg {\bar q_1}}
  \put(36,129)    {\pg {q_1}}
  \put(-16,79)    {\pl {\phi_\gamma}}
  \put(-2,177.6)  {\pl {\phi_\delta}}
  \put(-103,95)   {\longrsqarrow}
  \put(-103,205)  {\longrsqarrow}
  \put(-116,97)   {$ \arcT^2 $}
  \put(-116,207)  {$ \arcT^1 $}
  \put(65,111)    {\lsqarrow}
  \put(65,219)    {\lsqarrow}
  \put(92,113)    {$\vio \YT^2 $}
  \put(92,221)    {$\vio \YT^1 $}
  } } }
(Note that due to the identification of top and bottom the $q_1$- and $q_2$-ribbons
form a loop.) The identification of $\YS^\kappa$ with $\YT^\kappa$ is performed in 
such a manner that, for $\kappa\eq1,2$, the copy 
$\arcS^\kappa\,{\subset}\,\YS^\kappa$ of \arcS\ gets identified with the arc 
$\arcT^\kappa$ on the boundary of $\T_{\!q_1q_2\gamma\delta}$ at which an
$A$-ribbon ends, and also such that the pieces of $A$-ribbon that get connected
this way have matching 2-orientations.
\nxtp
Finally, $\M_{\X;q_1q_2\gamma\delta}$ is the resulting three-manifold, regarded
as a cobordism from the empty set to the two-manifold $\partial\M_{\X;q_1q_2\gamma\delta}
\eq \partial\MX$ which is homeomorphic to the double $\Xh$ of the world sheet.

\medskip

This prescription for $\M_{\X;q_1q_2\gamma\delta}$ can be understood as the result
of composing two cobordisms: First, the correlator $c(\X_{q_1q_2\gamma\delta})$ for
a world sheet $\X_{q_1q_2\gamma\delta}$ which is obtained by a cutting of \X\ 
and which as a result of the cutting procedure contains two additional bulk field
insertions $\PHi\gamma$ and $\PHi\delta$. And second, the \emph{gluing homomorphism}, 
which is the invariant of a cobordism from the double $\Xh_{q_1q_2\gamma\delta}$ of the
cut world sheet to the double \Xh\ of the original world sheet. Thus in particular it 
follows by construction that indeed $\partial\M_{\X;q_1q_2\gamma\delta}\,{\cong}\,\Xh$.
It must however be pointed out that the prescription above constitutes a shortcut 
that only describes the final result of the manipulations which are involved in the 
factorization. For instance, in order to understand the origin of the three-manifold
$\T_{\!q_1q_2\gamma\delta}$, which is the part of $\M_{\X;q_1q_2\gamma\delta}$
that in \cite{fjfrs} is displayed as the picture (5.9), it is essential that the 
actual cutting of \X\ is not performed along a circle, but rather involves an annular 
collar around that circle, as illustrated in picture (5.2) of \cite{fjfrs}. For a 
few more details see appendix \ref{app_tft}, while for a full derivation, the 
reader is invited to consult section 5 of \cite{fjfrs}.

Compatibility with factorization is an extremely tight constraint on correlators, 
as it relates correlators on world sheets with different topology. What renders the 
proof of factorization possible is the fundamental insight implicit in the proof in
\cite{fjfrs} that factorization is a local issue on the world sheet -- the required 
manipulations of three-manifolds only involve the fibers of \MX\ over some suitable
neighbourhood of the cutting circle
$S$. For our present purposes, however, we want to compare the actual values
of correlators on $\X$, which is a global issue on $\X$. As a consequence, the 
whole three-manifold $\M_{\X;q_1q_2\gamma\delta}$, and accordingly also the whole 
three-manifold corresponding to the gluing homomorphism, enter our analysis.

\medskip

Let us now apply the general prescription given above to the situation of our 
interest, i.e.\ to the correlator of $n$ bulk fields on the disk, see \erf{ctpd1n}. 
The cutting circle $S$ relevant for the bulk 
factorization runs parallel to the boundary ribbon, as indicated in
  \eqpic{S_for_disk} {350} {66} {
  \put(0,69)    {$ \MX ~= $}
  \put(54,0)   { \INcludepicclal{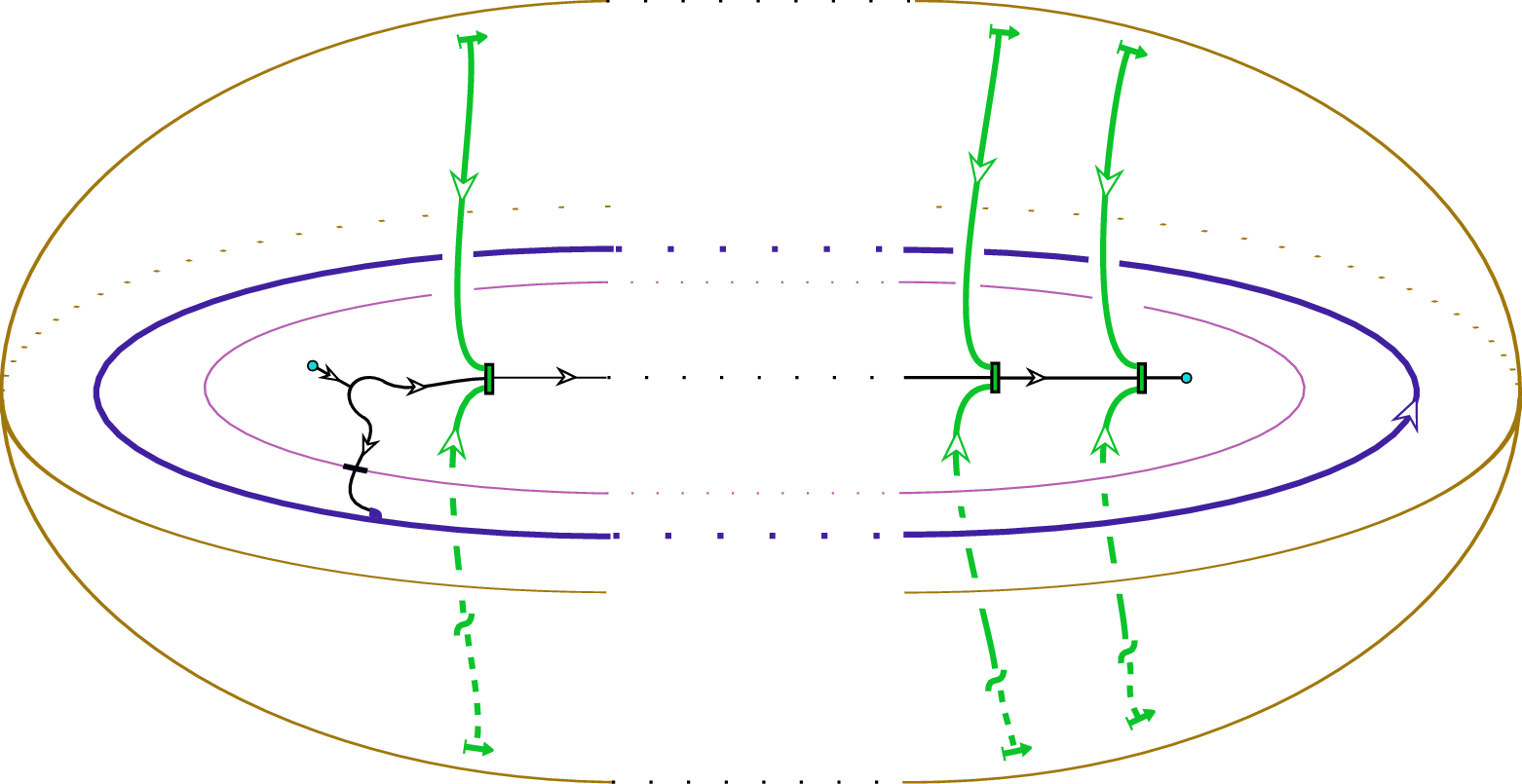} } 
  \put(74,12) {
  \put(81,48)     {\pl{\ia(S)}}
  \put(76,70)     {\pl{\phi_{\alpha_1}}}
  \put(145,70)    {\pl{\phi_{\alpha_{n-1}}}}
  \put(201,69)    {\pl{\phi_{\alpha_n}}}
  \put(79,-5.8)   {\pg{\ja_1^{}}}
  \put(78,128.7)  {\pg{\ia_1^{}}}
  \put(156,-6)    {\pg{\ja_{n-1}}}
  \put(155,128)   {\pg{\ia_{n-1}}}
  \put(204,.5)    {\pg{\ja_n}}
  \put(200,121)   {\pg{\ia_n}}
  \put(238,56)    {\pb M}
  } }
The corresponding annulus $\YS\,{\subset}\,\MX$ then looks as in figure 
\erf{YS_arcS_for_disk}. It follows that the three-manifold $\MX^\circ$, whose 
boundary contains two disjoint copies $\YS^{1,2}$ of \YS, is disconnected. Its 
two connected components are given by a `nibbled apple'
  \eqpic{MXcirc_disk_1} {329} {89} {
   \put(0,-6){
  \put(0,106)     {$ \MX^{\circ,1} ~= $}
  \put(65,0)  {  \Includepicclal{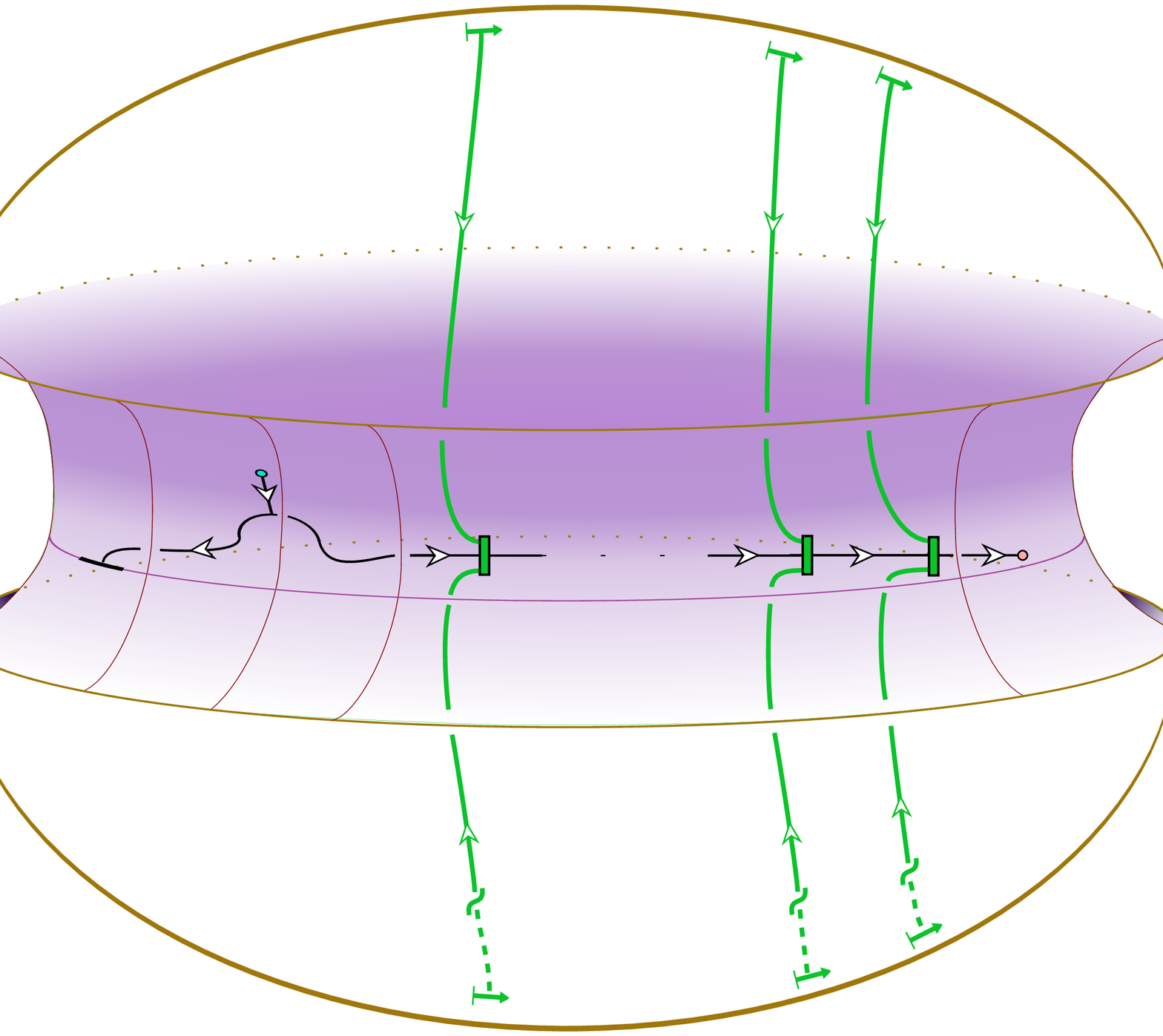}  
  \put(226,117)   {\lsqarrow}
  \put(253,119)   {$\vio \YS^1 $}
  \put(105,103)   {\pl{\phi_{\alpha_1}}}
  \put(166,103)   {\pl{\phi_{\alpha_{n-1}}}}
  \put(193,103)   {\pl{\phi_{\alpha_n}}}
  \put(103,-5)    {\pg{\ja_1^{}}}
  \put(103,209.9) {\pg{\ia_1^{}}}
  \put(163,-2)    {\pg{\ja_{n-1}^{}}}
  \put(161,206)   {\pg{\ia_{n-1}}}
  \put(191,7)     {\pg{\ja_n^{}}}
  \put(188,198)   {\pg{\ia_n}}
  \put(-14,75)    {\begin{turn}{25}\longrsqarrow\end{turn}}
  \put(-25,75)    {$ \arcS^1 $}
  } } }
and a `bigonal doughnut'
  \eqpic{MXcirc_disk_2} {330} {42} {
\setlength\unitlength{1.14pt}
  \put(0,44)      {$ \MX^{\circ,2} ~= $}
  \put(57,0)  {  \InCludepicclal{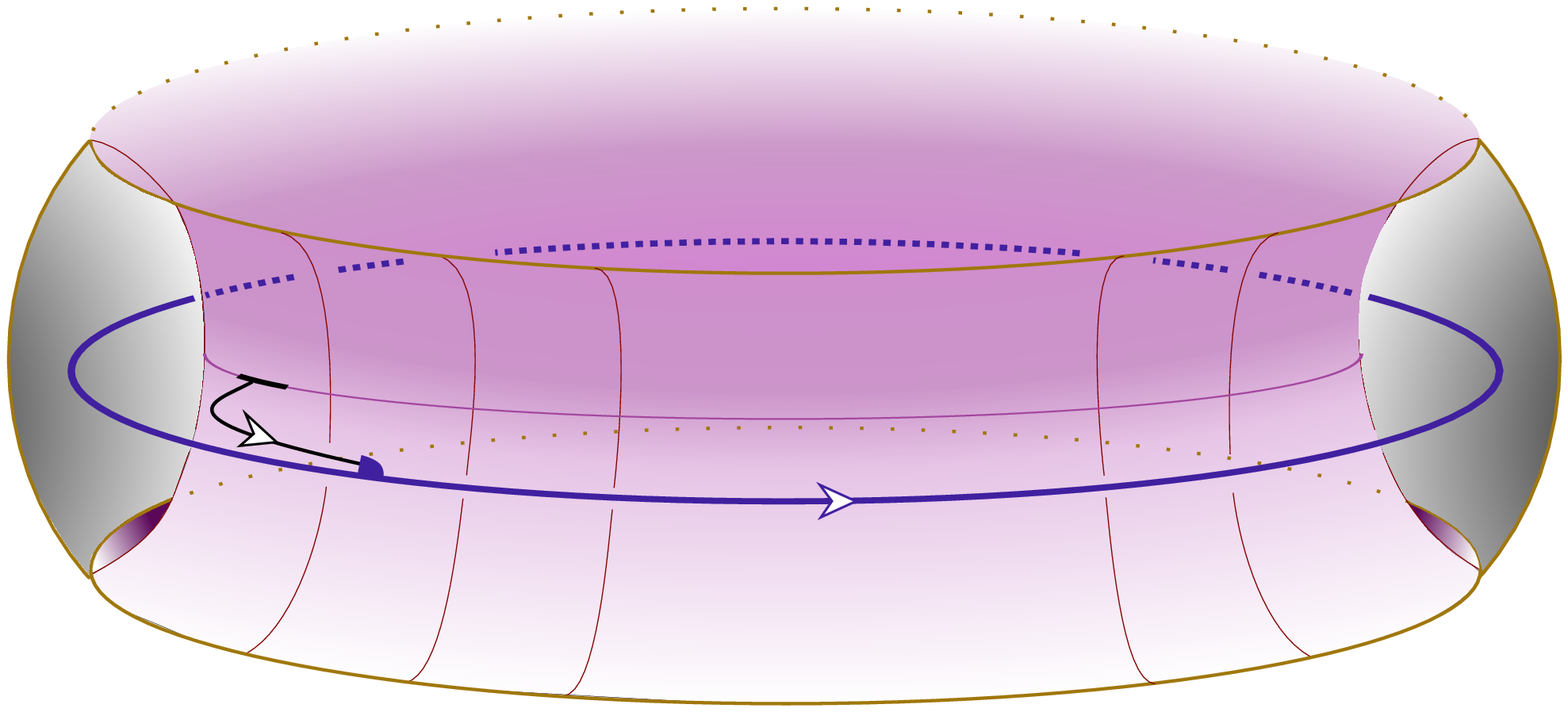}  
  \put(131,21.8)  {\pb M}
  \put(195,15)    {\lsqarrow}
  \put(219,17)    {$\vio \YS^2 $}
  \put(-5,29)     {\begin{turn}{25}\longrsqarrow\end{turn}}
  \put(-14,29)    {$ \arcS^2 $}
  } }
Here we indicate in particular the two copies $\YS^1$ and $\YS^2$ of \YS\ (they are 
shaded) as well as the arcs $\arcS^{1,2}$ contained in them. For completeness, let 
us also display schematically the intersection of $\MX^{\circ,1}$ and $\MX^{\circ,2}$ 
with a vertical plane containing the north and south poles of the three-ball:
  \eqpic{cut_cs} {230} {67} {
   \put(0,-8){
  \put(0,0)  { \includepicclal{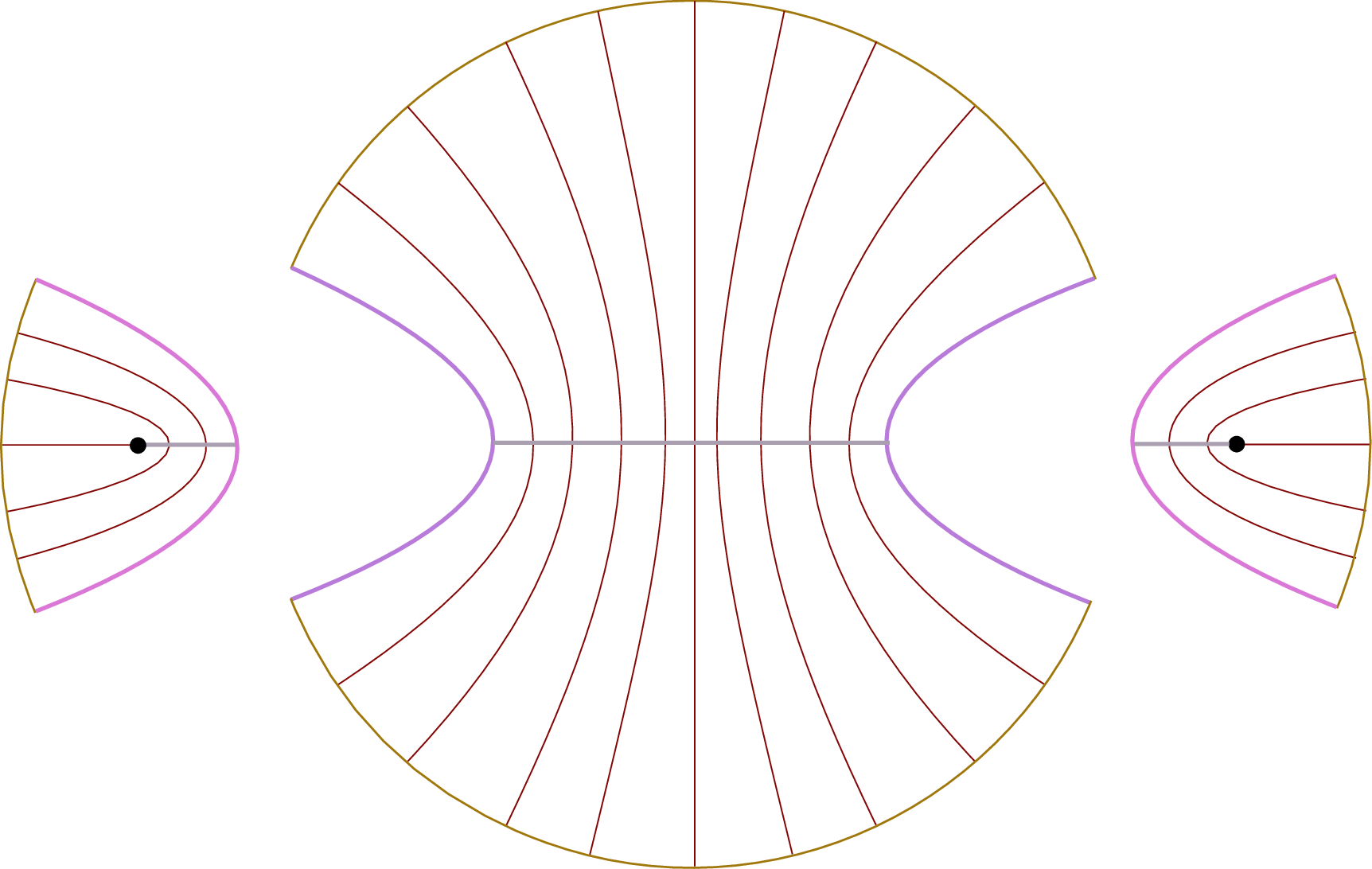} }
  \put(206,136.7) {$\vio \YS^1 $}
  \put(194,107.7) {\begin{turn}{60}\lsqarrow\end{turn}}
  \put(53,136.7)  {$\vio \YS^1 $}
  \put(55,130.7)  {\begin{turn}{-60}\rsqarrow\end{turn}}
  \put(34.5,21)   {$\vio \YS^2 $}
  \put(22.5,52)   {\begin{turn}{-60}\lsqarrow\end{turn}}
  \put(221,21)    {$\vio \YS^2 $}
  \put(222,29)    {\begin{turn}{60}\rsqarrow\end{turn}}
  } }

The manifold $\M_{\X;q_1q_2\gamma\delta}$ is now obtained by identifying
the annuli $\YS^1$ and $\YS^2$ with the corresponding parts $\YT^1$ and $\YT^2$ 
of the manifold $\T_{\!q_1q_2\gamma\delta}$ given in \erf{stickyft}.
Implementing this identification is easy in the case of the boundary $\YS^2$
of the bigonal doughnut \erf{MXcirc_disk_2}; it results
in the manifold ${\mathscr N}_{\X;q_1q_2\gamma\delta}$ that is
shown in the following picture (where top and bottom are still identified):
  \eqpic{YS2_to_YT2} {280} {130} {
   \put(0,-7){
  \put(0,137)   {$ {\mathscr N}_{\X;q_1q_2\gamma\delta} ~= $}
  \put(93,0)  { \Includepicclal{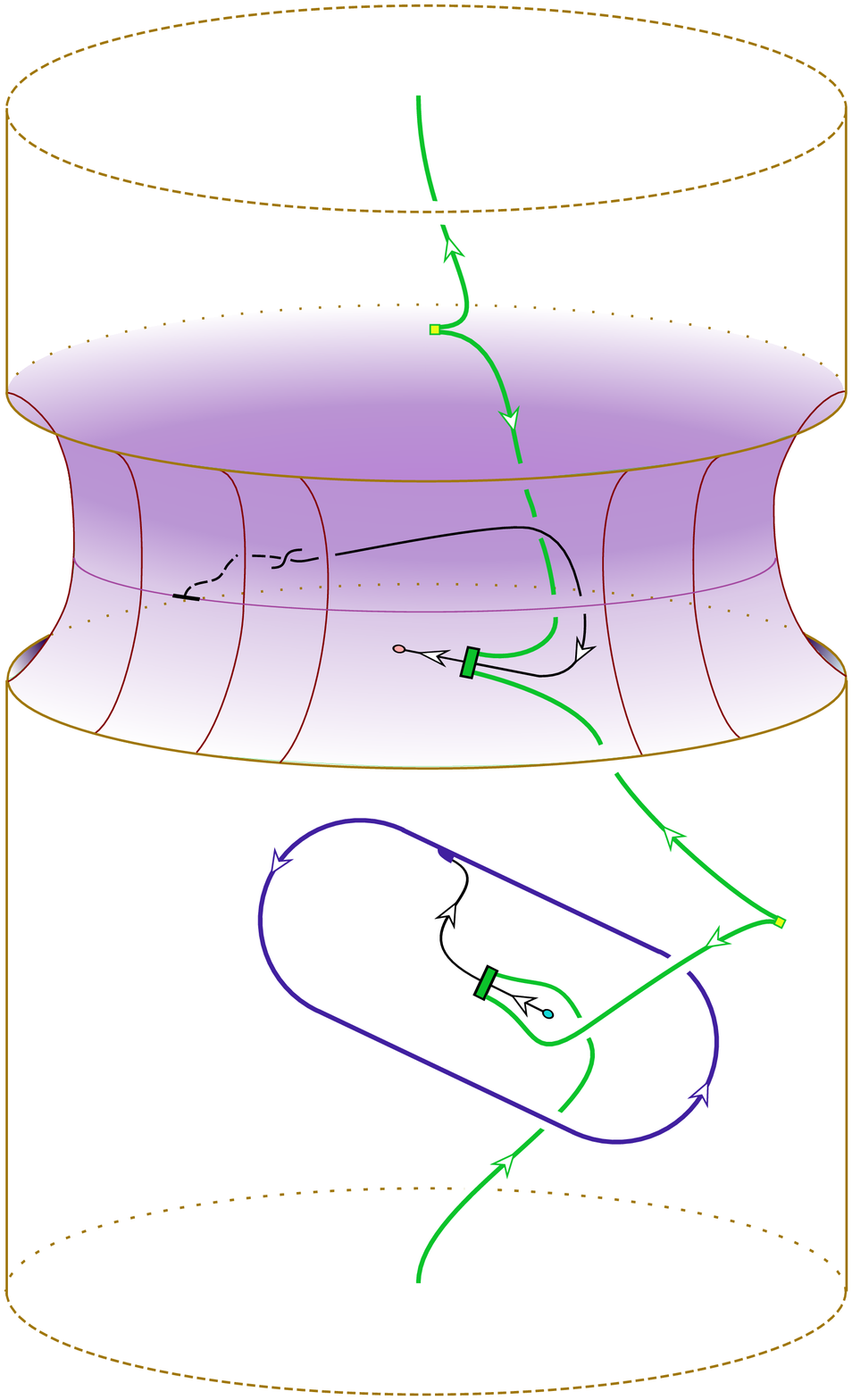}  
  \put(90,28)     {\pg {q_2}}
  \put(104,214)   {\pg {\bar q_2}}
  \put(147,112)   {\pg {\bar q_1}}
  \put(152,90)    {\pg {q_1}}
  \put(93,76)     {\pl {\phi_\gamma}}
  \put(91,142.3)  {\pl {\phi_\delta}}
  \put(69,66)     {\pb M}
  \put(163,190)   {\lsqarrow}
  \put(189.5,192) {$\vio \YT^1 $}
  \put(-7,163.5)  {\longrsqarrow}
  \put(-18,163.5) {$ \arcT^2 $}
  } } }
Identifying the boundary $\YS^1$ of the nibbled apple \erf{MXcirc_disk_1} with 
$\YT^1$ is direct as well, but presenting the resulting three-manifold pictorially 
without apparent self-intersections requires a little more care. The simplest way 
to achieve this is obtained by thinking of the manifold ${\mathscr N}_{\X;q_1q_2
\gamma\delta}$ of \erf{YS2_to_YT2} as describing the three-manifold $D\Times S^1$, 
with the disk $D$ lying in a horizontal plane and $S^1$ running vertically (that is, 
regarding $\YT^1$ as part of the boundary of ${\mathscr N}_{\X;q_1q_2\gamma\delta}$, 
rather than as a subset in the interior of ${\mathscr N}_{\X;q_1q_2\gamma\delta}$ 
beyond which the three-manifold continues), and moreover regarding $D\Times S^1$ as 
embedded in $S^2\Times S^1$. One may then redraw \erf{YS2_to_YT2} by
exchanging interior and exterior regions, i.e.\ by representing the manifold
$D\Times S^1$ of \erf{YS2_to_YT2} as the exterior of 
another copy of $D\Times S^1$ in $S^2\Times S^1$. This yields
  \eqpic{YS2_to_YT2_invert} {330} {135} {
   \put(0,-6){
  \put(0,144)   {$ {\mathscr N}_{\X;q_1q_2\gamma\delta} ~= $}
  \put(85,0)  {  \Includepicclal{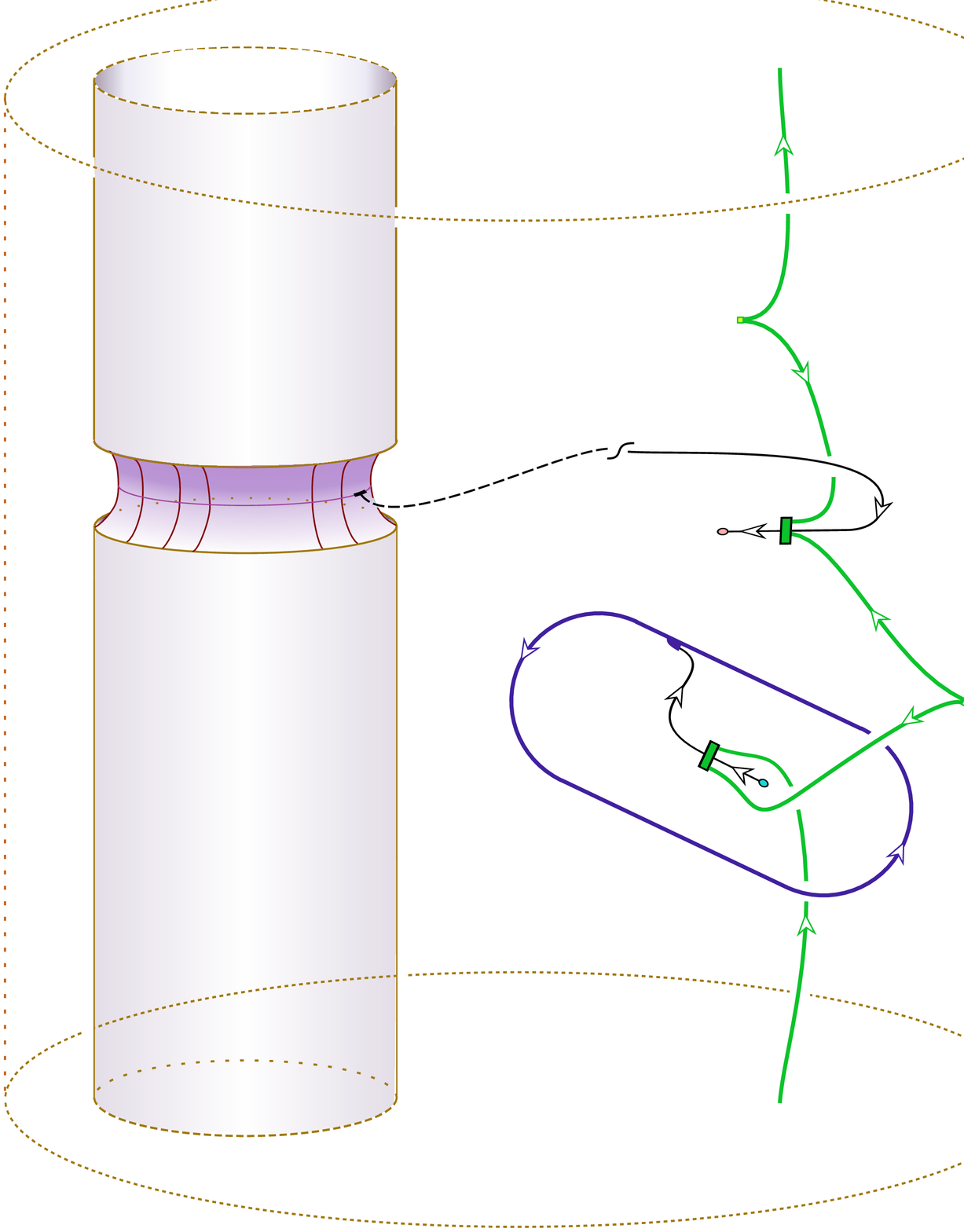} 
  \put(187,35)    {\pg {q_2}}
  \put(194,197)   {\pg {\bar q_2}}
  \put(213,141)   {\pg {\bar q_1}}
  \put(218,116)   {\pg {q_1}}
  \put(160,103.4) {\pl {\phi_\gamma}}
  \put(181,154)   {\pl {\phi_\delta}}
  \put(132,95)    {\pb M}
  \put(86,177)    {\begin{turn}{25}\lsqarrow\end{turn}}
  \put(112,190)   {$\vio \YT^1 $}
  } } }
where the shaded region is the boundary, i.e.\ now the interior of the inner cylinder 
(that is, full torus, since top and bottom are identified) is the part not belonging 
to the manifold.  Identification of $\YT^1$ in \erf{YS2_to_YT2_invert} with $\YS^1$ 
in \erf{MXcirc_disk_1} now yields the following description of the manifold 
$\M_{\X;q_1q_2\gamma\delta}$ (from here on we restrict to the case that the number 
of bulk fields in the nibbled apple \erf{MXcirc_disk_1} is $n\eq2$):
  \eqpic{im20} {320} {130} {
  \put(0,133) {$ \M_{\X;q_1q_2\gamma\delta} ~= $}
  \put(85,-2)  {  \Includepicclal{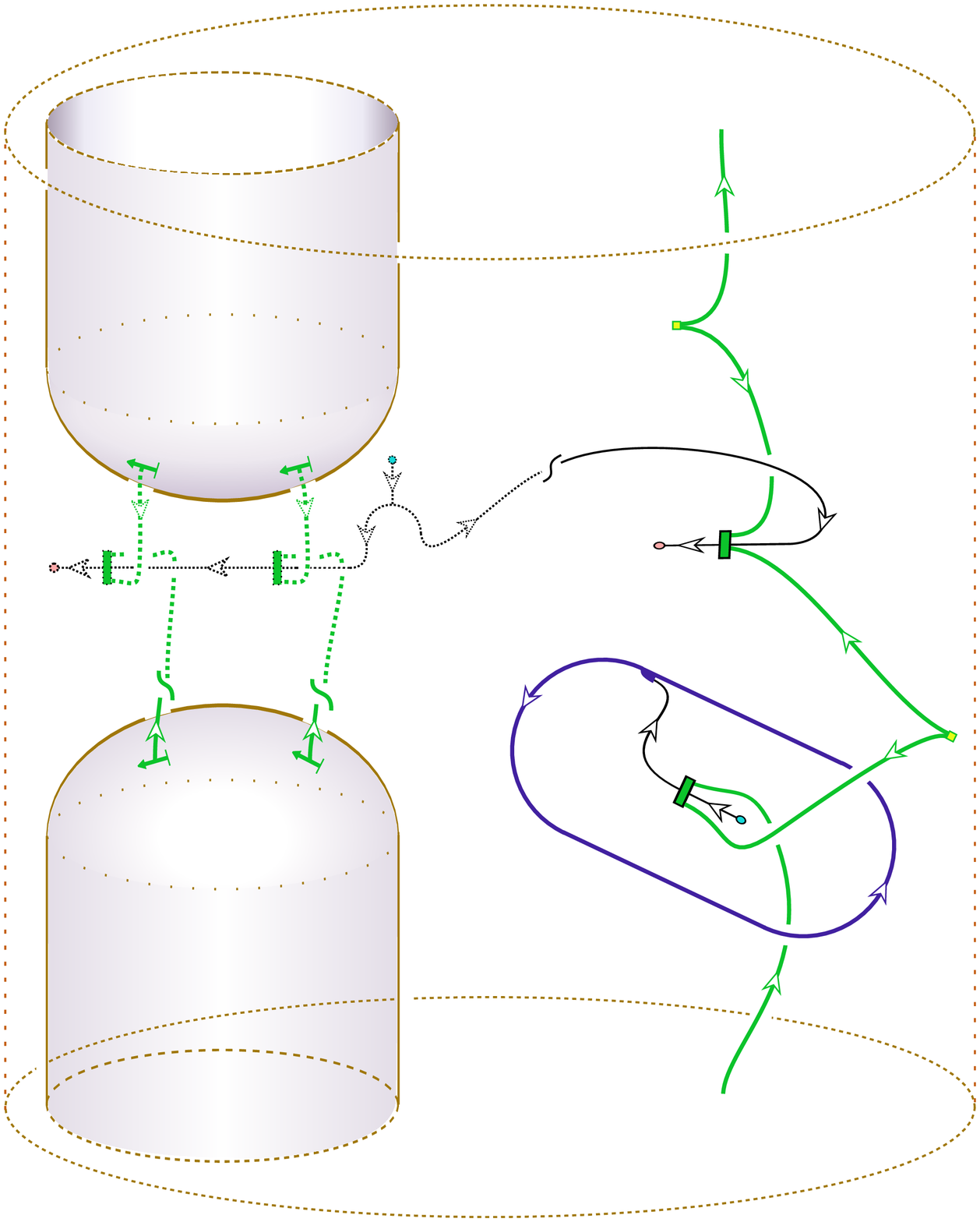} 
  \put(170,35)    {\pg {q_2}}
  \put(170,196)   {\pg {\bar q_2}}
  \put(202,127)   {\pg {\bar q_1}}
  \put(207,102)   {\pg {q_1}}
  \put(149,88)    {\pl {\phi_\gamma}}
  \put(161,144)   {\pl {\phi_\delta}}
  \put(128,77)    {\pb M}
  \put(59,138)    {\pl{\phi_\alpha}}
  \put(20,138)    {\pl{\phi_\beta}}
  \put(68,98)     {\pg\ia}
  \put(66,177)    {\pg\ja}
  \put(33,97)     {\pg k}
  \put(32,175)    {\pg l}
  } }
Note that the manifold $\M_{\X;q_1q_2\gamma\delta}$ is not simply connected (and 
also has nonvtrivial second homotopy group); so even though 
its boundary is a two-sphere, $\M_{\X;q_1q_2\gamma\delta}$ is not a three-ball.
Further, part of the ribbon graph in \erf{im20} is drawn with dashed lines, which
indicates that that part of the ribbons shows its `black' side. To make it show its
`white' side we rotate this part of the ribbon graph by $\pi$ around
a horizontal axis; this way $\M_{\X;q_1q_2\gamma\delta}$ can be redrawn as
  \eqpic{im21} {320} {130} {
  \put(0,133) {$ \M_{\X;q_1q_2\gamma\delta} ~= $}
  \put(85,0)  { \Includepicclal{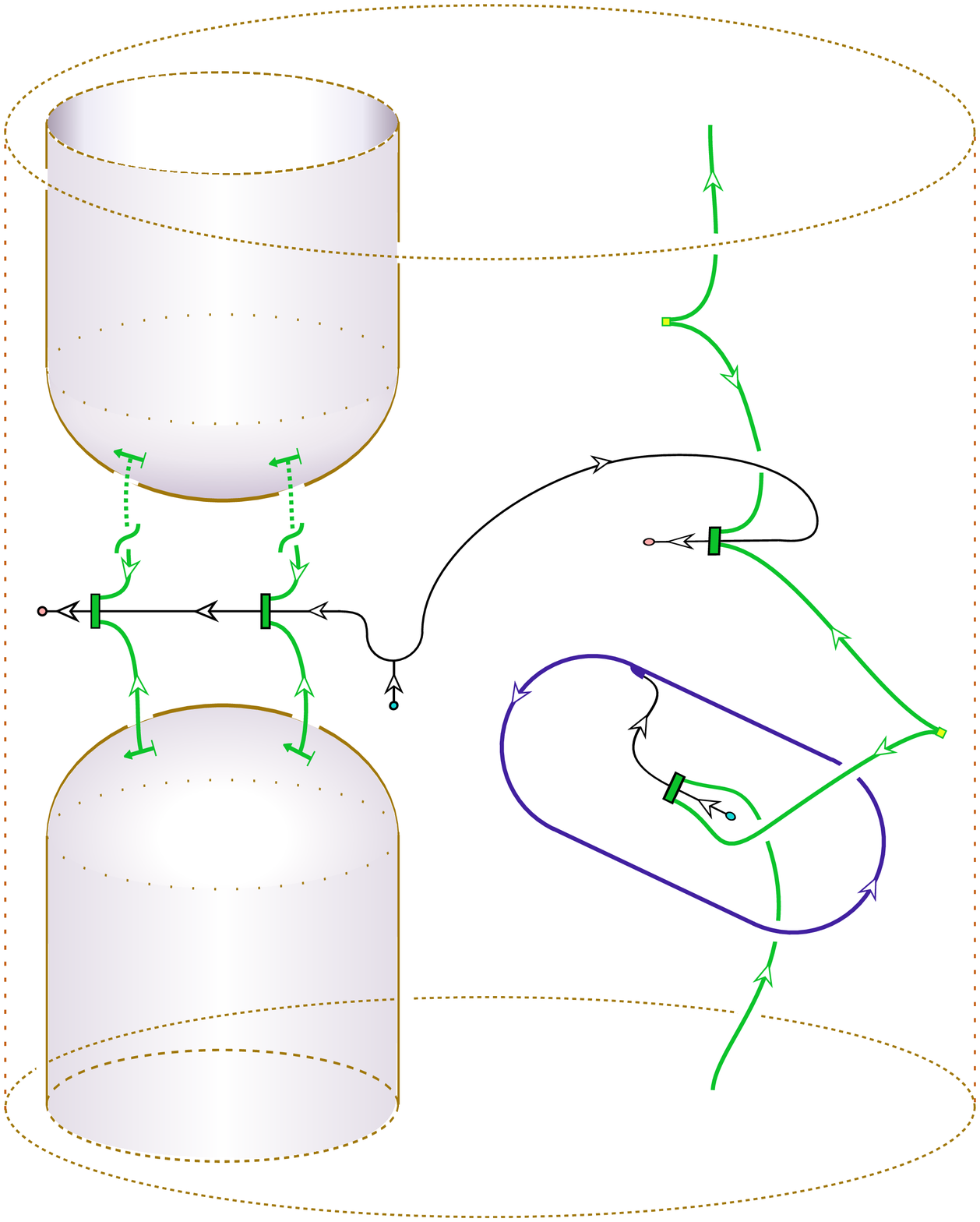} 
  \put(167,35)    {\pg {q_2}}
  \put(168,196)   {\pg {\bar q_2}}
  \put(200,127)   {\pg {\bar q_1}}
  \put(206,104)   {\pg {q_1}}
  \put(147,88.5)  {\pl {\phi_\gamma}}
  \put(159,145.2) {\pl {\phi_\delta}}
  \put(127,77)    {\pb M}
  \put(56,128)    {\pl{\phi_\alpha}}
  \put(17,128)    {\pl{\phi_\beta}}
  \put(66,99)     {\pg\ia}
  \put(63,178)    {\pg\ja}
  \put(30,99)     {\pg k}
  \put(28,178)    {\pg l}
  } }

\medskip

Analogously as in the case of boundary factorization, we are primarily interested
in the contribution of the vacuum channel to the invariant $Z(\M_{\X;q_1q_2\gamma
\delta})$. Extracting this contribution by applying the projetion $\pi_0$ we find
  \be
  \ctpdc{\Phi_\alpha}{\Phi_\beta}{M}0{\nl}{\nl} = \frac1{S_{0,0}}
  \sum_{q_1,q_2\in\I}\sum_{\gamma,\delta}\dim(U_{q_1})\,\dim(U_{q_2})\,
  (\cbulk_{q_1,q_2})\Inv_{\;\delta\gamma}\, Z(\Mb_{\X;q_1q_2\gamma\delta})
  \labl{ctpdc_1}
where $\Mb_{\X;q_1q_2\gamma\delta}$ is the closed three-manifold
$S^2\Times S^1$ with an embedded ribbon graph as shown in the following picture:
  \eqpic{im22} {320} {128} {
  \put(-4,132) {$ \Mb_{\X;q_1q_2\gamma\delta} ~= $}
  \put(85,0)  { \Includepicclal{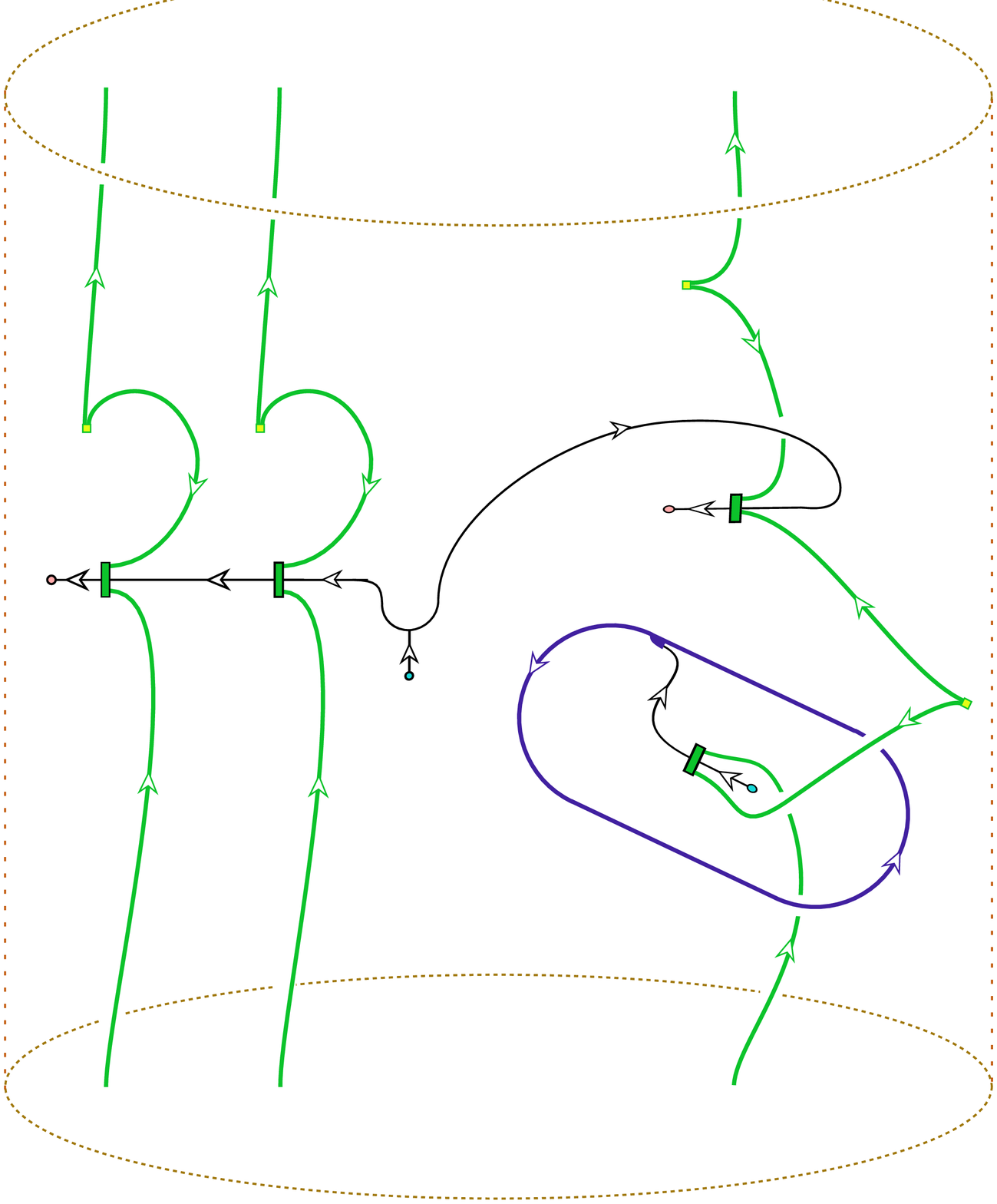} 
  \put(168,30)    {\pg {q_2^{}}}
  \put(168,199)   {\pg {\bar q_2^{}}}
  \put(202,127)   {\pg {\bar q_1}}
  \put(207,104)   {\pg {q_1}}
  \put(148,90)    {\pl {\phi_\gamma}}
  \put(160,145.7) {\pl {\phi_\delta}}
  \put(127,78)    {\pb M}
  \put(57,128.5)  {\pl{\phi_\alpha}}
  \put(18,128.5)  {\pl{\phi_\beta}}
  \put(64.4,30)   {\pg\ia}
  \put(75,183)    {\pg\ib}
  \put(25,30)     {\pg k}
  \put(35,183)    {\pg \kb}
  } }
As one ingredient, \erf{im22} contains, in the lower right quarter of the picture,
a morphism between the simple objects $U_{q_2}$ and $U_{\bar q_1}$. As a 
consequence, the number \erf{im22} is actually zero unless $q_1\eq\bar q_2
\,{=:}\,q$.  In the latter case, repeating the arguments relating the equalities 
\erf{C_disc_boundmorph} and \erf{C_disc_boundmorph1}, \erf{im22} can be rewritten as
  \eqpic{bulkfactcoeff_1} {320} {111} {
  \put(0,114) {$ \Mb_{\X;q\bar q\gamma\delta} ~= $}
  \put(85,0)  { \Includepicclal{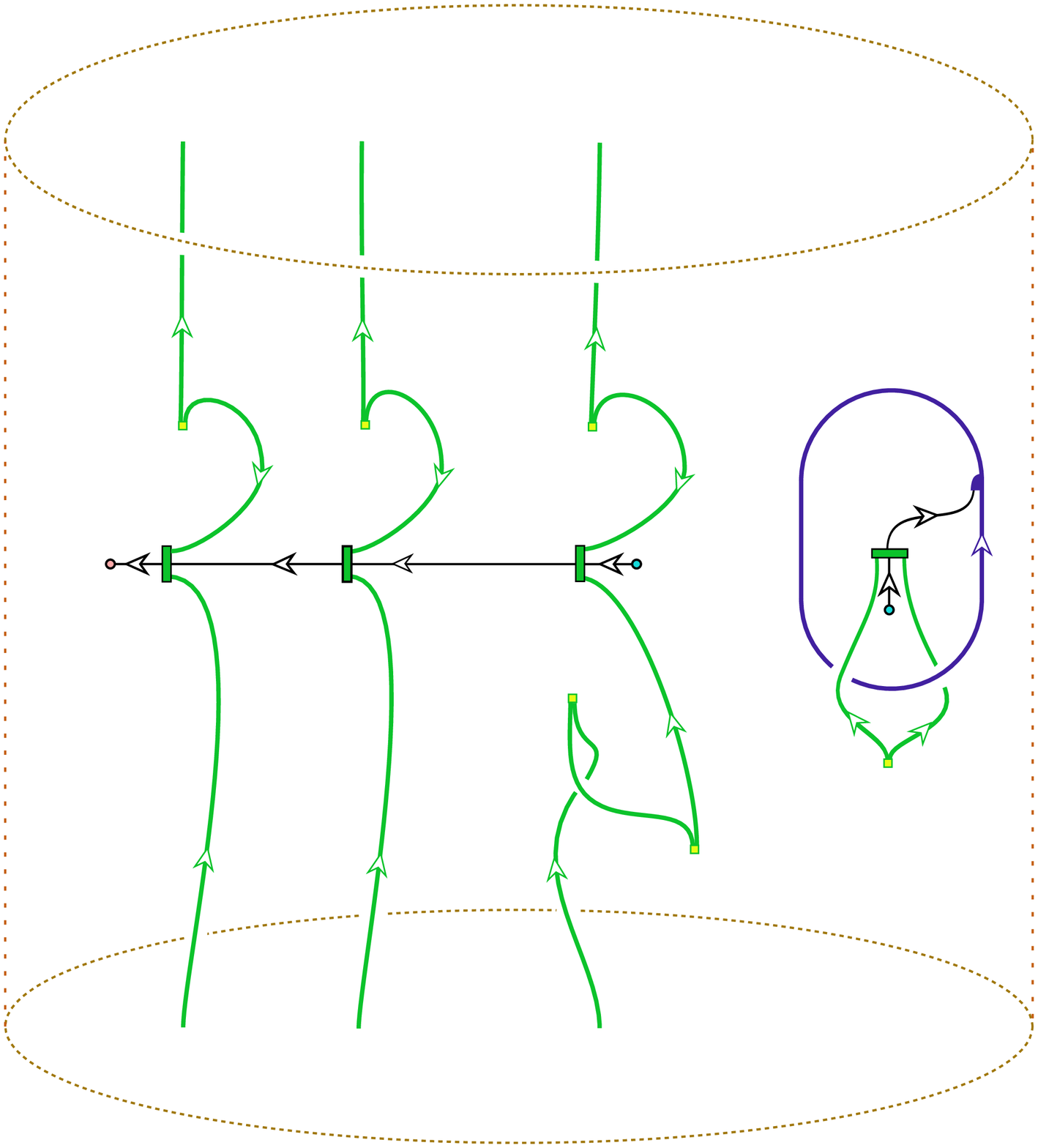} 
  \put(130,29)    {\pg {\qb}}
  \put(142,162)   {\pg q}
  \put(175.5,90)  {\pg q}
  \put(204,90)    {\pg {\qb}}
  \put(196.4,125) {\pl {\phi_\gamma}}
  \put(120.8,114) {\pl {\phi_\delta}}
  \put(162,131)   {\pb M}
  \put(70,114)    {\pl{\phi_\alpha}}
  \put(31.7,114)  {\pl{\phi_\beta}}
  \put(79.2,29)   {\pg\ia}
  \put(90,162)    {\pg\ib}
  \put(41,29)     {\pg k}
  \put(50,162)    {\pg \kb}
  } }

Now the disconnected part of the graph in \erf{bulkfactcoeff_1} that contains
the $M$-ribbon can be recognized as being nothing but the coefficient of the
one-point function on the disk (compare figure \erf{C_disc_boundmorph1} for $p\eq0$).
Moreover, in the disconnected part that contains the $U_\ia$- and $U_k$-ribbons
we can straighten out the $U_{\bar q}$-ribbon, which amounts to an additional
factor of $\theta_q{/}\dim(U_q)$. Then \erf{ctpdc_1} becomes
  \be
  \ctpdo{\PHi\alpha}{\PHi\beta}{M}0\nl\nl = \frac1{S_{0,0}}
  \sum_{q\in\I}\sum_{\gamma,\delta} \theta_q\,\dim(U_q)\,
  (\cbulk_{q\bar q})\Inv_{\;\delta\gamma}\, \almtps k\ia\qb\beta\alpha\delta \,
  \coronebd{\Phi_{\gamma}}M \,,
  \labl{ctpdc_2}
with 
  \be
  \almtps k\ia\qb\beta\alpha\delta = Z(\Almtps k\ia\qb\beta\alpha\delta)
  \ee
the invariant of the ribbon graph $\Almtps k\ia\qb\beta\alpha\delta$ 
shown in \erf{I_qdelta}.


\section{The classifying algebra}\label{s3}

We now combine the results of the boundary and bulk factorizations. Expressing 
the equalities \erf{res_boundfact} and \erf{ctpdc_2} in terms of the reflection 
coefficients $\onefsc\ia\alpha M\eq\coronebD{\Phi_\alpha}M$ 
that we introduced in \erf{refl_coeff}, we learn that
  \be
  \onefsc\ia\alpha M\, \onefsc k\beta M = \sum_{q\in\I} \sum_{\gamma=1}^{\Z_{q\qb}}
  \CAC \ia\alpha k\beta q\gamma\, \onefsc q\gamma M \,,
  \labl{CLA}
where $M$ is any elementary boundary condition and, 
for $\ia,\ja,k\iN\I$ and for $\alpha\iN\{1,2,...\,,\Z_{\ia\ib}\}$ etc.,
the coefficients $\CAC \ia\alpha k\beta q\gamma$ are defined as
  \be
  \CAC \ia\alpha \ja\beta k\gamma := \frac{\theta_k \dim(U_k)}{S_{0,0}}
  \sum_{\delta=1}^{\Z_{k\kb}} (c_{k\kb}^{\text{bulk}})^{-1}_{\delta\gamma}\,
  \almtps \ja\ia\kb\beta\alpha\delta \,.
  \labl{scdef}

The following interpretation of the formula \erf{CLA} suggests itself: In view of 
the description \erf{bulkfield} of bulk fields, consider the complex vector space
  \be
  \CA \,:=\, \bigoplus_{\ia\in\I} \Homaa{U_\ia\otip A\otim U_\ib}A \,\,
  \labl{CA_as_vect}
consisting of bulk fields which can have a non-vanishing one-point correlator on 
the disk. We endow this vector space in the following manner with the structure 
of an algebra. For each $\ia\iN\I$ choose a basis $\{\phi^{\ia,\alpha}\,|\,\alpha
\eq1,2,...\,,\Z_{\ia\ib}\}$ of $\Homaa{U_\ia\otip A\otim U_\ib}A$. In this basis, 
we put the structure constants of the algebra \CA\ to be the numbers \erf{scdef}, 
i.e.\ we define the multiplication of \CA\ by 
  \be
  \phi^{\ia,\alpha}\cdot \phi^{\ja,\beta} := \sum_{k,\gamma} \CAC \ia\alpha
  \ja\beta k\gamma\, \phi^{k,\gamma} .
  \ee
Then for any elementary boundary condition $M$, the reflection coefficients
$\onefsc\ia\alpha M$ are the \rep\ matrices, in the basis
$\{\phi^{\ia,\alpha}\}$, for a \onedim\ \rep\ $\rho_M$ of \CA, i.e.
  \be
  \onefsc\ia\alpha M = \rho_M(\phi^{\ia,\alpha}) \,.
  \ee

We will in fact show that \CA\ is a semisimple commutative associative algebra,
so that every \CA-\rep\ is a direct sum of irreducible \onedim\ \rep s, and that the
isomorphism classes of \onedim\ \rep s are in bijection with the inequivalent
elementary boundary conditions. In particular, \CA\ plays indeed the role of a
{\em classifying\/} algebra for boundary conditions -- its \rep\ theory determines
the space of boundary conditions completely.

\bigskip

We proceed to establish the relevant properties of \CA. We first note that 
according to theorem 5.18 of \cite{fuRs4} the dimension
  \be
  \dimc(\CA) = \sum_{\ia\in\I} \dimc(\Homaa{U_\ia\otip A\otim U_\ib}A)
  = \sum_{\ia\in\I} \Z_{\ia\ib}
  \ee
of \CA\ is equal to the number of inequivalent simple modules of the Frobenius
algebra $A$. Moreover, the proof of this equality rests on the non-degeneracy of a 
certain matrix $\widetilde S^A$ (defined in (5.97) of \cite{fuRs4}), and the entries 
of $\widetilde S^A$ are, in turn, proportional to the reflection coefficients 
$\onefsc\ia\alpha M$, see e.g.\ formula (4.21) of \cite{fuRs10}. This means in 
fact that, first, every simple $A$-module gives rise to a \onedim\ (and hence
in particular irreducible) \rep\ of \CA, and second, in this way inequivalent 
simple $A$-modules give non-isomorphic \CA-\rep s. 
It follows that the number $n_\text{simp}(\CA)$ of non-isomorphic irreducible 
\CA-\rep s is at least as large as the number of inequivalent simple $A$-modules, 
and thus at least as large as the dimension $\dimc(\CA)$: 
  \be
  n_\text{simp}(\CA) \ge \dimc(\CA) \,,
  \labl{ineq}

We will come back to this inequality below: once we have established associativity, 
we will see that the inequality is actually saturated. But for now, we switch to 
the proof of commutativity of \CA, which turns out to be easy:

\medskip\noindent
{\boldmath$(\CA1)$}~\,\emph{The algebra \CA\ is commutative.}

\smallskip\noindent
Indeed, by merely using the defining properties of the symmetric special 
Frobenius algebra $A$ and the fact that the morphisms $\phi_\alpha$ and 
$\phi_\beta$ are morphisms of $A$-bimodules, it follows that the ribbon graph 
$\Almtps k\ia\qb\beta\alpha\delta$ can be transformed into the graph
$\Almtps \ia k\qb\alpha\beta\delta$ without changing the value of the 
invariant. Thus $\almtps k\ia\qb\beta\alpha\delta 
\eq \almtps \ia k\qb\alpha\beta\delta$, which in turn implies that
  \be 
  \CAC \ia\alpha k\beta q\gamma = \CAC k\beta \ia\alpha q\gamma \,.
  \labl{sym}
This argument is in fact nothing else than the one used to show that the correlators 
constructed according to the rules of \cite{fuRs4} are independent of the choice of
dual triangulation of the world sheet. It also has the virtue of showing at 
the same time that actually $\almtps k\ia\qb\beta\alpha\delta$ is totally symmetric 
in all three pairs $(k,\beta)$, $(\ia,\alpha)$ and $(\qb,\delta)$ of indices.

As a step towards further properties we express the structure constants 
\erf{scdef} in a more accessible form. Consider the endomorphism that corresponds 
to the ribbon graph in the picture \erf{I_qdelta}. By performing on this morphism 
two moves of the form displayed in \erf{IV_67_for_A}, which each contribute an 
\FFA-coefficient as a factor, one can rewrite $\almtps k\ia\qb\beta\alpha\delta$ as
  \be
  \almtps k\ia\qb\beta\alpha\delta = \dim(A)\sum_{\zeta,\eta_1\eta_2}
  \FA k\ia\ib\kb\alpha\beta\zeta{q\qb}{\eta_1\eta_2}\;
  \FA q\qb q\qb\delta\zeta\nl{00}{\nl\nl}\;\IFAAnumb k\ia\qb{\eta_1}{\eta_2}
  \ee
with $\IFAAnumb k\ia\qb{\eta_1}{\eta_2}$ the morphism
  \eqpic{I_const_morph}{245}{63}{
  \put(0,65){$ \IFAAnumb k\ia\qb{\eta_1}{\eta_2} ~:=~ $} 
  \put(214,65){$ \in\End(\one) $} 
  \put(70,0){ 
  \put(-2,0) {   \Includepicclal{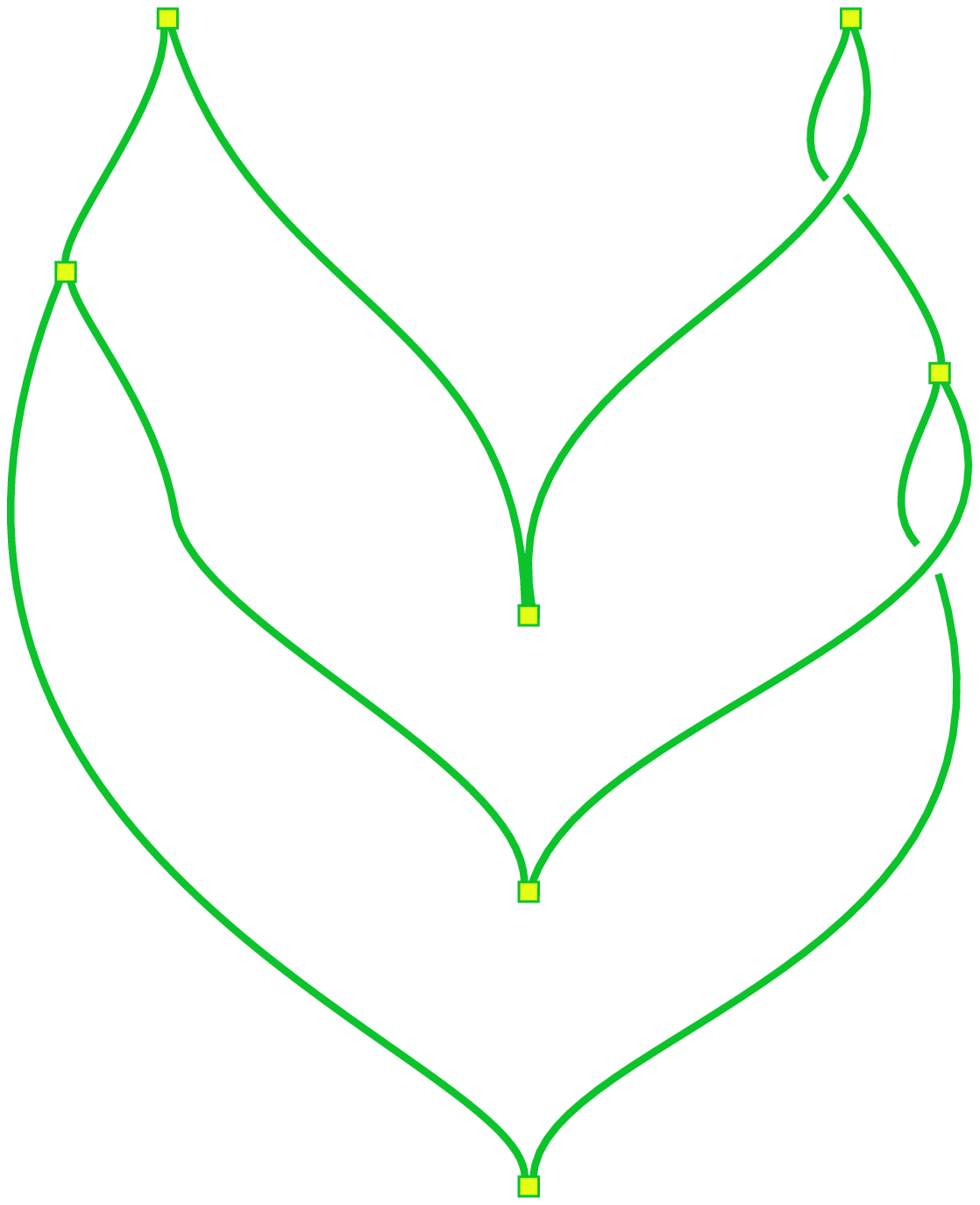}} 
  \put(122.4,102) {\pl {\eta_2^{}}} 
  \put(-4,91)     {\pg k}
  \put(24,91)     {\pg \ia}
  \put(-1.5,116)  {\pl {\eta_1^{}}}
  \put(9.6,132)   {\pg q}
  \put(32.3,132)  {\pg \qb}
  } } 
When inserted into \erf{scdef}, together with \erf{tpf_Faa} it follows that the
structure constants can be written as
  \be
  \CAC \ia\alpha k\beta q\gamma = \dim(U_q)\,\theta_q \sum_{\eta_1,\eta_2}
  \FA k\ia\ib\kb\alpha\beta\gamma{q\qb}{\eta_1\eta_2}\,
  \IFAAnumb k\ia\qb{\eta_1}{\eta_2} \,.
  \labl{SCFA}

Further it is easy to see that when one of the three bulk fields, say $\PHi\delta$, 
involved in $\almtps k\ia\qb\beta\alpha\delta$ is the identity field, we have
  \be
  \tildeI k\ia\beta\alpha \equiv \almtps k\ia0\beta\alpha\nl
  = \delta_{\kb,\ia}\, \dim(A)\, \FA\ib\ia\ib\ia\alpha\beta\nl{00}{\nl\nl}\,
  \R\ib\ia0\nl\nl\, \F{\ia\ib}\ia\ia00\nl\nl\nl\nl \,,
  \ee
where the last factor is a fusing coefficient as defined in \erf{Fmat} and 
$\R\ib\ia0\nl\nl$ is a braiding coefficient.
With the help of the identities \erf{tpf_Faa} and (see (2.2.46) of \cite{fuRs4}) 
$\R\ib \ia0\nl\nl\, \F{i\ib}ii00\nl\nl\nl\nl \eq (\theta_i\,\dim(U_i))^{-1}_{}$,
this can be rewritten as
  \be
  \almtps k\ia0\beta\alpha\nl
  = \delta_{\kb,\ia}\, \frac{S_{0,0}}{\theta_\ia\,\dim(U_\ia)} \,
  c^{\text{bulk}}_{\ib\ia,\beta\alpha} \,.
  \labl{almtps_cbulk}
We can now conclude:

\smallskip\noindent
{\boldmath$(\CA2)$}~\,\emph{The algebra \CA\ is unital.}

\smallskip\noindent
Indeed, using \erf{almtps_cbulk} together with the total symmetry of
$\almtps k\ia\qb\beta\alpha\delta$ in its three pairs of indices one checks that
  \be
  \CAC \ia\alpha 0\nl k\gamma = \delta_{k\ia}\,\delta_{\alpha\gamma}
  = \CAC 0\nl \ia\alpha k\gamma \,,
  \ee
i.e.\ the element $t^{0\nl}\iN\CA$, which corresponds to the identity
field, is a unit for the product of the algebra \CA.

\medskip

The proof of associativity turns out to be harder. Note in particular that 
associativity is not manifest when the structure constants are written in the form
\erf{SCFA}. A crucial observation is instead that our arguments leading to
the formula \erf{CLA} generalize to the case of a correlator of a general number 
$n$ (instead of just 2) of bulk fields. Rather than leading to an ordinary binary
product, one then obtains an $n$-ary product on the vector space \CA, and the
structure constants for this product in the basis $\{t^{\ia\alpha}\}$ are given by
  \be
  \CAC {\ia_1}{\alpha_1}{\ia_2}{\alpha_2,...,\ia_n\alpha_n} {~~~~~~~~~~~k}\gamma
  = \frac{\theta_k \dim(U_k)}{S_{0,0}}\sum_\delta
  (c_{k\kb}^{\text{bulk}})^{-1}_{\delta\gamma}\,
  \almtps {\ia_n}{\ia_{n-1}...\ia_1\,}\kb{\alpha_n}{\alpha_{n-1}...\alpha_1}\delta \,,
  \labl{n-scdef}
with $\almtps {\ia_n}{\ia_{n-1}...\ia_1}\kb{\alpha_n}{\alpha_{n-1}...\alpha_1}\delta$
the invariant of the ribbon graph
  \eqpic{I_n} {290} {97} {
  \put(-5,99) {$ \Almtps {\ia_n}{\ia_{n-1}...\ia_1\,}\kb{\alpha_n}{\alpha_{n-1}
                 ...\alpha_1}\delta ~= $}
  \put(95,0)    {\Includepicclal{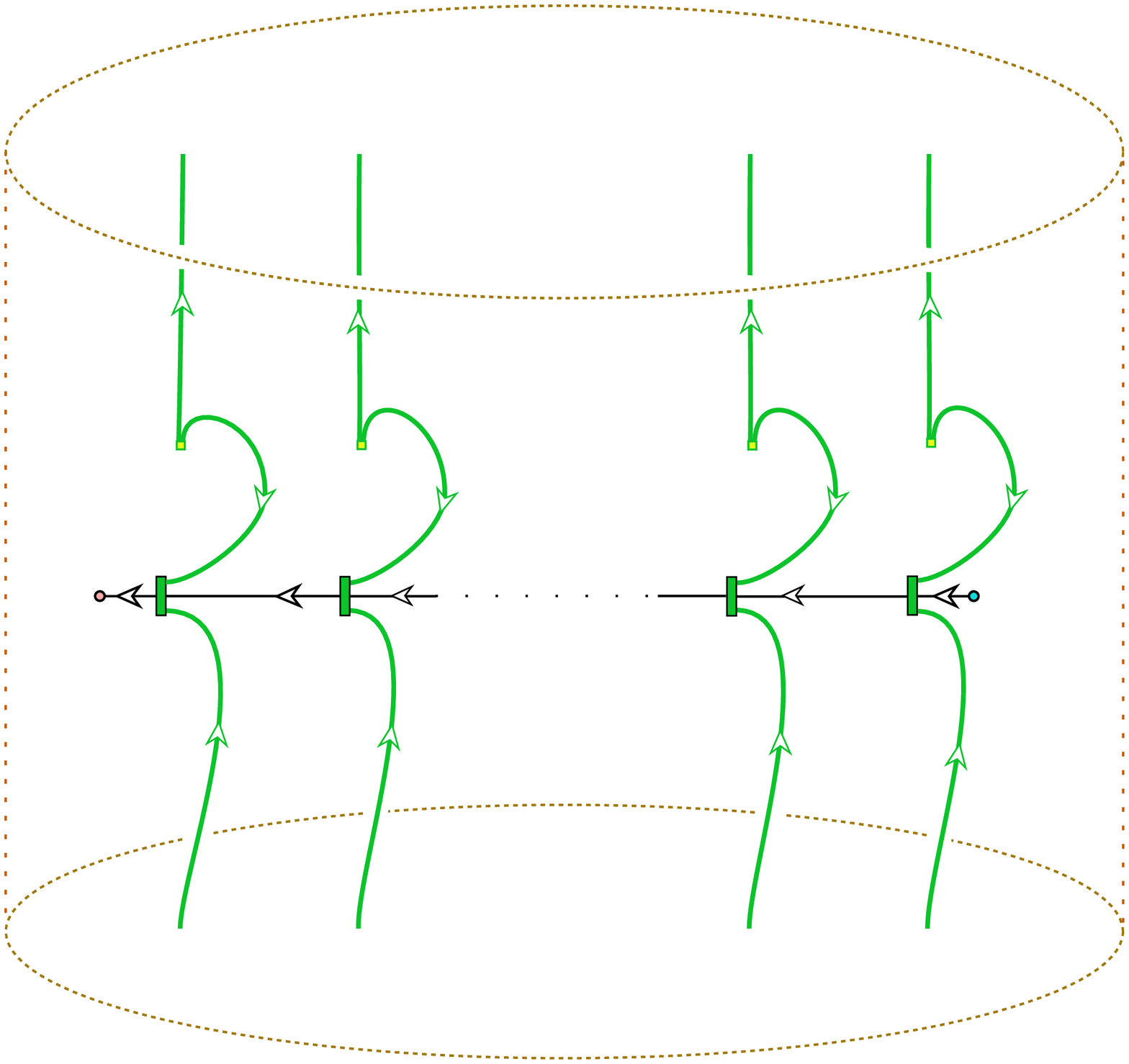}}
  \put(148,-45){
  \put(-5,171)   {\pg {\ib_n} }
  \put(-15,73)   {\pg {\ia_n} }
  \put(32,171)   {\pg {\ib_{n-1}} }
  \put(20,73)    {\pg {\ia_{n-1}} }
  \put(109,171)  {\pg {\ib_1^{}} }
  \put(97.5,73)  {\pg {\ia_1^{}} }
  \put(145,171)  {\pg k }
  \put(133,73)   {\pg \kb }
  \put(-26,126)  {\pl {\phi_{\!\alpha_n}^{}}}
  \put(3.9,133.8){\begin{turn}{320}\pl {\phi_{\!\alpha_{n-1}}^{}}\end{turn} }
  \put(87,126)   {\pl {\phi_{\!\alpha_1}^{}}}
  \put(123.5,126){\pl {\phi_{\!\delta}^{}}}
 } }
in $S^2\Times S^1$. Note that by the same argument as in the case of the binary product, 
the structure constants of the $n$-ary products are totally symmetric in all pairs of 
lower indices. In other words, the $n$-ary products are totally commutative.

The numbers \erf{I_n} can again be expressed in terms of 
\FFA-coefficients. In particular, for the case $n\eq3$ we find
  \be
  \almtps k{\ia\ja\,}\qb\gamma{\beta\alpha}\delta = \!\!
  \sum_{l_1,l_2,m_1,m_2,p} \sum_{\mu,\nu_1,\nu_2,\rho,\atop\sigma_1,\sigma_2,\tau_1,\tau_2}
  \FA k\ja\jb\kb\beta\gamma\mu {l_1l_2}{\nu_1\nu_2}\,
  \FA {l_1}\ia\ib{\ell_2}\alpha\mu\rho {m_1m_2}{\sigma_1\sigma_2}\,
  \FA {m_1}\qb q{m_2}\delta\rho\nl {p\bar p}{\tau_1\tau_2} \,
  L_{\mu\nu_1\nu_2\rho\sigma_1\sigma_2\tau_1\tau_2}
  ^{k\ia\ja\,\qb;l_1l_2m_1m_2p}
  \labl{FFFL}
with $L_{\mu\nu_1\nu_2\rho\sigma_1\sigma_2\tau_1\tau_2}
^{k\ia\ja\,\qb;l_1l_2m_1m_2p}$ the invariant of the ribbon graph
  \eqpic{I_3FA} {370} {137} {
  \put(10,144)  {$ \mathcal L_{\mu\nu_1\nu_2\rho\sigma_1\sigma_2\tau_1\tau_2}
                            ^{k\ia\ja\,\qb;l_1l_2m_1m_2p} ~:= $}
  \put(133,0)    {\Includepicclal{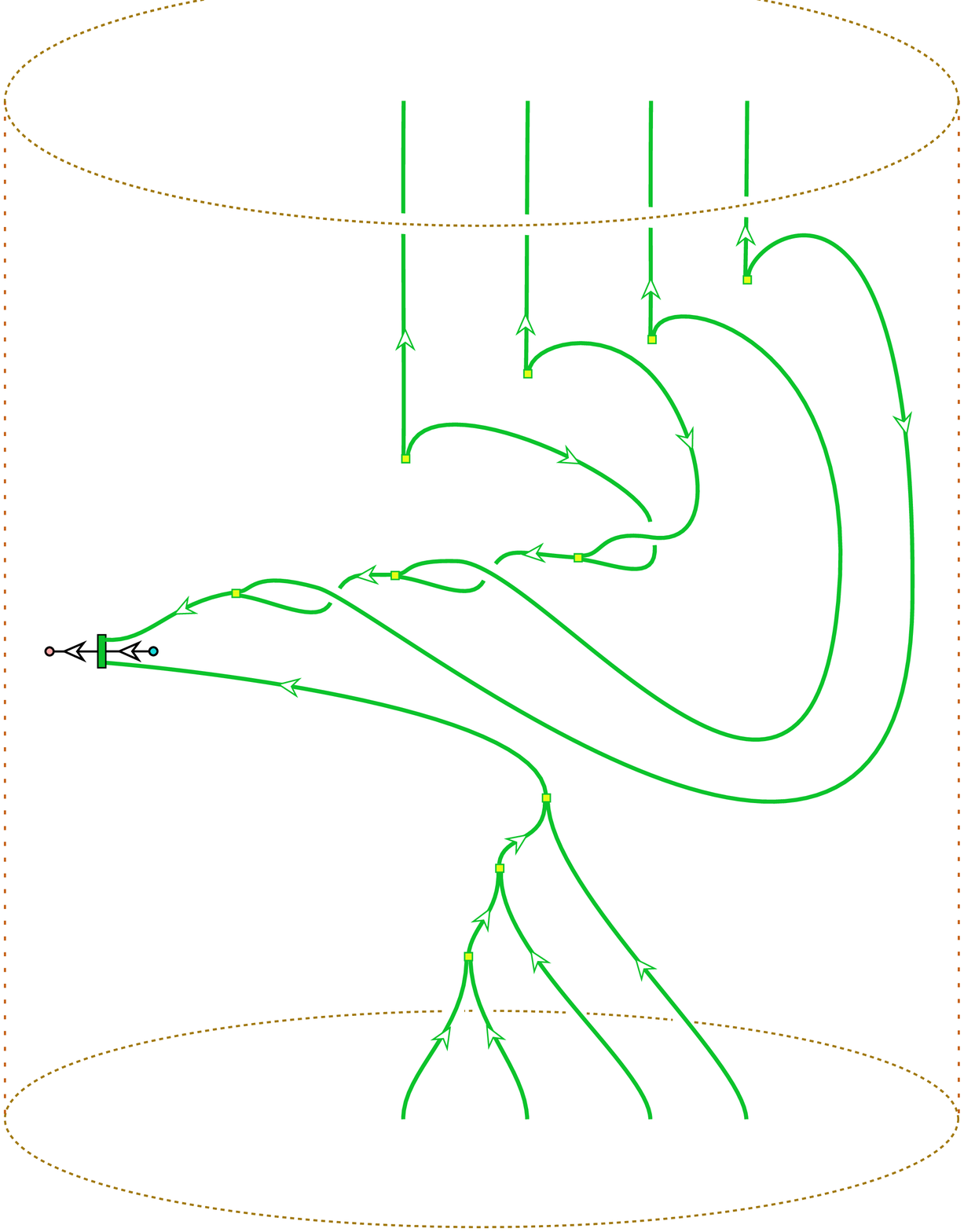}}
  \put(189,-55){
  \put(118,82)    {\pg \qb}
  \put(119,312)   {\pg \qb}
  \put(39,311)    {\pg k}
  \put(68,312)    {\pg \ja}
  \put(97,312)    {\pg \ia}
  \put(39,81)     {\pg k}
  \put(67.4,82)   {\pg \ja}
  \put(95.6,82)   {\pg \ia}
  \put(54,117)    {\pl {\nu_1^{}}}
  \put(48.9,130.3){\pg {l_1^{}}}
  \put(62,137)    {\pl {\sigma_1^{}}}
  \put(55,149)    {\pg {m_1}}
  \put(72.3,153.5){\pl {\tau_1^{}}}
  \put(24,171)    {\pg p}
  \put(-25,197)   {\pg {\bar p}}
  \put(-5.5,208)  {\pl {\tau_2^{}}}
  \put(8,193)     {\pg {m_2^{}}}
  \put(31.2,212)  {\pl {\sigma_2^{}}}
  \put(47,196.5)  {\pg {l_2^{}}}
  \put(73.4,216)  {\pl {\nu_2^{}}}
 } }
in $S^2\Times S^1$. This invariant vanishes unless $p\eq0$, in which case \erf{FFFL} 
simplifies to (using also \erf{tpf_Faa})
  \be
  \almtps k{\ia\ja\,}\qb\gamma{\beta\alpha}\delta
  = S_{0,0}\, \sum_{l_1,l_2} \sum_{\mu,\nu_1,\nu_2,\atop\rho,\sigma_1,\sigma_2}
  \FA k\ja\jb\kb\beta\gamma\mu {l_1l_2}{\nu_1\nu_2}\,
  \FA {l_1}\ia\ib{\ell_2}\alpha\mu\rho {q\qb}{\sigma_1\sigma_2}\,
  c^{\text{bulk}}_{q\qb,\rho\delta}\;\tilde\varpi^{kji\qb,l_1l_2}_{\nu_1\nu_2,\sigma_1\sigma_2}
  \ee
with $\tilde\varpi^{k\ja\ia\qb,l_1l_2}_{\nu_1\nu_2,\sigma_1\sigma_2}$ the morphism
  \eqpic{I_const_morph3} {210} {88} {
  \put(0,95)  {$ \tilde\varpi^{k\ja\ia\qb,l_1l_2}_{\nu_1\nu_2,\sigma_1\sigma_2} ~:= $}
  \put(91,0)    {\put(2,0){\Includepicclal{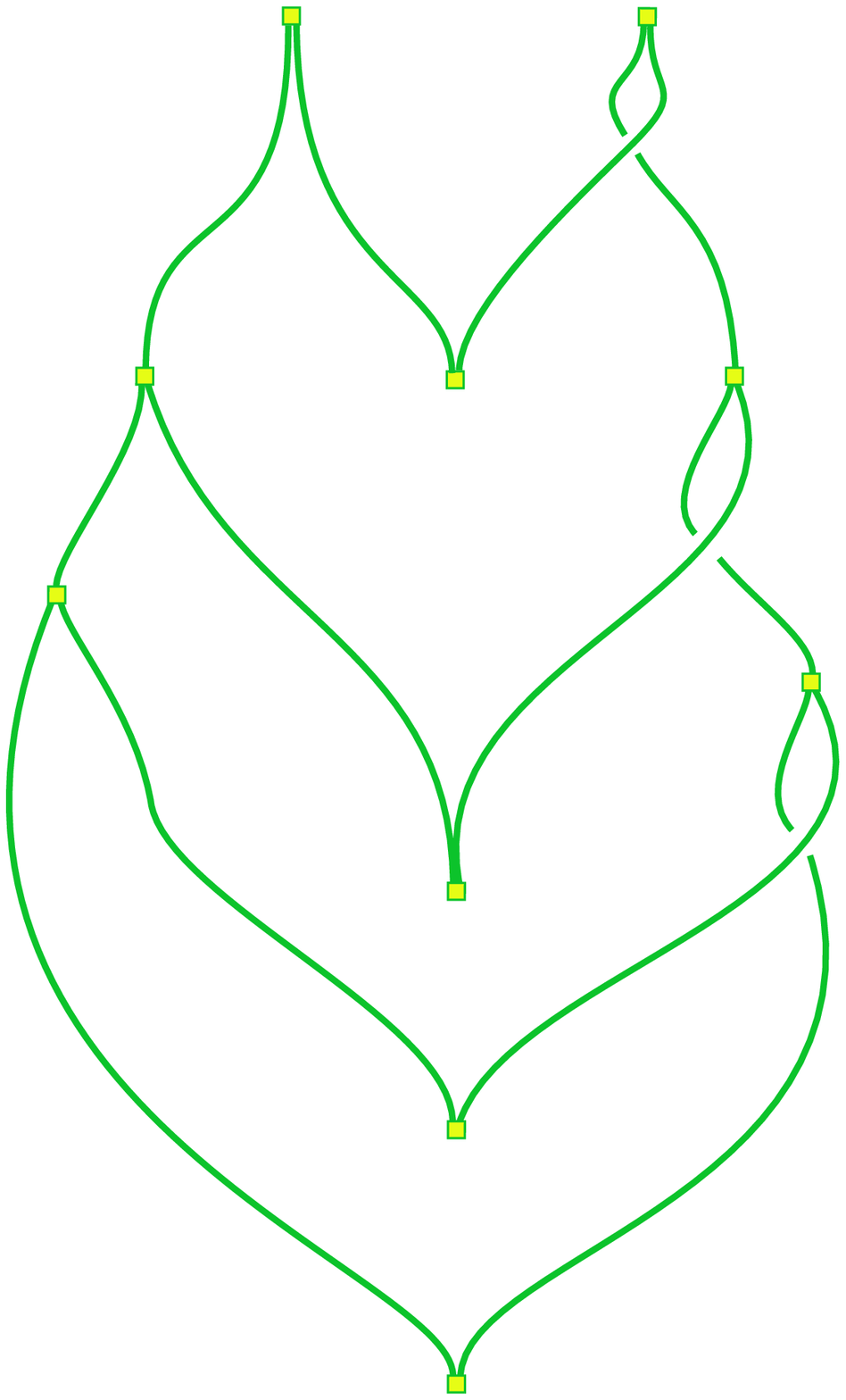}}
  \put(121.7,102.7) {\pl {\nu_2^{}}}
  \put(93,134)    {\pg {l_2^{}}}
  \put(111.2,146) {\pl {\sigma_2^{}}}
  \put(-4,92)     {\pg k}
  \put(24.5,92)   {\pg \ja}
  \put(-.8,116)   {\pl {\nu_1^{}}}
  \put(9,132)     {\pg {l_1}}
  \put(33,132)    {\pg \ia}
  \put(11.4,147)  {\pl {\sigma_1}}
  \put(27.3,173)  {\pg q}
  \put(52.3,173)  {\pg \qb}
 } }
After using dominance in $\Hom{U_{l_1}\oti U_{l_2}}{U_{l_1}\oti U_{l_2}}$, only the
contribution from the tensor unit survives in the resulting summation. As a
consequence \erf{I_const_morph3} can be factorized as
  \be  
  \tilde\varpi^{kji\qb,l_1l_2}_{\nu_1\nu_2,\sigma_1\sigma_2}
  = \delta_{l_2,\bar l_1}\, \dim(U_{l_1})\, \theta_{l_1}\,
  \IFAAnumb {l_1}\ia\qb{\sigma_1}{\sigma_2}\, \IFAAnumb k\ja{\bar l_1}{\nu_1}{\nu_2}
  \ee
with the factors $\varpi$ as defined in \erf{I_const_morph}.

We now insert this result into \erf{n-scdef} and also perform the $\delta$-summation;
this yields the structure constants of the ternary product as
  \be
  \CAC \ia\alpha \ja{\beta, k\gamma} {~~~q}\delta
  = \dim(U_q)\,\theta_q\, \sum_{l\in\I} \sum_{\mu,\nu_1,\nu_2,\sigma_1,\sigma_2}\!
  \dim(U_l)\,\theta_l\; \FA k\ja\jb\kb\beta\gamma\mu {l\bar l}{\nu_1\nu_2}\,
  \FA l\ia\ib{\bar l}\alpha\mu\delta {q\qb}{\sigma_1\sigma_2}\, 
  \IFAAnumb l\ia\qb{\sigma_1}{\sigma_2}\, \IFAAnumb k\ja{\bar l}{\nu_1}{\nu_2} \,.
  \ee
Thus by comparison with \erf{SCFA} it follows that
  \be
  \CAC \ia\alpha \ja{\beta, k\gamma} {~~~q}\delta
  = \sum_{\ell,\mu} \CAC \ja\beta k\gamma \ell\mu\, \CAC \ia\alpha \ell\mu q\delta \,.
  \labl{ass1}
On the other hand, using in addition the total commutativity of the products, one
also has
  \be
  \CAC \ia\alpha \ja{\beta, k\gamma} {~~~q}\delta
  = \CAC k\gamma \ja{\beta,\ia\alpha}{~~~q}\delta
  = \sum_{\ell,\mu} \CAC \ja\beta \ia\alpha \ell\mu\, \CAC k\gamma \ell\mu q\delta
  = \sum_{\ell,\mu} \CAC \ia\alpha \ja\beta \ell\mu\, \CAC \ell\mu k\gamma q\delta \,.
  \labl{ass2}
Together, \erf{ass1} and \erf{ass2} establish:

\smallskip\noindent
{\boldmath$(\CA3)$}~\,\emph{The algebra \CA\ is associative.}

\medskip

Once we know that \CA\ is associative, we are finally in a position to address the 
issue of semisimplicity. Namely, as any \findim\ associative algebra, 
\CA\ is isomorphic, as an \CA-module, to a direct sum of all inequivalent
indecomposable projective \CA-modules, each one occuring with a multiplicity
given by the dimension of the corresponding simple \CA-module (see e.g.\ Satz G.10 
of \cite{JAsc}). Comparing dimensions, this tells us in particular that the dimension 
$\dimc(\CA)$ is at least as large as the number $n_\text{simp}(\CA)$of 
non-isomorphic irreducible \CA-\rep s. In view of the opposite inequality \erf{ineq},
this means that these two numbers are atually equal,
  \be
  n_\text{simp}(\CA) = \dimc(\CA) \,.
  \ee
This is only possible if atually each of these irreducible \rep s is \onedim\ as
well as projective, and thus \CA\ is isomorphic to the direct sum 
of the corresponding \onedim\ \CA-modules. In particular this means:

\smallskip\noindent
{\boldmath$(\CA4)$}~\,\emph{The algebra \CA\ is semisimple.}

\medskip

Thus we have established the claim that the classifying algebra \CA\ is a 
semisimple commutative associative algebra.

\medskip
Let us also add the following observations:
\nxtp
When the Frobenius algebra $A$ is Morita equivalent to the tensor unit $\one$ -- 
the so-called Cardy case -- the expression \erf{CA_as_vect} for \CA\ reduces to
$\bigoplus_{\ia\in\I}\Hom{U_\ia\oti U_\ib}\one\eq\bigoplus_{\ia\in\I} \complex$,
and the structure constants become the fusion rules,
$\CAC\ia{\,}{\,\ja}{}{\!\!k}{}\eq\dimc\Hom{U_\ia\oti U_\ja\oti U_k}\one$.
Thus in this case the classifying algebra is nothing else than the Verlinde
algebra $K_0(\C)\,{\otimes_\zet}\,\complex$ of the category \C. In particular,
it then has a canonical basis.
\nxtp
The structure constants furnish the matrix elements of the regular \rep\ of \CA. 
Since \CA\ is semisimple, the regular \rep\ is fully reducible, hence its
\rep\ matrices can be diagonalized simultaneously.
\nxtp
By \erf{scdef} and \erf{almtps_cbulk} we have
  \be
  \CAC \ia\alpha \ja\beta 0\nl = \frac1{S_{0,0}}\,
  (c_{00}^{\text{bulk}})^{-1} \almtps \ja\ia0\beta\alpha\nl 
  = [\theta_\ia\dim(U_\ia)\,c_{00}^{\text{bulk}}]^{-1} \delta_{\jb,\ia}\,
  c^{\text{bulk}}_{\ia\ib,\alpha\beta} \,.
  \labl{CACij0}
Thus by the non-degeneracy of the bulk two-point functions on the sphere
these structure constants provide a non-degenerate bilinear form on \CA.
This allows us to introduce structure constants with only lower indices:
  \be
  \CACl \ia\alpha \ja\beta k\gamma := \sum_{\ell,\delta}
  \CAC \ia\alpha \ja\beta \ell\delta\, \CAC \ell\delta k\gamma 0\nl \,.
  \ee
Inserting the expressions \erf{scdef}, \erf{CACij0} and \erf{tpf_00}, these 
are simply given by
  \be
  \CACl \ia\alpha \ja\beta k\gamma = \frac1{\dim(A)}\,
  \almtps \ja\ia k\beta\alpha\gamma \,.
  \ee
In particular, they are totally symmetric in the three pairs of indices.
\\
Also note that by associativity of \CA\ the bilinear form is invariant,
while by commutativity it is symmetric. Thus \CA\ has the structure of a
symmetric Frobenius algebra.
\nxtp
In the same way as in \erf{ass1} above it follows that for any $n$-ary product 
and any $\ell\eq1,2,...\,,n\,{-}\,1$ an analogous factorization of the $n$-ary 
product into $(\ell{+}1)$- and $(n{-}\ell{+}1)$-ary products holds, and all 
sequences of concatenations of arbitrary products are 
compatible. Thus the products defined by \erf{n-scdef} provide a
representation of the associative operad.

\medskip

With the previous results at hand, we are finally also in a position to discuss
the compatibility with the mapping class group \rep s\ on the spaces of conformal
blocks on the sphere. The mapping class group $G_n$ of 
the $n$-punctured sphere is generated by braiding homeomorphisms which interchange 
any two insertion points and by Dehn twists around each insertion point. Now 
consider the ribbon graph $\widehat{\mathcal K}$ shown in figure \erf{I_qdelta}, 
or rather its generalization $\widehat{\mathcal K}_n$ to the case of any number $n$ 
of bulk field insertions. The invariant $\widehat{K}_n\eq Z(\widehat{\mathcal K}_n)$
of this ribbon graph constitutes an endomorphism of the space of $n$-point conformal 
blocks on the sphere. 

Now one can easily convince oneself that commuting any 
generator of $G_n$, acting on the horizontal factor $S^2$ of the three-manifold, 
through $\widehat{\mathcal K}_n$ merely amounts to a change of the dual
triangulation of $S^2$ that underlies the horizontal part of the ribbon graph. 
By the independence from the choice of dual triangulation of the world sheet as
already invoked above, it therefore follows that the invariant $\widehat{K}_n$ 
commutes with the action of $G_n$ on the space of conformal blocks, i.e.\ that
$\widehat{K}_n$ intertwines the action of the mapping class group on the space of 
blocks. This argument is similar to the one showing invariance of the correlators 
under the mapping class group, compare section 3 of \cite{fjfrs}.

\vskip 4.5em

\noindent{\sc Acknowledgments:}
We thank Valya Petkova for urging us to use the results of the TFT approach
to RCFT correlators to revamp our understanding of the classifying algebra.
\\
JF is partially supported by VR under project no.\ 621-2006-3343.
CSc is partially supported by the Collaborative Research Centre 676 ``Particles,
Strings and the Early Universe - the Structure of Matter and Space-Time'',
and would like to thank Karlstad University for hospitality.
CSt gratefully acknowledges a K\"ahler research fellowship from the Center for 
Mathematical Physics Hamburg.

\newpage
\appendix

\section{Appendix}

\subsection{Chiral data}\label{app_chiral}

The chiral symmetries of a CFT can be realized in the structure of a conformal
vertex algebra. From a conformal vertex algebra one can, in turn, construct 
(see e.g.\ \cite{FRbe2}) a chiral CFT. Crucial information about the chiral 
theory is contained in the \rep\ category \C\ of the vertex algebra. For a 
\emph{rational} CFT \C\ is, by definition, a (semisimple) modular tensor category.

Despite its name, a chiral CFT is not a conventional quantum field theory: instead 
of local correlation functions it supplies a system of conformal blocks, i.e.\ 
vector bundles with connection over the moduli spaces of curves with marked 
points. A chiral CFT can, however, be used as an ingredient in the construction of
a full local CFT. For these quantum field theories, many aspects, including the 
computation of structure constants of operator product expansions, can
be discussed on the basis of \C\ as an abstract category (which can in particular 
be taken to be strict, i.e.\ with strictly associative tensor product and with 
tensoring by the tensor unit $\one$ acting as the identity on objects), without 
reference to its concrete realization as the \rep\ category of a vertex algebra.

A modular tensor category is in particular semisimple and \complex-linear and 
has a finite number of isomorphism classes of simple objects. We denote 
representatives for these classes by $U_i$, with $i$ taking values in a finite 
index set \I, and choose $\I\,{\ni}\,0$ such that $U_0\eq\one$ is the tensor 
unit. For any simple object $U$ its space of endomorphisms is isomorphic to \C,
$\End(U) \,{\equiv}\, \Hom U U \eq \complex\id_U$;
we identify the space $\End(\one,\one)\eq\complex\id_\one$ with \complex.
The (quantum) dimension of an object $V$ is denoted by $\dim(V)$.

\medskip

As \C\ is semisimple and \complex-linear, any morphism between objects $V$ and
$W$ of \C\ can be written as a sum of morphisms between simple subobjects of 
$V$ and of $W$; this property is known as {\em dominance\/}. Also,
the essential properties of the tensor product are already 
captured by the spaces of morphisms between tensor products of simple objects.
In particular we may express any morphism space as a direct sum of tensor 
products (over \complex) of spaces of three-point couplings of simple objects,
i.e.\ of the morphism spaces $\hom(U_i\oti U_j,U_k)$ and their duals
$\hom(U_k,U_i\oti U_j)$ with $i,j,k\iN\I$. This reduces the choice of bases in 
arbitrary morphism spaces to the choice of bases of all such spaces of 
three-point couplings. We select bases $\{\lambda_{ij}^{k\alpha}\}$ of all spaces 
$\hom(U_i\oti U_j,U_k)$ and keep them fixed throughout the paper; for the dual 
spaces $\hom(U_k,U_i\oti U_j)$ we choose the dual bases. We depict these bases as
  \eqpic{I:2.29}{240}{23} {
  \put(0,0)         {\Includepicclal{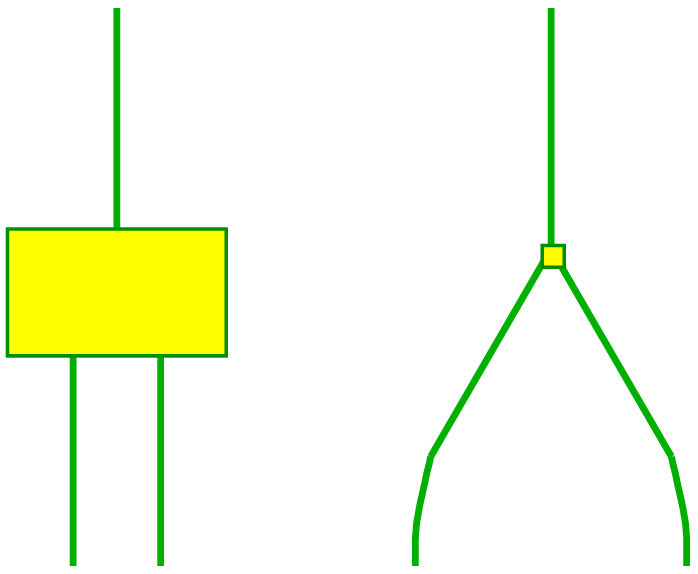} } 
  \put(114,20)       {and}
  \put(170,0)       {\Includepicclal{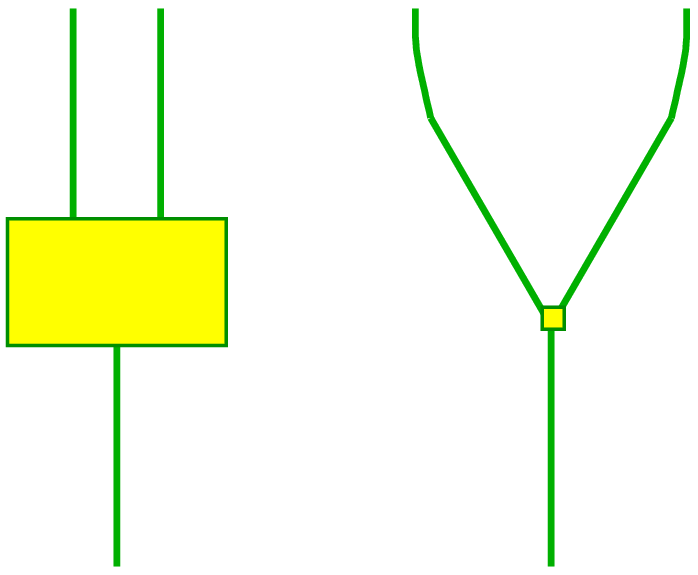} } 
  \put(5.1,29.5)  {\footnotesize$\lambda_{ij}^{k\alpha}$}
  \put(5.8,-7)    {\pg i}
  \put(10.7,65.3) {\pg k}
  \put(15.1,-7)   {\pg j}
  \put(33.7,28)   {$=$}
  \put(43.8,-7)   {\pg i}
  \put(58.7,65.3) {\pg k}
  \put(64.2,33.6) {\tiny$\alpha$}
  \put(73.1,-7)   {\pg j}
  \put(30,0){
  \put(145.5,29.4){\footnotesize$\bar\lambda^{ij}_{k\bar\alpha}$}
  \put(146.2,65.3){\pg i}
  \put(150.3,-7)  {\pg k}
  \put(156.6,66.2){\pg j}
  \put(174.7,30)  {$=$}
  \put(183.3,65.3){\pg i}
  \put(198.2,-7)  {\pg k}
  \put(203.4,27.3){\tiny$\bar\alpha$}
  \put(214.6,66.2){\pg j}
  } }
Lines labeled by $0\iN\I$ can be omitted, as they correspond to the tensor unit 
$U_0 \eq \one$. Further, taking $\bar k$ to be the unique
element of $\I$ for which $U_{\bar k}$ is isomorphic to $U_k^\vee$, we abbreviate
  \eqpic{picV-06}{250}{18} {
  \put(50,0)         {\Includepicclal{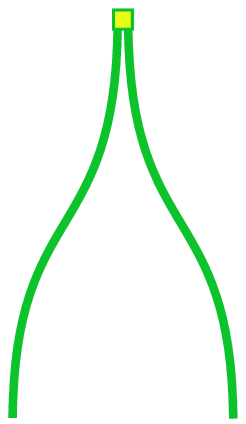} }
  \put(250,0)        {\Includepicclal{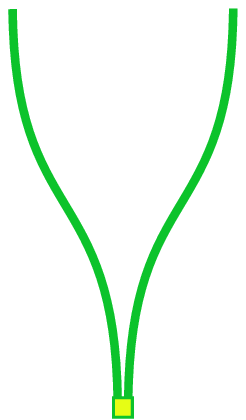} }
  \put(114,20)       {and}
  \put(49,-8)        {\pg \ia}
  \put(74,-8)        {\pg \ib}
  \put(70,0){
  \put(179,50)       {\pg \ia}
  \put(203,50)       {\pg \ib}
  }
  \put(-33,20)       {$ \lambda_{i\ib} \,\equiv\, \lambda_{i\ib}^{0\nl} ~= $}
  \put(168,20)       {$ \bar\lambda^{\ia\ib}_{} \,\equiv\, \bar\lambda^{\ia\ib}_{0\nl} ~= $}
  }
In this case the spaces are \onedim, so that the label takes only a single
value, for which we choose the symbol `$\circ$'. Moreover, just like in
the pictures just displayed we often suppress this label in our notations.

When expressing the bases of morphism spaces involving multiple tensor products
in terms of the chosen bases of three-point couplings, several choices are
possible, corresponding to the different possible orders in which to perform
the tensor products. The two possibilities for a triple tensor product are
related by a {\em fusing move\/}, according to
  \eqpic{Fmat} {220}{40}{
  \put(0,0)   {\Includepicclal{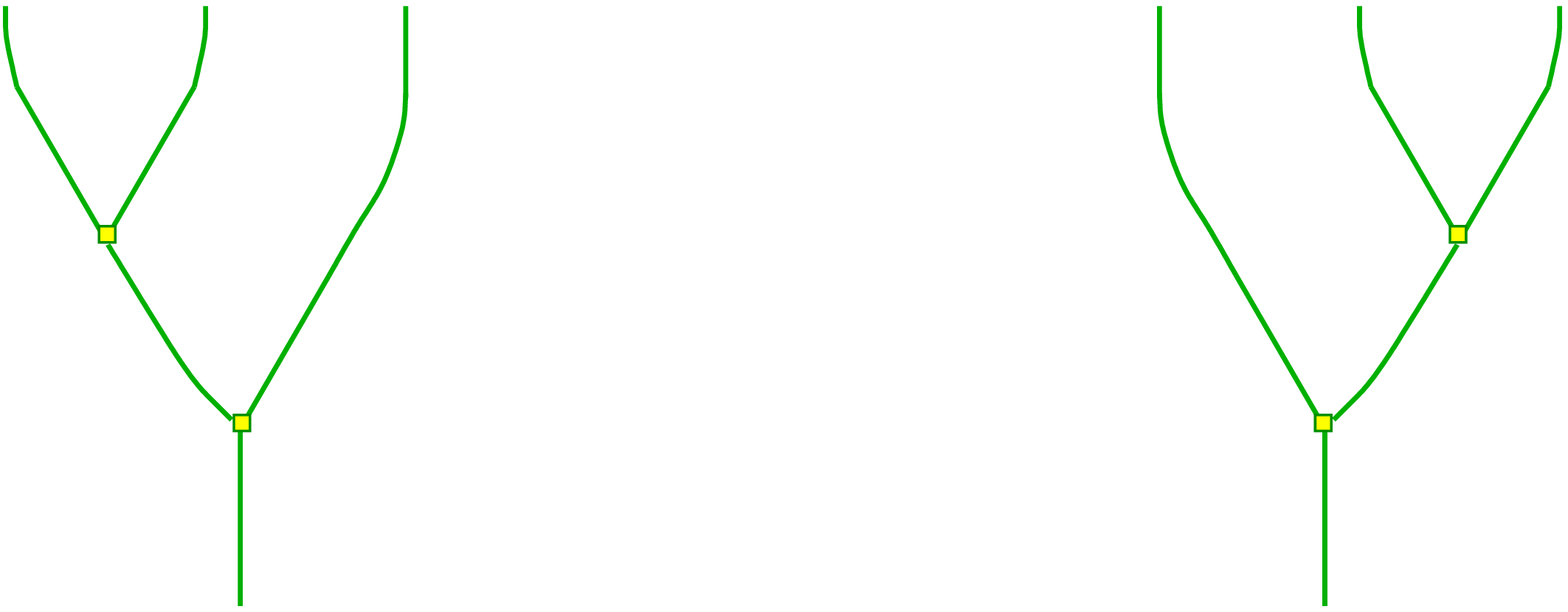}}
  \put(-0.5,93.5) {\pg i}
  \put(18.6,54.9) {\tiny$\bar\gamma$}
  \put(34.5,-8)   {\pg l}
  \put(17.3,38.9) {\pg q}
  \put(29.4,94.9) {\pg j}
  \put(38.8,25.7) {\tiny$\bar\delta$}
  \put(58.3,93.5) {\pg k}
  \put(74,42)    {$=~ \dsty\sum_{p\in\I}\sum_{\alpha,\beta}\; \F{i\, j}klpq\alpha\beta\gamma\delta$}
  \put(171.7,93.5){\pg i}
  \put(195.9,-8)  {\pg l}
  \put(200.1,25.7){\tiny$\bar\alpha$}
  \put(201.2,94.9){\pg j}
  \put(211.1,38.2){\pg p}
  \put(220.2,54.1){\tiny$\bar\beta$}
  \put(230.7,93.5){\pg k}
  }
The fusing matrices $\FF$ satisfy a pentagon identity, assuring that any two
different orderings of performing any multiple tensor product 
are related by a suitable sequence of fusing moves.
 
\medskip

To specify a standard basis of the space $B(V_1,...\,,V_m)$ of $m$-point chiral 
blocks on the sphere, with insertions labeled by objects $V_1,V_2,...\,,V_m$ of \C, 
we note that this space is isomorphic to the space $\Hom\one{V_1\oti V_2\,{\otimes}
\,{\cdots}\,{\otimes}\,V_m}$ of morphisms of \C. And in accordance with 
the comments about dominance made above we decompose the latter space as
  \be
  \hspace*{-.5em}\begin{array}r\dsty
  \Hom{\one}{V_1\oti V_2\,{\otimes}{\cdots}{\otimes}\,V_m}
  = \!\bigoplus_{p_0,p_1,p_2,...,p_{m-3}\in\I} \!\!\Big[\,
  \Hom{\one}{V_1\oti U_{p_0}} \otic \Hom{U_{p_0}}{V_2\oti U_{p_1}}
  \hspace*{1.7em}\\{}\\[-.8em]
  \otic \Hom{U_{p_1}}{V_3\oti U_{p_2}} \otic \cdots \otic 
  \Hom{U_{p_{m-4}}}{V_{m-2}\oti U_{p_{m-3}}}
  \hspace*{.8em}\\{}\\[-.7em]
  \otic \Hom{U_{p_{m-3}}}{V_{m-1}\oti V_m} \Big] .
  \eear\labl{Hom1,12m}
Taking the objects $V_k\eq U_{\ia_k}$ for $k\eq1,2,...\,,m$ to be simple, 
a basis of $B(U_{\ia_1},...,U_{\ia_m})$ corresponding to this decomposition 
is given by the elements \erf{npbasis}.


\subsection{Algebras in tensor categories}

Besides the category \C, the construction of a full CFT needs one other
piece of algebraic data: an object $A$ of \C\ endowed with the
structure of a symmetric special Frobenius algebra. Here by the term 
algebra one means a unital associative algebra; in complete analogy with
a conventional algebra in the category of vector spaces, an object of a
tensor category is an algebra iff there exist a multiplication morphism
$m\iN\Hom{A\oti A}A)$ and a unit morphism $\eta\iN\Hom\one A$ obeying 
associativity and unitality relations, respectively. An algebra is Frobenius 
iff it is also a coalgebra, with the coproduct being an isomorphism of
$A$-bimodules. For a Frobenius algebra $A$ there exist two distinguished 
isomorphisms between $A$ and its dual $A^\vee$; $A$ is symmetric iff these
two isomorphisms are equal. (Equivalently, a Frobenius algebra may be
characterized by the categorical analogue of a nondegenerate invariant
bilinear form; then the algebra is symmetric iff that form is symmetric,
see e.g.\ \cite{fuSt}.)
A Frobenius algebra is special iff the counit is a left-inverse of the unit 
and the product is a left-inverse of the coproduct. For more details about 
algebras in \C\ see e.g.\ section 3.1 of \cite{fuRs4}.  

Modules and bimodules over an algebra in a tensor category can again be
studied in full analogy with the classical case of algebras in the
category of vector spaces. In the case of a modular tensor category \C,
the braiding of \C\ can be used to obtain, for any object $U$ of \C, 
two $A$-bimodule structures $U\,{\otimes^\pm}A$ and $A\,{\otimes^\pm}\,U$
on the objects $U\oti A$ and $A\oti U$, respectively.  
For details see e.g.\ sections 4.1 and 5.4 of \cite{fuRs4}.

The dimensions
  \be
  \Z_{pq} := \dimc(\HomAA(U_p\otip A\otim U_q,A))
  \labl{ZZ}
of the morphism space $\HomAA(U_p\otip A\otim U_q,A)$ involving such specific 
bimodules give the entries of the matrix $\Z$ that describes the torus partition
function $Z$ in the standard basis of characters. Also, for such spaces of 
bimodule morphisms we have isomorphisms
  \be \bearl
  \Homaa{U_j\otip A\otim U_k}A \otic \Homaa{U_i\otip A\otim U_l}A
  \Nxl4 \qquad\dsty
  \stackrel\cong\longrightarrow \bigoplus_{q,q'\in\I} \Hom{U_i\oti U_j}{U_q} \otic
  \Hom{U_k\oti U_l}{U_{q'}} \otic \Homaa{U_q\otip A\otim U_{q'}}A \,.
  \eear \ee
We denote the coefficients of these isomorphisms with respect to a standard choice of 
bases by $\FA ijkl\alpha\beta\gamma {qq'}{\delta\delta'}$, i.e. 
  \eqpic{IV_67_for_A}{400}{45}{
  \put(0,0){     {\Includepicclal{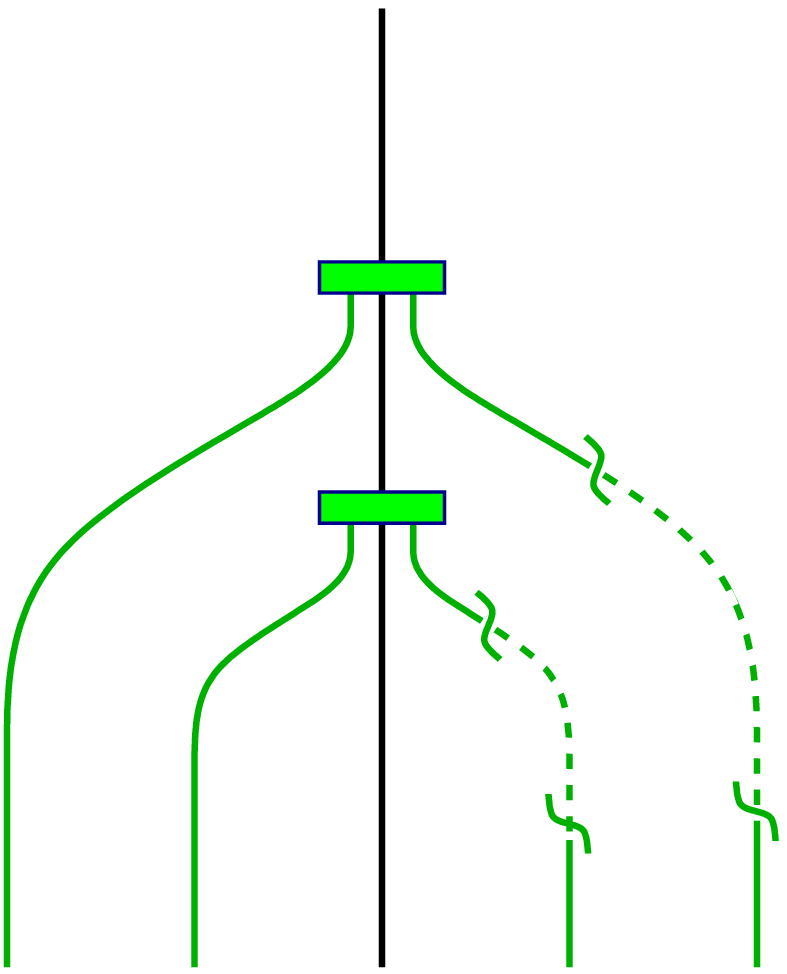}}
  \put(-0.9,-8.8){\pg i}
  \put(18.8,-8.8){\pg j}
  \put(37.5,-8.8){\pl A}
  \put(38.4,110) {\pl A}
  \put(49.8,49.6){\tiny$\alpha $} 
  \put(49.8,74.6){\tiny$\beta $}
  \put(59.9,-8.8){\pg k}
  \put(81.6,-8.8){\pg l}
 }
  \put(114,47)   {$ =~ \dsty\sum_{q,q'\in\I} \sum_{\gamma=1}^{\Z_{qq'}}
                 \sum_{\delta=1}^{\N ijq}\sum_{\delta'=1}^{\N ij{q'}}\,
                 \FA ijkl\alpha\beta\gamma {qq'}{\delta\delta'} $}
  \put(294,0){   {\Includepicclal{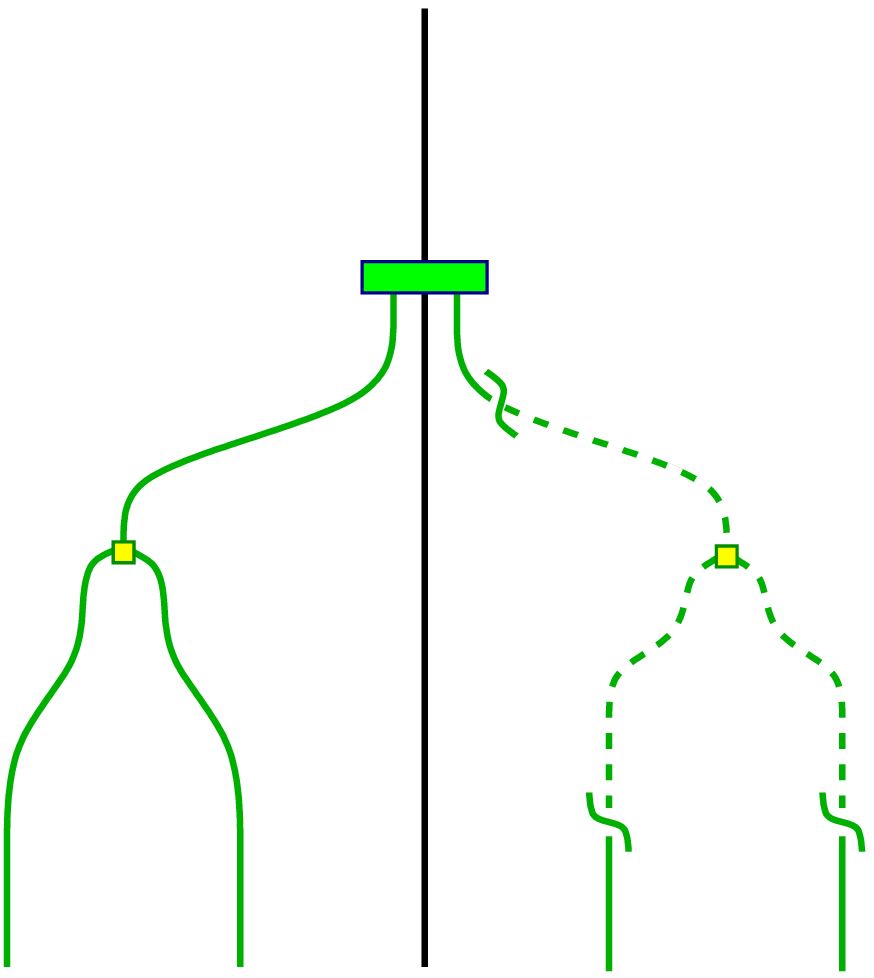}}
  \put(-0.8,-8.8){\pg i}
  \put(24.0,-8.8){\pg j}
  \put(24.6,63.2){\pg q}
  \put(42.8,-8.8){\pl A}
  \put(43.3,110) {\pl A}
  \put(54.8,75.9){\scriptsize$ \gamma $}
  \put(61.8,63.2){\pg {q'}}
  \put(64.6,-8.8){\pg k}
  \put(17.2,46.3){\tiny$\delta $}
  \put(82.5,46.3){\tiny$\delta' $}
  \put(91.3,-8.8){\pg l}
 } }
These coefficients are special cases (namely for all involved defect lines $X$ being
the trivial defect line $X_0\eq A$) of the entries of fusion matrices for defect fields,
see section 2.2 of \cite{fuRs10}.


\subsection{Three-dimensional topological field theory}

For studying rational chiral CFT, tools from three-dimensional topological 
quantum field theory (TFT) can be used. By definition \cite[chap.\,IV.7]{TUra},
a TFT \tftC\ functorially associates a finite-dimensional
vector space $\tftC(E)$ to any extended surface $E$, and a linear map
from $\tftC(E)$ to $\tftC(E')$ to any extended cobordism $\M\colon E\To E'$;
the image $\tftC(\M)$ is also called the \emph{invariant} of $\M$ and denoted 
by $Z(\M)$. The surfaces $E$ and cobordisms $\M$ form the objects and morphisms 
of the decorated cobordism category $\Cob_\C$. The modular tensor category \C\ 
provides \emph{decoration data} for $\Cob_\C$. In more detail, an extended
surface $E$ is, roughly, a compact closed oriented two-manifold with a finite 
set of embedded arcs, each arc being marked by an object of \C. An extended 
cobordism $E \To E'$ is a compact oriented three-manifold $\M$ satisfying
$\partial\M\eq(-E)\,{\sqcup}\,E'$, together with an oriented ribbon graph
\GammaM\ in $\M$ such that at each marked arc of $(-E)\,{\sqcup}\,E'$
a ribbon of \GammaM\ is ending. Each ribbon of \GammaM\ is labeled by an 
object of \C, while each coupon of \GammaM\ is
labeled by an element of the morphism space of \C\ that corresponds
to the objects of the ribbons which enter and leave the coupon.


\subsection{The TFT construction of full CFT}\label{app_tft}

The construction of full CFTs using tools from TFT, for brevity to be called the 
\emph{TFT construction}, has been accomplished in \cite{fuRs4,fuRs10,fjfrs}. 
Here we summarize some pertinent features of the construction. (We suppress 
various ingredients, such as the role of the orientation of
ribbons or of boundary components, and restrict our attention to oriented world 
sheets. For a few more details see section 4.3 of \cite{fnsw}; a comprehensive 
exposition can e.g.\ be found in appendix B of \cite{fjfrs}.)

While a full CFT is defined on a world sheet \X\ -- a surface with a conformal
structure, which may in particular have non-empty boundary -- the
corresponding chiral CFT is defined on the complex double \Xh\ of \X, 
i.e.\ the orientation bundle over \X\ mo\-du\-lo identification of
the two points in the fibre over each boundary point of \X. 
The double \Xh\ can be endowed with the structure of an extended surface. 
\X\ can be obtained from its double \Xh\ as the quotient by an 
orientation-reversing involution $\tau$. In particular, the double of the 
disk is the two-sphere, with $\tau$ 
acting, in standard complex coordinates, as $z\,{\mapsto}\,\overline z^{-1}$.

Further, when \X\ comes with field insertions, that is, embedded arcs
labeled either by objects of \C\ (for boundary fields, corresponding to arcs on 
$\partial\X$) or by pairs of objects of \C\ (for bulk fields, corresponding to
arcs in the interior of \X), then there are corresponding 
arcs on \Xh\ labeled by objects of \C.
A correlation function \CX\ of the full CFT on \X\ is a specific element in 
the resulting space of conformal blocks of the chiral CFT on the double \Xh. 
The TFT construction gives this element as
  \be
  \CX = \tftC(\MX)\,1 ~\in \tftC(\Xh)
  \label{fnsw:CX}\ee
with $1\iN\complex\eq\tftC(\emptyset)$.

Here $\MX\;{\equiv}\;\emptyset {\stackrel\MX\longrightarrow}\Xh$, called the 
{\em connecting manifold\/} for the world sheet \X, is an extended cobordism 
constructed from the data of \X, the category \C, and a symmetric special 
Frobenius algebra $A$ in \C. \MX\ is assembled as follows.
\def\leftmargini{1.12em}~\\[-1.8em]\begin{itemize}\addtolength\itemsep{-4pt}
\itx
As an oriented three-manifold, \MX\ is the interval bundle over \X\ modulo a 
$\mathbb{Z}_2$-identification of the intervals over $\partial\X$: 
$\MX \eq \big( \Xh\Times[-1,1] \big) /{\sim}$ with $([x,\orr],t)
\,{\sim}\,([x,-\orr],-t)$, where $\orr$ denotes the local
2-orientation. It follows in particular that $\partial\MX\eq\Xh$ and that \X\ 
is naturally embedded in \MX\ as $\imbed\colon \X\,{\stackrel\simeq\to}\,
\X\Times\{t{=}0\}\, {\hookrightarrow}\,\MX$. 
The fiber in \MX\ over a point of \X\ is called a \emph{connecting interval}.
\\
Compare picture \erf{Sphere_CI}, in which 
the connecting intervals are displayed for the case of the disk.
\itx
Contained in \MX\ there is a the ribbon graph $\Gamma_{\!\!\MX}\,{\subset}\,\MX$.
A crucial part of $\Gamma_{\!\!\MX}$ is a (dual) oriented triangulation $\Gamma$ 
of $\imbed(\X)$. The edges and vertices of $\Gamma\,{\setminus}\,\imbed(\partial\X)$ 
are labeled by objects and morphisms of \C, respectively, and it is here that
the Frobenius algebra $A$ enters: an edge of $\Gamma\,{\setminus}\,
\imbed(\partial\X)$ is covered with a ribbon labeled by the object $A$, 
while each (three-valent) vertex is covered with a coupon labeled by the
multiplication morphism $m$ of $A$.\,%
  \footnote{~Standing alone, these assignments can be in conflict with the 
  orientations of some of the edges of $\Gamma$. This conflict is resolved by
  inserting along each such edge a coupon with a suitable morphism, in either
  $\hom(A\oti A,\one)$ or $\hom(\one,A\oti A)$, coming from the invariant
  bilinear form of $A$.}
\\
That \CX\ does not depend on the choice of triangulation $\Gamma$ is
equivalent to the assertion that the object $A$ carries the structure
of a symmetric special Frobenius algebra.
\itx
If \X\ has non-empty boundary, each edge $e$ of $\Gamma\,{\cap}\,\imbed
(\partial\X)$ is covered with a ribbon labeled by a left $A$-module $N \eq N(e)$,
while each vertex of $\Gamma$ lying on $\imbed(\partial\X)$ is covered with a 
coupon labeled by the representation morphism $\rho_N \iN\hom(A\oti N,N)$. The 
$A$-module $N$ formalizes the notion of the boundary condition associated to a 
segment of $\partial\X$. 
\itx
Bulk and boundary fields are implemented by coupons lying in the interior,
respectively covering a boundary arc, of $\imbed(\X)$, which are
labeled by appropriate morphisms. These are described in more detail in the
next subsection.
\end{itemize}

In the picture \erf{n_point_cor} the cobordism resulting from this description
is shown for the case of a disk with $n$ bulk field insertions.

The correlators \erf{fnsw:CX} obtained by the prescription above
satisfy \cite{fjfrs} all consistency constraints. Conversely, for a CFT obeying 
some natural non-degeneracy assumptions, any consistent set of correlators can 
be obtained through the TFT construction \cite{fjfrs2}.

\medskip

When displaying ribbon graphs arising in the TFT construction, for simplicity 
we draw ribbons just as lines or, rather, framed lines, with blackboard framing
being implicit. Occasionally the ribbon graphs also involve half-twists of ribbons,
which change the visible side of a ribbon from its `white' to its `black' side or 
vice versa. In blackboard framing, the white and black sides of a ribbon are 
indicated by a solid and a dashed line, respectively; see e.g.\ the picture 
\erf{IV_67_for_A} above. 

In the process of calculating invariants of ribbon graphs, one projects the 
graphs in a non-singular manner to a plane and interprets the resulting pictures
as morphisms of the category \C. These resulting morphisms can in particular
involve the duality, braiding and twist morphisms of \C; for these we employ
the pictorial conventions summarized e.g.\ in appendix A.1 of \cite{fjfrs}.


\subsection{Bulk and boundary fields}\label{app_fields}

As explained in detail in section 3.2.1 of \cite{fuRs10}), a boundary field insertion
$\Psi$ can be characterized as 
  \be
  \Psi = (M,N,V,\psi,p,[\gamma])
  \ee
by the following data: two boundary conditions (i.e.\ left $A$-modules) $M$ and $N$, 
an object $V$ of the category \C\ of chiral data, a morphism 
$\psi$ in the space $\HomA(M\oti V,N)$ of module morphisms, a point $p$ on
the boundary $\partial\X$ of the world sheet, and finally a germ $[\gamma]$ 
of arcs aligned to $\partial \X$ at $\gamma(0)\eq p$.
Also, for every component of $\partial \X$ an orientation must be specified.
All correlators can be obtained from a set of basic correlators for which
$V\eq U_i$ is a simple object and $\psi\eq\psi_\alpha$ is a basis element of
the morphism space $\HomA(M\oti U_i,N)$.

Similarly, as described in e.g.\ section 3.3.1 of \cite{fuRs10} (and in accordance
with formula \erf{ZZ}) a bulk field insertion
  \be
  \Phi = (V,V',\phi,p,[\gamma_\phi],\orr)
  \labl{bulkfield}
is determined by the following data: two objects $V$ and $V'$ of \C, a bimodule
morphism $\phi$ in $\HomAA(V\oT+A\ot-V',A)$, a point $p$ in the interior of \X, 
a germ $[\gamma]$ of arcs at $p$, and a local orientation $\orr$ of \X\ around 
$p$. One can obtain all correlators from those which involve bulk fields for
which $V\eq U_i$ and $V'\eq U_j$ are simple objects and $\phi\eq\phi_\alpha$
is a basis element of the bimodule morphism space $\HomAA(U_i\oT+A\ot-U_j,A)$.

In addition, both for boundary fields and for bulk fields one must divide out
a certain equivalence relation (see definitions 3.3 and 3.6 of \cite{fuRs10}).
In the main text the insertion points, arc-germs and local orientations are 
suppressed, as they do not play a direct role in our considerations. Moreover, 
to reduce the notational burden, we usually suppress the boundary labels 
$M,N$ as well as the chiral labels $i,j,...\iN\I$ when they can be recovered from 
the rest of the information. In particular we write
  \be
  \Psi_\alpha^{} \equiv \Psi_\alpha^{MN;j} = (M,N,U_j,\psi_\alpha)  \qquand
  \Phi_\alpha^{} \equiv \Phi_\alpha^{ij} = (U_i,U_j,\phi_\alpha)
  \ee
for boundary and bulk fields, respectively.


\subsection{Two-point functions}

The standard two-point correlation function of boundary fields on the unit
disk with boundary segments carrying boundary conditions $M$ and $N$,
as defined e.g.\ in formula (2.22) of \cite{fjfrs}, is an element in the
space of two-point chiral blocks on the sphere. We consider the case that
the chiral insertions are given by simple objects and, correspondingly,
select chiral labels $k$ and $l\eq \bar k$ for the boundary fields, so that
this space of blocks is one-dimensional (for $l\,{\ne}\,\bar k$ the space is
zero). For given boundary fields, the coefficient of the two-point function
with respect to a basis in the \onedim\ space of blocks is denoted by
$c^{\rm bnd}_{M,N,k;\psi^+,\psi^-}$,
where the morphisms $\psi^\pm$ specify the boundary fields in question.
Choosing bases $\{ \psi^+_\alpha \}$ and $\{ \psi^-_\beta \}$ in the spaces
$\HomA(M\oti U_k,N)$ and $\HomA(N\oti U_{\bar k},M)$ of boundary fields with
fixed chiral labels, these numbers are arranged into a square matrix with entries
  \be
  {(c^{\rm bnd}_{M,N,k})}^{}_{\alpha\beta}
  := c^{\rm bnd}_{M,N,k;\psi^+_\alpha,\psi^-_\beta} \,.
  \labl{bnd_2pfu}
As shown in appendix C.1 of \cite{fjfrs}, for any fixed values of $M$, $N$
and $k$ this matrix furnishes a non-degenerate bilinear form
$\,\HomA(M\oti U_k,N) \Times \HomA(N\oti U_\kb,M) \to \Hom{U_k\oti U_\kb}\one$.

In the particular case that $M\eq N$ is a simple $A$-module and $k\eq0$, the
space $\HomA(M\otimes U_0, M)$ has the unit constraint as a canonical basis.
In this basis, we denote the value of the bilinear form by
  \be
  ({c^{\text{bnd}}_{M,M,0}})\Inv_{\;\nl\nl} =: c^{\text{bnd}}_{M,0} \,.
  \labl{A13}
As one sees directly from formula (C.3) of \cite{fjfrs},  we have
  \be
  c^{\text{bnd}}_{M,0} = \dim(M) \,.
  \labl{A14}

\medskip

The standard two-point function of bulk fields on the sphere, as defined e.g.\
in formula (2.41) of \cite{fjfrs}, is an element in the tensor product of two
copies of the space of two-point chiral blocks on the sphere. We assume that
the chiral labels of the two bulk fields $\Phi$ and $\Phi'$ are
conjugate to each other, say $\Phi\eq\Phi_{\ia\ja}$, $\Phi'\eq\Phi_{\ib\jb}$,
in which case this space is \onedim\ (while it is zero otherwise). With this
choice of chiral labels, the coefficient of the two-point function with respect
to a basis of the space of blocks is denoted by $\cbulk_{i,j,\phi,\phi'}$.
Selecting bases $\{ \phi^{}_\alpha \}$ of $\HomAA(U_i \oT+ A \ot- U_j, A)$ and 
$\{\phi'_\beta\}$ of $\HomAA(U_\ib \oT+ A \ot- U_\jb,A)$ for the relevant spaces 
of bimodule morphisms, the coefficients arrange into a matrix $\cbulk_{i,j}$ by
  \be
  \cbulk_{i,j, \phi^{}_\alpha,\phi'_\beta} =:
  {(\cbulk_{i,j})}^{}_{\alpha\beta} \,.
  \labl{bulk_2pfu}
For fixed $i,\,j$ this furnishes a non-degenerate bilinear form
$\,\HomAA(U_i \oT+ A \ot- U_j, A) \Times \HomAA
   $\linebreak[0]$  
(U_\ib \oT+ A \ot- U_\jb,A) \to \Hom{U_\ia\oti U_\ib}\one
\otic \Hom{U_\ja\oti U_\jb}\one$, see appendix C.2 of \cite{fjfrs}.

A graphical description of the numbers ${(\cbulk_{i,j})}^{}_{\alpha\beta}$ is
given in formula (C.14) of \cite{fjfrs}. This actually constitutes one instance
in which the coefficients \FFA\ introduced in \erf{IV_67_for_A} occur naturally.
Namely, by performing an \FFA-move on the expression (C.14) of \cite{fjfrs}
one finds
  \be
  c^{\text{bulk}}_{q\qb,\alpha\beta}
  = \frac{\dim(A)}{S_{0,0}}\, \FA q\qb q\qb\beta\alpha\nl{00}{\nl\nl}.
  \labl{tpf_Faa}
Analogously as for boundary fields, there exists a canonical basis in the
case that the bulk field is the tensor unit, i.e.\ $q\eq0$. In this basis we have
  \be
  c^{\text{bulk}}_{00} = \frac{\dim(A)}{S_{0,0}}\,.
  \labl{tpf_00}

\newpage

 \newcommand\wb{\,\linebreak[0]} \def\wB {$\,$\wb}
 \newcommand\Bi[1]    {\bibitem{#1}}
 \newcommand\Erra[3]  {\,[{\em ibid.}\ {#1} ({#2}) {#3}, {\em Erratum}]}
 \newcommand\J[5]     {{\em #5}, {#1} {#2} ({#3}) {#4} }
 \newcommand\K[6]     {{\em #6}, {#1} {#2} ({#3}) {#4} {\tt[#5]} }
 \newcommand\Prep[2]  {{\em #2}, preprint {\tt #1} }
 \newcommand\PhD[2]   {{\em #2}, Ph.D.\ thesis (#1)}
 \newcommand\BOOK[4]  {{\em #1\/} ({#2}, {#3} {#4})}
 \newcommand\inBO[7]  {{\em #7}, in:\ {\em #1}, {#2}\ ({#3}, {#4} {#5}), p.\ {#6}}
 \def\adma  {Adv.\wb in Math.}
 \newcommand\ahxxxi[2] {\inBO{Theory of Elementary Particles} { H.\ Dorn et al.,
            eds.} {Wiley-VCH}{Berlin}{1998} {{#1}}{{#2}}}
 \def\ajse  {Arabian Journal for Science and Engineering} 
 \def\atmp  {Adv.\wb Theor.\wb Math.\wb Phys.}
 \def\comp  {Com\-mun.\wb Math.\wb Phys.}
 \def\inma  {Invent.\wb math.}
 \def\jgap  {J.\wb Geom.\wB and\wB Phys.}
 \def\jhep  {J.\wb High\wB Energy\wB Phys.}
 \def\jopa  {J.\wb Phys.\ A} 
 \def\nupb  {Nucl.\wb Phys.\ B}
 \def\phlb  {Phys.\wb Lett.\ B}
 \def\phrl  {Phys.\wb Rev.\wb Lett.}
 \def\taac  {Theo\-ry\wB and\wB Appl.\wb Cat.}
 \def\trgr  {Transform.\wb Groups}

\end{document}